\documentclass[aps,reprint,superscriptaddress,floatfix,amsfonts,amssymb,amsmath]{revtex4-1}
\usepackage[english]{babel}
\usepackage{mathptmx}
\usepackage{xcolor}
\usepackage[utf8]{inputenc}
\usepackage[T1]{fontenc}
\usepackage{graphicx}
\usepackage{grffile}
\usepackage{longtable}
\usepackage{multirow}
\usepackage{array}
\newcolumntype{P}[1]{>{\raggedright\arraybackslash}p{#1}}
\usepackage{wrapfig}
\usepackage{rotating}
\usepackage[normalem]{ulem}
\usepackage{amsmath}
\usepackage{textcomp}
\usepackage{amssymb}
\usepackage{capt-of}
\usepackage{hyperref}
\usepackage{booktabs}
\usepackage{float}
\usepackage{bm}

\newcommand{\Tr}{\ensuremath{\operatorname{Tr}}}
\newcommand\ket[1]{\ensuremath{|#1\rangle}}
\newcommand\bra[1]{\ensuremath{\langle #1|}}
\newcommand\ketbra[2]{\ensuremath{|#1\rangle\langle #2|}}

\newcommand\mrm[1]{\mathrm{#1}}

\makeatletter
\newcommand{\closure}[2][3]{%
{}\mkern#1mu\overline{\mkern-#1mu#2}}
\newcommand*{\textoverline}[1]{\ensuremath{\closure[1]{\hbox{#1}}\m@th}}
\newcommand*{\ov}[1]{\textoverline{#1}}
\makeatother

\date{\today}
\hypersetup{colorlinks,linkcolor={red!50!black},citecolor={blue!50!black},urlcolor={blue!80!black}}
\begin{document}
\title{Theory of rotationally-resolved two-dimensional infrared spectroscopy including polarization dependence and rotational coherence dynamics}
\author{Grzegorz Kowzan}
\email[Electronic address: ]{gkowzan@umk.pl}
\affiliation{Institute of Physics, Faculty of Physics, Astronomy and Informatics, Nicolaus Copernicus University in Toru\'{n}, ul. Grudzi\k{a}dzka 5, 87-100 Toru\'{n}, Poland}
\affiliation{Department of Chemistry, Stony Brook University, Stony Brook, NY 11790-3400, USA}
\author{Thomas K. Allison}
\affiliation{Department of Chemistry, Stony Brook University, Stony Brook, NY 11790-3400, USA}
\affiliation{Department of Physics and Astronomy, Stony Brook University, Stony Brook, NY 11790-3400, USA}

\begin{abstract}
  
Two-dimensional infrared (2DIR) spectroscopy is widely used to study molecular dynamics but it is typically restricted to solid and liquid phase samples and modest spectral resolution.
Only recently, its potential to study gas-phase dynamics is beginning to be realized.
Moreover, the recently proposed technique of cavity-enhanced 2D spectroscopy using frequency combs and developments in multi-comb spectroscopy are expected to dramatically advance capabilities for acquisition of rotationally-resolved 2DIR spectra. This demonstrates the need for rigorous and quantitative treatment of rotationally-resolved, polarization-dependent third-order response of gas-phase samples.
In this article, we provide a rigorous and quantitative description of rotationally-resolved 2DIR spectroscopy using density-matrix, time-dependent perturbation theory and angular momentum algebra techniques.
We describe the band and branch structure of 2D spectra, decompose the molecular response into polarization-dependence classes, use this decomposition to derive and explain special polarization conditions and relate the liquid-phase polarization conditions to gas-phase ones.
Furthermore, we discuss the rotational coherence dynamics during waiting time.
\end{abstract}
\maketitle

\section{Introduction}
\label{sec:intro}
Two-dimensional infrared (2DIR) spectroscopy is a nonlinear all-optical technique using ultrashort broadband pulses to study dynamics of molecules after photoexcitation~\cite{Cho2008,Ogilvie2009,Elsaesser2009,Kim2009,Hamm2011a}.
In a common experimental arrangement, three ultrashort pulses ($<$1 ps) are made to interact with the sample in sequence, and the molecular response is recorded as a function of delays between pulses ($t_{1}$, $t_{2}$) and of the spectrum of the third pulse ($\omega_{3}$).
Fourier-transforming the response along $t_{1}$ axis produces 2D correlation maps between the pump excitation ($\omega_{1}$) and the probe interaction ($\omega_{3}$).
Conventional 2DIR spectrometers provide only modest spectral resolution ($>$1 cm$^{-1}$), as they are commonly used to study liquid and solid state samples with broad spectral features and large optical densities.

Recent work has begun to explore 2DIR spectroscopy of gas-phase molecules and the resulting rotational structure.
Mandal \textit{et al.}~\cite{Mandal2018, Pack2019} measured 2DIR gas-phase spectra of N$_{2}$O in SF$_{6}$ to study gas-to-liquid transition in supercritical fluids.
Due to high pressure of the sample and insufficient experimental resolution, individual rovibrational resonances were not resolved.
The presented theoretical description did not consider individual third-order rovibrational pathways separately, instead it modeled whole branches as single excitations with individual lines treated as perturbations of the central branch frequencies.
Nonetheless, this study clearly demonstrated the notable differences between liquid-phase and gas-phase 2DIR molecular response, in particular the correlation between P-branch excitation and R-branch detection and \textit{vice versa}.
Very recently, Gronborg \textit{et al.}~\cite{Gronborg2022} presented 2DIR gas-phase spectra of optically thick pure CO$_{2}$ at atmospheric pressure using a standard 2DIR spectrometer operated in pump-probe geometry.
Here, individual rotationally-resolved resonances (RR2DIR) were observed but the resolution was still insufficient to record the details of their line shapes.

This previous pioneering experimental work has used conventional ultrafast optical technology, however recent advances in frequency comb-based methods promise to dramatically advance the capabilities for acquisition of RR2DIR spectra. 
For example, Allison and co-workers have used frequency comb techniques to enhance ultrafast transient absorption spectroscopy (a third-order response) in dilute gases with detection limits approximately 4 orders of magnitude lower than conventional methods~\cite{Reber2016, Silfies2021}.
Allison also framed the more general coherent 2D spectroscopy in terms of wave mixing of multiple frequency combs and described methods for cavity-enhancing 2D spectroscopy signals~\cite{Allison2017}.
Lomsadze and Cundiff have demonstrated rapid, high-resolution coherent multidimensional spectroscopy in optically thick Rubidium vapors with multiple frequency combs~\cite{Lomsadze2018a}. In parallel with these technique developments, there has been rapid progress in the bandwidth and power of mid-IR and long-wave IR frequency comb light sources \cite{Gaida2018b,Catanese2020,Xing2021,Ru2021,Lesko2021,Nakamura2022}.

Sensitive, broadband, high-resolution ($<$0.1 cm$^{-1}$) RR2DIR spectroscopy in the fingerprint region ($\lambda = 3$--$20$ $\mu$m) will enable analysis of complex mixtures of polyatomic gases with unprecedented specificity. 
As is well known from conventional 2DIR spectroscopy~\cite{Hamm2011a} and 2D NMR spectroscopy~\cite{Giraudeau2017}, adding another dimension to molecular spectra enables easier discrimination of individual molecular resonances, simplifying detection of constituent species in complex mixtures.
Such mixtures commonly occur in human breath~\cite{Henderson2018a}, in flames, and in detection of explosives and narcotics, but their linear spectra are highly congested and difficult to interpret.
The new spectroscopic capabilities will also benefit fundamental chemical physics studies on problems such as intramolecular vibrational redistribution~\cite{Nesbitt1996} and collisional processes in gases~\cite{Chen2007,Hartmann2018a,Pack2019}.
For example, observing intensities of off-diagonal resonances that emerge as $t_{2}$ increases will enable accurate measurements of collisional population and coherence transfer~\footnote{By collisional coherence transfer we mean collision-induced transitions between density matrix states $|\alpha\rangle\langle\beta|\leftrightarrow|\alpha'\rangle\langle\beta'|$, where $\alpha\neq\beta$ and $\alpha'\neq\beta'$}, which is responsible for line mixing in linear absorption spectra, and tracking the evolution of diagonal and anti-diagonal line widths will directly report on the effect on velocity-changing collisions. 
With a large information density of rotationally resolved 2DIR (RR2DIR) spectra, there are likely many unforeseen applications as well.

To illustrate the difference between liquid-phase and gas-phase dynamics, let us compare the orientational dynamics in these two environments.
Considerations of molecular orientation in liquid-phase 2D spectroscopy usually begin with a classical propagator for rotational diffusion written in terms of spherical harmonics $Y_{lm}$ \cite{Hamm2011a,Berne1975}
\begin{equation}
  \label{eq:1}
  G(\Omega_j t_{j}|\Omega_{i}) = \sum_{lm} e^{-l(l+1)Dt_{j}} Y_{lm}^{\ast}(\Omega_j)Y_{lm}(\Omega_i).
\end{equation}
The propagator (Green's function) describes the distribution of molecules with respect to orientation $\Omega_j = (\theta_j, \phi_j) $ after initial localization at an orientation $\Omega_i$, with $G(\Omega_j, t=0 | \Omega_i) = \delta(\Omega_j - \Omega_i)$. 
In a third-order spectroscopy experiment, each interaction with the pulse ($\epsilon_{i}\cdot\hat{\mu}$) excites and aligns the sample along the polarization axis, with diffusive evolution of the distribution described by the propagator in between interactions.
The dynamics are terminated by polarized detection of the third-order polarization ($\epsilon_{4}\cdot\hat{\mu}$).
Denoting polarization unit vectors as $\hat{\epsilon}_i$, $i=1,2,3$ for pulses and $i=4$ for detection, the ensemble-averaged orientational part of the response is given by:
\begin{equation}
  \label{eq:2}
  \begin{split}
    \langle (\hat{\epsilon}_{4}&\cdot\hat{\mu} )(\hat{\epsilon}_{3}\cdot\hat{\mu} )(\hat{\epsilon}_{2}\cdot\hat{\mu} )(\hat{\epsilon}_{1}\cdot\hat{\mu} )\rangle =\\
    \int & d\Omega_1 \int d\Omega_2 \int d\Omega_3 \int d\Omega_4\; 
    (\hat{\epsilon}_{4}\cdot\hat{\mu} ) G(\Omega_{3}t_{3}|\Omega_{2})(\hat{\epsilon}_{3}\cdot\hat{\mu}) \\
    &\times G(\Omega_{2}t_{2}|\Omega_{1}) (\hat{\epsilon}_{2}\cdot\hat{\mu} )  G(\Omega_{1}t_{1}|\Omega_{0}) (\hat{\epsilon}_{1}\cdot\hat{\mu} ) p_{0}(\Omega_{0})
  \end{split}
\end{equation}
where $p_{0}(\Omega_{0})$ is an initially isotropic distribution. Hochstrasser~\cite{Hochstrasser2001} used this formalism to derive expressions for the polarization dependence of third-order spectroscopy signals produced by correlated vibrational transition dipoles in liquids.
These results are widely used to simplify 2DIR spectra, for example by removing diagonal peaks~\cite{Zanni2001} involving a single vibrational mode to emphasize off-diagonal peaks due to the coupling of different vibrational modes.

In the gas phase, the situation is drastically different, since orientational dynamics are dominated by coherent evolution of quantum rotors unperturbed by collisions.
A quantum expression analogous to Eq.~\eqref{eq:2} uses the unitary time evolution operator $U_0(t) = e^{-iH_0 t/\hbar}$ instead of the classical propagator and an initial density matrix $\rho^{(0)}(-\infty)$, instead of $p_{0}(\Omega_{0})$, viz.
\begin{equation}
  \label{eq:3}
  G(\Omega_{1}t_{1}|\Omega_{0})p_{0}(\Omega_{0})  \to e^{-\frac{i}{\hbar}H_{0}t_{1}} \rho^{(0)}(-\infty) e^{\frac{i}{\hbar}H_{0}t_{1}}.
\end{equation}
As a consequence, the effect of rotational dynamics on 2DIR lineshapes (or branch structures in the case of rotationally-resolved spectra) are very different than in the liquid phase. 
Furthermore, the standard liquid-phase polarization conditions are no longer as useful and instead new ones, unique to the gas phase, bring more clarity to the 2D spectra. 

Here we provide a comprehensive treatment of the theory of rotationally-resolved two-dimensional infrared spectroscopy.
Our work builds on previous work studying state-resolved four-wave mixing in gases~\cite{Williams1994,Williams1994a,Williams1995,Williams1996,Williams1997,Wasserman1998,Bracamonte2003,Murdock2009,Wells2015}, but is more general in several ways, including the consideration of rotationally coherent pathways (Feynman diagrams with rotational coherence during the waiting time between the second and third pulses) and the discovery of new polarization conditions for suppressing branches of the 2DIR spectrum.
In this longer article, we aim to provide a comprehensive guide to the theory of rotationally-resolved 2DIR spectroscopy with many example signals.
In a shorter article~\cite{Kowzan2022b}, we present three new polarization conditions and demonstrate their power with simulations of multi-isotopologue spectra of methyl chloride.
Both papers are accompanied by a software package~\cite{Kowzan2022rotsim2d} which can be used to simulate rotationally-resolved 2DIR spectra.
In both papers, we restrict the discussion to third-order excitations occurring within a single vibrational mode in order to focus on the unique aspects of the spectrum produced by rotational-state resolution.
However, our theory is easily generalized to spectra involving multiple vibrational modes with corresponding cross peaks, as often considered in solution-phase 2DIR spectroscopy.

The paper is structured as follows: Section~\ref{sec:general-framework} presents the theoretical framework of time-dependent perturbation theory.
Section~\ref{sec:simulation} describes the general features and branch structure of 2D IR spectra.
Section ~\ref{sec:fourfold} describes the polarization dependence of RR2DIR spectroscopy signals while section~\ref{sec:suppression} applies the presented theory to controlling molecular response with polarization.
Section~\ref{sec:interstate-coherences} describes rotational coherences and time-dependent interference between pathways after excitation with two broadband pulses and finally section~\ref{sec:conclusions} concludes the article.
Appendix~\ref{app:fourfold} provides details on spherical tensor decomposition of the four-fold dipole operator, appendix~\ref{app:comp-polar-tens} gives explicit formulas for the components of the polarization tensor, and appendix~\ref{app:magic-angle} discusses the magic angle and population-alignment canceling angle conditions in more detail.
Appendix~\ref{sec:example} gives an example calculation of the R-factor and compares the results to direct evaluation of the four-fold dipole operator for a particular Feynman pathway.

\section{Macroscopic polarization and time-dependent perturbation theory}
\label{sec:general-framework}

The main purpose of this section is to relate the source term in electromagnetic wave equation, the macroscopic third-order polarization $\vec{P}^{(3)}(t)$, to microscopic light-matter interaction and to establish the experimental conditions and approximations within which our results are applicable.
Throughout the article, we exclusively use SI units.
The third-order polarization is related to the incident field by:
\begin{multline}
  \label{eq:4}
  \vec{P}^{(3)}(\vec{r},t) = \int_{0}^{\infty} \mrm{d}t_{3}\int_{0}^{\infty} \mrm{d}t_{2}\int_{0}^{\infty} \mrm{d}t_{1} \;
  \mathbf{R}(t_{3}, t_{2}, t_{1}) :  \big[\vec{E}(\vec{r},t-t_{3})\\ \vec{E}(\vec{r},t-t_{3}-t_{2}) \vec{E}(\vec{r},t-t_{3}-t_{2}-t_{1})\big],
\end{multline}
where $\mathbf{R}(t_{3}, t_{2}, t_{1})$ is the third order nonlinear response function, which is a fourth order tensor, and ``$:$'' denotes tensor contraction---here, threefold contraction with electric field terms.
The incident field is assumed here to take the form of three optical pulses:
\begin{align}
  \label{eq:5}
  \vec{E}(\vec{r},t) &= \frac{1}{2} \sum_{i=1}^{3} \hat{\epsilon}_{i} E_{i}(\vec{r}, t) + \mrm{c.c.},\\
  E_{i}(\vec{r}, t) &= \mathcal{E}_{i}(t-\tau_{i}) e^{i[\vec{k}_{i}\cdot\vec{r}-\omega_{i} (t-\tau_{i})]}
\end{align}
where $\mrm{c.c.}$ is the complex conjugate of the preceding term, $\hat{\epsilon}_{i}$ is unit polarization vector and $\mathcal{E}_{i}(t-
\tau_{i})$ is a slowly varying envelope of the pulse centered at $\tau_{i}$.
The exponential term specifies the carrier frequency, $\omega_{i}$, and the angular wavevector, $\vec{k}_{i}$, which determines the propagation direction.
We limit our analysis to linearly polarized beams, expressed in terms of the polarization vector:
\begin{equation}
  \label{eq:6}
  \hat{\epsilon}_{i} = \cos \theta_{i} \hat{x} + \sin \theta_{i} \hat{y},
\end{equation}
where $\theta_{i}$ is the linear polarization angle.  The presented formalism can be extended to elliptically polarized beams by considering a more general expression for the polarization vector:
\begin{equation}
  \label{eq:elliptical}
  \hat{\epsilon}_{i} = \cos \theta_{i} \hat{x} + e^{i\delta_i}\sin \theta_{i} \hat{y}.
\end{equation}

Substituting Eq.~\eqref{eq:5} into Eq.~\eqref{eq:4} produces $6^{3}$ terms, corresponding to three-fold product of six electric field terms from Eq.~\eqref{eq:5}.
Here, we're only considering $2^{3}$ terms corresponding to the sample interacting with $E_{1}$, $E_{2}$ and $E_{3}$ in a time-ordered sequence.
Picking one of these terms, we obtain:
\begin{equation}
  \label{eq:7}
  \begin{split}
  \vec{P}^{(3)}&(\vec{r}, t) = \frac{1}{8} e^{i(\vec{k}_{s}\cdot\vec{r}-\omega_{s}t)}
  e^{i(\kappa_{1}\omega_{1}\tau_{1} + \kappa_{2}\omega_{2}\tau_{2} + \kappa_{3}\omega_{3}\tau_{3})}\\
  \times \iiint_{0}^{\infty}& \mrm{d}t_{3}\, \mrm{d}t_{2}\, \mrm{d}t_{1}\,
  \mathbf{R}(t_{3}, t_{2}, t_{1}) : \widetilde{\epsilon}_{3} \widetilde{\epsilon}_{2}\widetilde{\epsilon}_{1} \widetilde{\mathcal{E}}_{3}(t-t_{3}-\tau_{3}) \\
  &\widetilde{\mathcal{E}}_{2}(t-t_{3}-t_{2}-\tau_{2}) \widetilde{\mathcal{E}}_{1}(t-t_{3}-t_{2}-t_{1}-\tau_{1})\\
  &e^{i(\kappa_{3}\omega_{3}+\kappa_{2}\omega_{2}+\kappa_{1}\omega_{1})t_{3}} e^{i(\kappa_{2}\omega_{2}+\kappa_{1}\omega_{1})t_{2}} e^{i\kappa_{1}\omega_{1}t_{1}}.
  \end{split}
\end{equation}
The $\kappa_{i}$ terms in the exponents are equal to $\pm 1$ and select either positive or negative frequency components of fields from Eq.~\eqref{eq:5}.
Tilded polarization vectors and pulse envelopes, $\widetilde{\epsilon}_{i}$ and $\widetilde{\mathcal{E}}_{i}(t)$, implicitly depend on $\kappa_{i}$ and they are either equal to $\hat{\epsilon}_{i}$ [$\mathcal{E}_{i}(t)$] for $\kappa_{i}=+1$ or to $\hat{\epsilon}^{\ast}_{i}$ [$\mathcal{E}^{\ast}_{i}$(t)] for $\kappa_{i}=-1$.
The wavevector and frequency of molecular polarization are defined as:
\begin{align}
  \label{eq:8}
  \vec{k}_{s} &= \kappa_{1}\vec{k}_{1}+\kappa_{2}\vec{k}_{2}+\kappa_{3}\vec{k}_{3},\\
  \omega_{s} &= \kappa_{1}\omega_{1}+\kappa_{2}\omega_{2}+\kappa_{3}\omega_{3}.
\end{align}
Following previous work~\cite{Scheurer2001,Khalil2003}, we label the directions associated with $\boldsymbol{\kappa}=(\kappa_{1}, \kappa_{2}, \kappa_{3})=(-1, 1, 1)$ as $S_{I}$ (rephasing), with $(1, -1, 1)$ as $S_{II}$ (nonrephasing) and with $(1, 1, -1)$ as $S_{III}$ (double quantum).

The nonlinear response function for dipole interaction can be compactly written as \cite{Hamm2011a}:
\begin{multline}
  \label{eq:9}
  \mathbf{R}(t_{3},t_{2},t_{1}) =\\ -\left( \frac{i}{\hbar} \right)^{3}
  \Tr \left\{ \vec{\mu}(t_{3}) [\vec{\mu}(t_{2}), [\vec{\mu}(t_{1}), [ \vec{\mu}(0), \rho^{(0)}(-\infty) ]]] \right\},
\end{multline}
where $\Tr \{\cdot\}$ is the trace over internal molecular degrees of freedom and $\rho^{(0)}(-\infty)$ is the density matrix of the molecule prior to the interaction.
The interaction picture dipole operator is given by:
\begin{equation}
  \label{eq:10}
  \vec{\mu}(t) = e^{\frac{i}{\hbar}H_{0}t} \vec{\mu} e^{-\frac{i}{\hbar}H_{0}t},
\end{equation}
where $H_{0}$ is the field-free Hamiltonian. 
So far, the formalism we have laid out is a standard treatment of third-order spectroscopy, common to both liquid-phase and gas-phase spectroscopy.
In rotationally-resolved 2DIR spectroscopy, we will consider Feynman pathways involving different rotational states, so we expand the density matrix in a basis of the molecule's rotational eigenstates.
\begin{equation}
  \label{eq:11}
  \rho = \sum_{\substack{\alpha',J',m'\\\alpha'',J'',m''}} \bra{\alpha' J'M'}\rho\ket{\alpha'' J''M''} \ket{\alpha'' J''M''}\bra{\alpha' J'M'},
\end{equation}
where $J', J''$ and $M', M''$ are the rotational quantum numbers and their projections on laboratory-fixed axis, and $\alpha', \alpha''$ denote the remaining quantum state labels: projection of $J$ on the molecular axis ($K_{m}$), vibrational and electronic state, etc.
The field-free and collision-free time evolution in this basis is given trivially by:
\begin{equation}
  \label{eq:12}
  \begin{split}
    U_{0}(t)&\ket{\alpha'' J''M''}\bra{\alpha' J'M'}U^{\dagger}_{0}(t)\\ &=e^{-\frac{i}{\hbar}H_{0}t}\ket{\alpha'' J''M''}\bra{\alpha' J'M'}e^{\frac{i}{\hbar}H_{0}t} \\ &=e^{-\frac{i}{\hbar}(E_{\alpha''J''M''}-E_{\alpha'J'M'})t}\ket{\alpha'' J''M''}\bra{\alpha' J'M'}.
  \end{split}
\end{equation}
In thermal equilibrium the initial density matrix $\rho^{(0)}(-\infty)$ is a diagonal matrix of Boltzmann population factors:
\begin{multline}
  \label{eq:13}
  \rho^{(0)}(-\infty) = \sum_{\alpha'', J''} [\rho^{(0)}(-\infty)]_{\alpha'',J''}\\
  = \sum_{\alpha'',J''} \sum_{M''} \frac{e^{-E_{\alpha''J''M''}/kT}}{Q} \ket{\alpha'' J''M''}\bra{\alpha''J''M''} ,
\end{multline}
where $Q$ is the total internal partition function including rotations and vibrations.
The zeroth-order density matrix does not evolve prior to interaction with light, i.e. $U_{0}(t)\rho^{(0)}(-\infty)U_{0}^{\dagger}(t) = \rho^{(0)}(-\infty)$.
Expanding the commutators in Eq.~\eqref{eq:9} produces $2^{3}$ sequences of three-fold ket-side or bra-side interactions.
Selecting any one of them, fixing a phase-matching condition [Eq.~\eqref{eq:8}] and substituting some initial rotational energy level $[\rho^{(0)}(-\infty)]_{\alpha_{i},J_{i}}$ for the full density matrix $\rho^{(0)}(-\infty)$ in Eq.~\eqref{eq:9}, will produce a number of Feynman pathways (i.e. sequences of excitations of the density matrix that can be described with a double-sided Feynman diagram).
The details will depend on the selected initial level and on the range of levels accessible by dipole interaction under specified phase-matching condition.
Nevertheless, the contribution to the molecular response from each third-order rovibrational pathway can be written as:
\begin{widetext}
  \begin{equation}
    \label{eq:14}
    \begin{split}
      \widetilde{\epsilon}_{4}\cdot \big[&\mathbf{R}(t_{3},t_{2},t_{1}): \widetilde{\epsilon}_{3} \widetilde{\epsilon}_{2} \widetilde{\epsilon}_{1}\big]^{ijkl}_{\alpha_{j}J_{j},\alpha_{k}J_{k},\alpha_{l}J_{l},\alpha_{i}J_{i}} =
      i\frac{(-1)^{\lambda}}{\hbar^{3}} \mathcal{I}(t_{1},t_{2},t_{3}) \sum_{\substack{M_i,M_j, M_{k},M_{l}}} \rho^{(0)}(-\infty)_{\alpha_{i},J_{i},M_{i}}\\&
      \times  \langle \alpha_iJ_iM_i|\widetilde{\epsilon_i}\cdot\vec{\mu_i}|\alpha_jJ_jM_j\rangle
\langle \alpha_jJ_jM_j |\widetilde{\epsilon_j}\cdot\vec{\mu_j}|\alpha_kJ_kM_k\rangle
\langle \alpha_kJ_kM_k |\widetilde{\epsilon_k}\cdot\vec{\mu_k}|\alpha_lJ_lM_l\rangle
\langle \alpha_lJ_lM_l |\widetilde{\epsilon_l}\cdot\vec{\mu_l}|\alpha_iJ_iM_i\rangle,
\end{split}
\end{equation}
\end{widetext}
where the indices $ijkl$ in the superscript are for beam polarization vectors, $\lambda$ is equal to the number of bra-side interactions in the selected sequence and $\mathcal{I}(t_{1},t_{2},t_{3})$ contains purely the time dependence of the response.

The time dependence of the response for each time delay is given by Eq.~\eqref{eq:12}. We additionally include phenomenological population relaxation and coherence dephasing-decay to obtain:
\begin{equation}
  \label{eq:15}
  \mathcal{I}(t_{1},t_{2},t_{3}) = e^{-(i\Omega_{1}+\Gamma_{1})t_{1}} e^{-(i\Omega_{2}+\Gamma_{2})t_{2}} e^{-(i\Omega_{3}+\Gamma_{3})t_{3}},
\end{equation}
where the molecular coherence frequencies $\Omega_{i}$ are given by energy differences from Eq.~\eqref{eq:12}.
In gas-phase molecular samples, dephasing and relaxation are primarily caused by collisions, therefore the $\Gamma_{i}$ coefficients for population states should be identified with inelastic collision rates and for coherences with collisional pressure broadening widths.
In a more complete description of the molecular response, a collision operator would be included in the field-free operator $\mathcal{U}_{0}$ to explicitly model collisional effects, including transfer of population and coherence~\cite{Ben-Reuven1966a, Koszykowski1985}.
Collisional transfer of coherence will produce additional off-diagonal peaks in the spectrum and significantly increase the number of rovibrational pathways that need to be considered.
We further discuss the appropriateness of neglecting collisional transfer of coherence at the end of Sec.~\ref{sec:simulation}, after describing the structure of RR2DIR spectra.

With regards to shapes of individual resonances, a fully rigorous treatment would additionally include the Doppler effect and the effect of velocity-changing collisions.
This can be done by replacing the density matrix with the velocity distribution of the density matrix and solving an appropriate quantum Boltzmann equation~\cite{Shapiro2001}.
Alternatively, more approximate models could be used that include only Doppler broadening~\cite{Wasserman2002} or phenomenological models of velocity-changing collisions~\cite{Tran2005}.
Inclusion of Doppler broadening would produce lineshapes elongated along the diagonal~\cite{Hamm2011a}.
At $t_{2}=0$, the diagonal linewidth would be given by the Voigt linewidth, whereas the antidiagonal linewidth would correspond to the intrinsic Lorentzian width.
With increasing $t_{2}$ the total lineshape would symmetrize as the velocity-changing collisions thermalize nonequilibrium velocity distribution.
Indeed, rotationally-resolved 2DIR spectroscopy can be an excellent platform for detailed studies of molecular collisions and sophisticated lineshape theories.
However, to focus on the essential features of 2D spectra, here we adopt the simple model of Eq.~\eqref{eq:15}.

In the usual semi-impulsive limit, $\mathcal{E}_{i}(t-\tau_{i})\to \mathcal{E}_{i} \delta(t-\tau_{i})$, the contribution to macroscopic polarization from a single pathway can be compactly written as
\begin{multline}
  \label{eq:16}
  \hat{\epsilon}_{4}\cdot\vec{P}^{(3),ijkl}_{\alpha_{j}J_{j},\alpha_{k}J_{k},\alpha_{l}J_{l},\alpha_{i}J_{i}}(\vec{r}, t_{3}; t_{2}, t_{1})\\
  =i\frac{(-1)^{\lambda}}{8\hbar^{3}} \langle O_{ijkl} \rangle
  e^{ik_{s}z} \mathcal{I}(t_{1}, t_{2}, t_{3})
  \mathcal{E}_{1} \mathcal{E}_{2} \mathcal{E}_{3}
\end{multline}
where the previous integrations over times between interactions $t_{1}, t_{2}, t_{3}$ have collapsed to times between pulses in the impulsive limit, with $t_{3}=t-\tau_{3}$, $t_{2}=\tau_{3}-\tau_{2}$, $t_{1}=\tau_{2}-\tau_{1}$ in Eq.~\eqref{eq:16} and going forward.
The $\mathcal{E}_{i}$  are the areas of electric field pulse envelopes and the four-fold sum over degenerate $M$-states has been compactly represented with the expectation value of an operator $\langle O_{ijkl} \rangle$ for reasons that will become apparent in section~\ref{sec:fourfold}.
Assuming negligible depletion of the first two pulses, weak absorption of the third pulse and perfect phase-matching~\cite{Mukamel1995a}, the absorption coefficient for the probe beam is given by:
\begin{equation}
  \label{eq:17}
  \alpha_{I}(\omega_{3}; t_{1}, t_{2}) = \frac{N}{\pi}\mathcal{E}_{1} \mathcal{E}_{2}A^{(3)}(t_1, t_2, \omega_3) ,
\end{equation}
where $N$ is the number density of molecules, and $A^{(3)}(t_1, t_2, \omega_3)$ is the amplitude given by a sum over the contribution of each Feynman pathway to the spectrum:
\begin{equation}
  \label{eq:124}
  A^{(3)}(t_{1}, t_{2}, \omega_{3}) =  \sum_{\mathrm{pathways}} \mathcal{I}(t_{1}, t_{2}, \omega_{3}) S^{(3)},
\end{equation}
with pathway amplitudes $S^{(3)}$:
\begin{equation}
  \label{eq:18}
  S^{(3)} = \frac{(-1)^{\lambda}\pi\omega_{3}}{8n\epsilon_{0}\hbar^{3}c} \langle O_{ijkl} \rangle.
\end{equation}
It is easier to illustrate the strength various resonances in the 2D spectrum using the resonance amplitude
\begin{equation}
  \label{eq:66}
	A^{(3)}_{\Omega_1,\Omega_3}(t_2) = \sum_{\mathrm{pathways}} S^{(3)} e^{-i\Omega_2 t_2}.
\end{equation}
The discrete resonance amplitudes $A^{(3)}_{\Omega_1,\Omega_3}(t_2)$ do not contain lineshape information.
They do, however, capture the $t_2$ dependence due to rotational coherences.

\begin{figure}
  \includegraphics{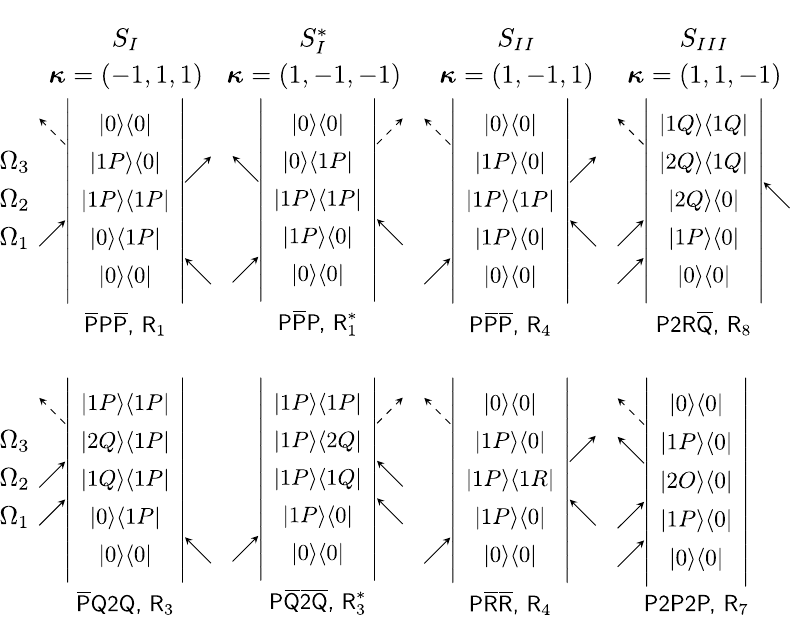}
  \caption{Diagrams of third-order rovibrational pathways phase-matched in direction: $S_{I}$, first column; $S_{I}^{\ast}$, second column; $S_{II}$, third column; $S_{III}$, fourth column.
First row contains pathways not in a rotationally-coherent (RC) state during waiting time.
Second row contains RC pathways.\label{fig:diagrams}}
\end{figure}

In the following, we specialize our description to rovibrational transitions within a single vibrational mode.
The label for remaining quantum numbers becomes $\alpha \equiv \eta \nu$, where $\nu$ is the vibrational quantum number and we will henceforth omit the $\eta$ label.
Any particular third-order pathway starting in the ground vibrational state can in this case be represented by a double-sided Feynman diagram such as those in Fig.~\ref{fig:diagrams}.
The double-sided diagrams represent time evolution of the density matrix and follow the usual conventions~\cite{Hamm2011a}: time flows from bottom  to top; solid arrows represent interactions with the light fields ($\widetilde{\epsilon}_{1}$, $\widetilde{\epsilon}_{2}$ or $\widetilde{\epsilon}_{3}$, in order) and a dashed arrow represents taking the expectation value ($\widetilde{\epsilon}_{4}$); arrow on the left (right) represents action of the dipole operator on the ket (bra) part of the density matrix; arrows pointed to the right stand for the positive frequency part of the field ($\hat{\epsilon}_{i}$) and arrows pointed to the left stand for the negative frequencies ($\hat{\epsilon}^{\ast}_{i}$).
Within the rotating-wave approximation that we use here, absorption corresponds to positive-frequency ket-side interaction or negative-frequency bra-side interaction, and \textit{vice versa} for emission~\footnote{Whether a dashed arrow corresponds to emission or absorption has no physical meaning in the theory.
All physical quantities would be left unchanged if the expectation value arrows were moved from ket side to bra side and \textit{vice versa}, which is a direct consequence of the trace operation being invariant to cyclic permutation of its arguments, see Eq.~\eqref{eq:9}}.

It is useful to adopt some compact notations for labelling pathways.
For individual density matrix elements within a pathway, we use the notation previously introduced by \citet{Mandal2018}.
In this shorthand, the initial rovibrational state is labeled by $\ket{\nu=\nu_{i},J=J_{i}}\equiv\ket{0}$ and following states are identified by the vibrational quantum number and the difference in angular referenced to initial state, \textit{e.g.} $\ket{\nu=1,J=J_{i}-1}\equiv\ket{1P}$, $\ket{\nu=0,J=J_{i}+2}\equiv\ket{0S}$ etc., where the rotational letters follow standard spectroscopic notation~\cite{Bernath2016}. 
In many cases, it is useful to classify pathways instead by the \emph{changes} in angular momentum occurring at each step. 
For this, we identify each pathway by the three dipole transitions that produce it.  
Each step is assigned a term consisting of a letter, P for $\Delta J=-1$, Q for $\Delta J=0$ and R for $\Delta J=+1$, depending on the $J$-number difference between higher and lower vibrational state.
$|\Delta J| > 1$ is not allowed in the dipole approximation, so the step labels are restricted to P, Q, and R, whereas density matrix \emph{elements} can also include S and O, since these states can be reached after multiple dipole interactions.
For the step-based pathway labeling, if the upper vibrational state lies above the first excited state, then the rotational transition letter is preceded by the larger vibrational quantum number, \textit{e.g.} 2P. 
Bra-side transitions are distinguished from ket-side ones by marking the term with an overline representing complex conjugation, \textit{e.g.} \ov{2P}.
Both these notations will allow us to discuss more the features of RR2DIR spectra that depend on the changes of rotational quantum numbers but not on their specific values.
We show the correspondence between double-sided diagrams, pathway labels and perturbation theory expressions by writing out Eq.~\eqref{eq:14} for P\ov{Q}\ov{2Q} pathway, also shown in Fig.~\ref{fig:diagrams}:
\begin{widetext}
  \begin{equation}
    \label{eq:19}
    \begin{split}
      \widetilde{\epsilon}_{4}&\cdot\big[\mathbf{R}(t_{3},t_{2},t_{1}): \widetilde{\epsilon}_{3} \widetilde{\epsilon}_{2} \widetilde{\epsilon}_{1}\big]_{\mathrm{P\closure[1]{Q}\closure[1]{2Q}}} = i\frac{(-1)^{\lambda}}{\hbar^{3}} e^{-(i\Omega_{\ket{1P}\bra{0}}+\Gamma_{\ket{1P}\bra{0}})t_{1}} e^{-(i\Omega_{\ket{1P}\bra{1Q}}+\Gamma_{\ket{1P}\bra{1Q}})t_{2}} e^{-(i\Omega_{\ket{1P}\bra{2Q}}+\Gamma_{\ket{1P}\bra{2Q}})t_{3}}  \sum_{\substack{M_i,M_j, M_{k},M_{l}}} \rho^{(0)}(-\infty)_{0,J_{i},M_{i}}\\&
      \times \langle 0,J_i,M_i|\hat{\epsilon}^{\ast}_2\cdot\vec{\mu_2}|1,J_i,M_j\rangle
\langle 1,J_i,M_j |\hat{\epsilon}^{\ast}_3\cdot\vec{\mu_3}|2,J_i,M_k\rangle
\langle 2,J_i,M_{k} |\hat{\epsilon}_4\cdot\vec{\mu_4}|1,J_i-1,M_l\rangle
\langle 1,J_i-1,M_l |\hat{\epsilon}_1\cdot\vec{\mu_1}|0,J_i,M_i\rangle.
\end{split}
\end{equation}
\end{widetext}

Figure~\ref{fig:diagrams} illustrates the usage of both notations on eight third-order pathways.
We also associate each pathway with an $R_i, i=1,\dots,8$, label following the convention in \citet{Hamm2011a}.
In Fig.~\ref{fig:diagrams}, we show only pathways starting with P-branch excitation to minimize the differences between them and to simplify the discussion in following paragraphs.
For reference, Fig.~S1 in the Supplemental Material shows analogous pathways starting with R-branch excitation.
Comparing \ov{P}P\ov{P} and P\ov{P}P or \ov{P}Q2Q and P\ov{Q}\ov{2Q} pathways, it is clear that conjugating a pathway corresponds to conjugating all the terms of its label.
Rotationally-coherent (RC) pathways (bottom row) are those for which ket and bra sides of the density matrix element produced by the second interaction have different rotational quantum numbers.
For $S_{I}$ and $S_{II}$, only RC pathways are in a coherent state during waiting time, while the remaining pathways are in a population state.
Using the three-letter notation for pathways, all RC ones have different first two terms, while all non-RC ones have the same two terms, see P\ov{P}P and P\ov{P}\ov{P} vs P\ov{Q}\ov{2Q} and P\ov{R}\ov{R} in Fig.~\ref{fig:diagrams}.
For $S_{III}$, all pathways oscillate at overtone frequency during waiting time, but the subset that is also RC isn't so trivially distinguishable by the three-letter notation.

Rotationally coherent pathways unbalance rephasing ($S_{I}$) and nonrephasing ($S_{II}$) pathways, preventing acquisition of purely absorptive spectra without resorting to the magic angle (MA) condition \cite{Kowzan2022b}.
This can be explained most easily by comparing $S_{I}^{\ast}$ with $S_{II}$ pathways shown in Fig.~\ref{fig:diagrams}.
The non-RC P\ov{P}P and P\ov{P}\ov{P} pathways differ only by the conjugate of the last transition and the sign of the third molecular coherence, $\Omega_{3}=\Omega_{\ket{0}\bra{1P}}=-\Omega_{\ket{1P}\bra{0}}$, and constitute a balanced pair.
For all $S_{I}^{\ast}$ ($S_{II}$) non-RC pathways we can obtain the balancing $S_{II}$ ($S_{I}^{\ast}$) counterpart by conjugating the last transition term.
Note that to transform \ov{P}P\ov{P} into P\ov{P}\ov{P} one needs to conjugate the first two interactions instead of just the $\omega_{3}$ one, which makes the symmetry less apparent and which is the reason why in the rest of the article we discuss $S_{I}^{\ast}$ instead of $S_{I}$ pathways.

\begin{figure}
  \includegraphics{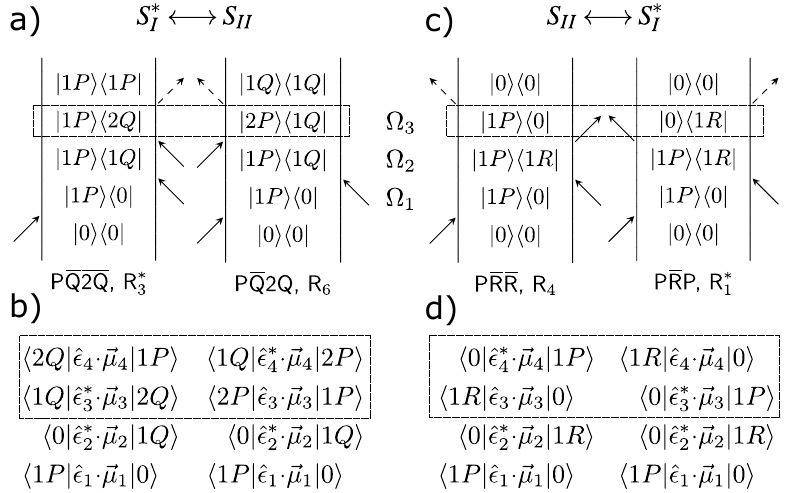}
  \caption{Correspondence between $S_{I}^{\ast}$ and $S_{II}$ RC pathways.
Top row (a, c) shows two pairs of $S_{I}^{\ast}$ and $S_{II}$ pathways connected by the bijective map.
Bottom row (b, d) shows the associated transition matrix elements.
Dashed rectangles highlight the change of $\Omega_{3}$ coherences in a) and c), and of transition matrix elements in b) and d).\label{fig:unbalancing_diagrams}}
\end{figure}

As for RC pathways, $S_{I}^{\ast}$ pathways can be mapped bijectively to $S_{II}$ RC pathways by conjugating the last term of the label.
Additionally, if the term contains letter Q, then the letter is left unchanged but if the letter is R, then it is replaced with P and \textit{vice versa}.
For example, the P\ov{Q}\ov{2Q} pathway becomes P\ov{Q}2Q and the $\Omega_{3}$ coherence \ket{1P}\bra{2Q} is replaced with \ket{2P}\bra{1Q} (Fig.~\ref{fig:unbalancing_diagrams}a).
The P\ov{R}\ov{R} pathway becomes P\ov{R}P and \ket{1P}\bra{0} is replaced with \ket{0}\bra{1R} (Fig.~\ref{fig:unbalancing_diagrams}c).
For both RC pairs, the $\Omega_{3}$ coherences are not merely conjugates of each other as for non-RC pathways, with their frequencies differing only in sign.
The \ket{1P}\bra{2Q} coherence lies in the R branch, while \ket{2P}\bra{1Q} lies in the P branch and similarly for the other pair.
Moreover, while P\ov{R}\ov{R} and P\ov{R}P contain the same set of transitions but in different sequence, see Fig.~\ref{fig:unbalancing_diagrams}d), this is not the case for pathways P\ov{Q}\ov{2Q} and P\ov{Q}2Q, see Fig.~\ref{fig:unbalancing_diagrams}d).
It is worth noting that the pathway unbalancing described here is entirely analogous to the presence of so-called diagonal cross-peaks in spectra of coupled anharmonic modes~\cite{Hamm2011a}, except here only a single vibrational mode is involved.
Further differences between spectra phasematched in $S_{I}$ and $S_{II}$ direction will be discussed in Sec.~\ref{sec:simulation}.
The spectral separation of $S_{I}$ and $S_{II}$ RC pathways will be further illustrated when discussing the polarization decomposition of third-order pathways in Sec.~\ref{sec:suppression}.

Restricting ourselves to symmetric top parallel transitions and assuming the usual selection rules---$\Delta J= \pm 1$ for $J=0$ or $K_{m}=0$ and $\Delta J= 0, \pm 1$ otherwise---we obtain 152 third-order pathways contributing to macroscopic polarization for each initial $\ket{\eta,\nu_{i}=0,J_{i}}$ state.
Explicitly summing over degenerate $M_{a}$ states in Eq.~\eqref{eq:14} would increase this number by a factor of $\sim 3^{3}(2J_{i}+1)$, which is prevented by the use of spherical tensor operator techniques discussed in section~\ref{sec:fourfold}.
Categorizing the pathways by phase-matching direction, there are 57 for $S_{I}$ and $S_{II}$ directions each and 38 for $S_{III}$.
For each $\vec{k}_{s}$, each pathway can be assigned to a 2D resonance.
For $S_{I}$ there are 28 distinct $(\Omega_{1}, \Omega_{3})$ pairs, for $S_{II}$---34, and for $S_{III}$---28.
Clearly, multiple pathways may contribute to the same 2D resonance.
When several of these undergo rotational coherence evolution during waiting time, we will observe interference between them, as illustrated by several examples in section~\ref{sec:interstate-coherences}.

\section{Branch structure of 2D spectra}
\label{sec:simulation}

\begin{figure}
  \includegraphics{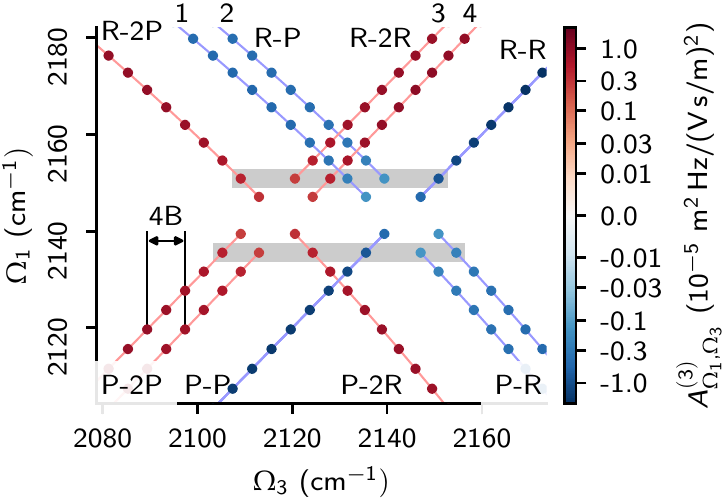}
  \caption{2D resonance structure of diatomic molecule transitions based on the CO vibrational mode.
The figure shows only pathways phase-matched in $S_{II}$ direction.
Branches of 2D resonances are labeled in analogy to usual linear spectroscopy labeling of rovibrational transitions.
Blue and red thin solid lines connecting the resonances are drawn as guides for an eye.
The shaded regions highlight all resonances obtained starting from $\ket{0}=\ket{\nu=0,J_{i}=1}$.\label{fig:resonance_structure_co}}
\end{figure}

\begin{figure}
  \includegraphics{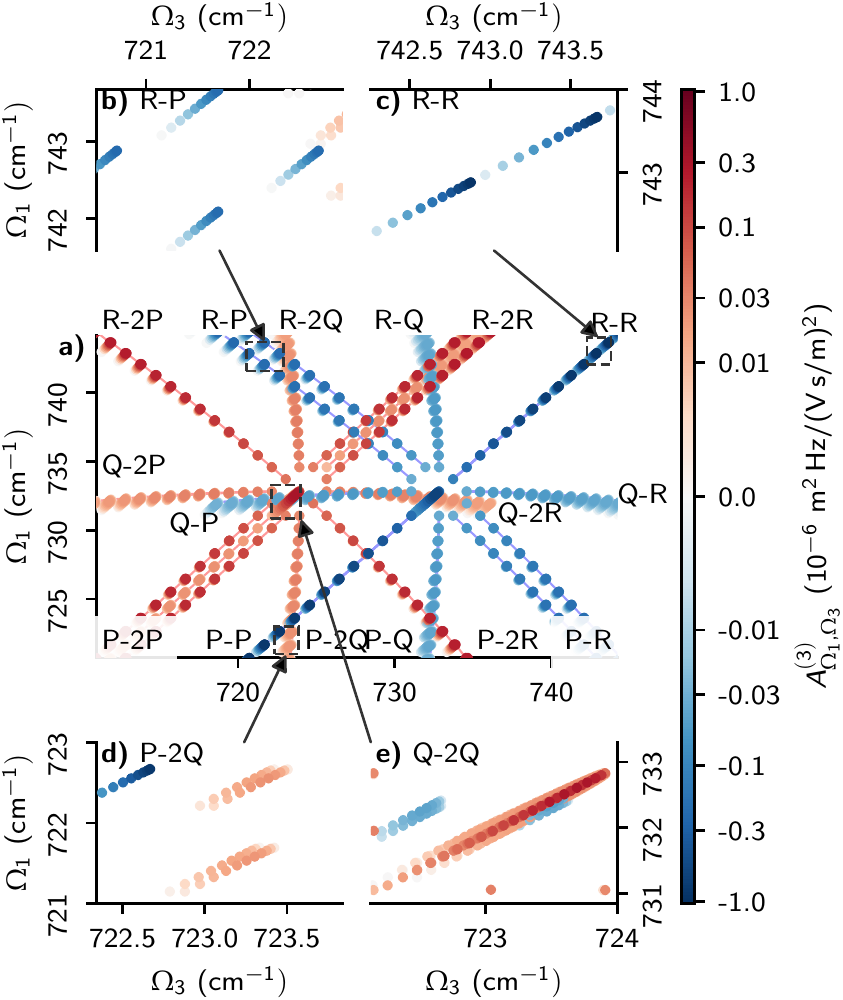}
  \caption{\label{fig:resonance_structure}2D resonance structure of symmetric top transitions based on the CH$_{3}$Cl $\nu_{3}$ mode.
The figure shows only pathways phase-matched in $S_{II}$ direction.
a) Branches of 2D resonances are labeled in analogy to usual linear spectroscopy labeling of rovibrational transitions.
Blue and red thin solid lines connecting the resonances are drawn as guides for an eye.
b) The congestion is reduced within the R-P branch because $J$ and $K_{m}$ dependence evolves along different spectral axes.
c) The diagonal R-R branch is highly congested, similar to linear spectra.
d) All Q-X and X-Q branches split into 2 closely lying subbranches.
e) Unlike the Q-Q branch, the Q-2Q branch splits into 3 subbranches.}
\end{figure}

We illustrate the structure of RR2DIR spectra with 2D resonance maps of CO (Fig. \ref{fig:resonance_structure_co}) and CH$_{3}$$^{35}$Cl $\nu_{3}$ (Fig. \ref{fig:resonance_structure}).
To generate 2D resonance maps we extracted the required energy levels, reduced matrix elements, quantum state degeneracies and partition functions for these molecules from the HITRAN database~\cite{Gordon2017,Nikitin2005,Hashemi2021,Li2015}.
We assume thermal equilibrium at $T=296$ K and include pathways with $J_{i}$ values up to 37 for CH$_{3}$$^{35}$Cl and up to 15 for CO.
The figures show resonance amplitudes, Eq.~\eqref{eq:66}, limited to the $S_{II}$ direction.
Similar to notation for pathways described in sec.~\ref{sec:general-framework}, 2D branches are labeled using the standard spectroscopic notation for rovibrational transitions, except here the labels are determined by the \textit{coherences} and not by the \textit{transitions} that produce them.
Any given pathway can be associated with a Y-X branch by examining its $\Omega_1$ and $\Omega_3$ coherences.
The branch label is determined by the $J$-number difference between coherences' higher vibrational state and lower vibrational state.
For example, \ov{P}P\ov{P}, P\ov{P}P and P\ov{P}\ov{P} in Fig.~\ref{fig:diagrams} have as $\Omega_1$ and $\Omega_3$ coherences either \ketbra{0}{1P} or \ketbra{1P}{0}, hence they lie in the P-P branch, as well as P\ov{R}\ov{R} and P2P2P pathways.
Perhaps less intuitively, the P\ov{Q}\ov{2Q} pathway with $(\Omega_1,\Omega_3)=(\ketbra{1P}{0}, \ketbra{1P}{2Q})$ lies in the P-2R branch since the $J$-difference between the higher vibrational state, $\bra{2Q}$, and the lower one, $\ket{1P}$, is $+1$.
We encourage the reader to further explore the structure of RR2DIR spectra by using our \textsc{peak\_picker} computer application~\cite{Kowzan2022rotsim2d}, which displays double-sided diagrams and other relevant information associated with 2D resonances.

The simpler example of CO branch structure is shown in Fig.~\ref{fig:resonance_structure_co}.
For easier interpretation, resonances within each branch are connected by guiding lines.
Moreover, all the resonances obtained starting from $\ket{\nu=0, J_i=1}$ state are highlighted by the shaded regions.
The branch structure can be thought to grow outward from the shaded  ``seed'' pattern with increasing $J_i$.
The diagonal branches P-P and R-R include (one-color) degenerate four-wave mixing (DFWM) and two-color pathways.
Each resonance in these branches is associatied with two DFWM pathways and two two-color RC pathways.
The antidiagonal branches R-P and P-R have one pathway per each peak and are split into two subbranches.
The subbranches can be distinguished by specifying $\Omega_3$ coherences relative to common $\Omega_1$ coherence (and $J_i$ reference number).
For R-P, these are \ketbra{1R}{0S} (lower frequency, 1) and \ketbra{1P}{0} (higher frequency, 2); and in the same order for P-R, \ketbra{1P}{0O} and \ketbra{1R}{0}.
Neither of the subbranches include RC pathways.
The lower frequency subbranches include stimulated-emission pumping (SEP) pathways with $\Omega_2$ in higher vibrational manifold, whereas the higher frequency subbranches include ground-state hole-burning (GSHB) pathways.

The excited-state absorption (ESA) resonances lie in R-2P, P-2R, R-2R, P-2P
branches, of which the latter two are split into subbranches.
The lower frequency subbranches (3) are RC and involve three different excitation
wavelengths, whereas the higher frequency ones (4) are two-color and non-RC.
For
R-2R, they are $\Omega_3=\ketbra{2Q}{1P}$ (lower) and $\Omega_3=\ketbra{2S}{1R}$
(higher); for P-2P, they are \ketbra{2Q}{1R} and \ketbra{2O}{1P}, in the same order.
As shown in Fig.~\ref{fig:resonance_structure_co}, the spacing between
subbranches in P-2P, R-2R, P-R and R-P branches is approximately $4B$, where $B$
is the rotational constant.

\begin{figure}
  \includegraphics{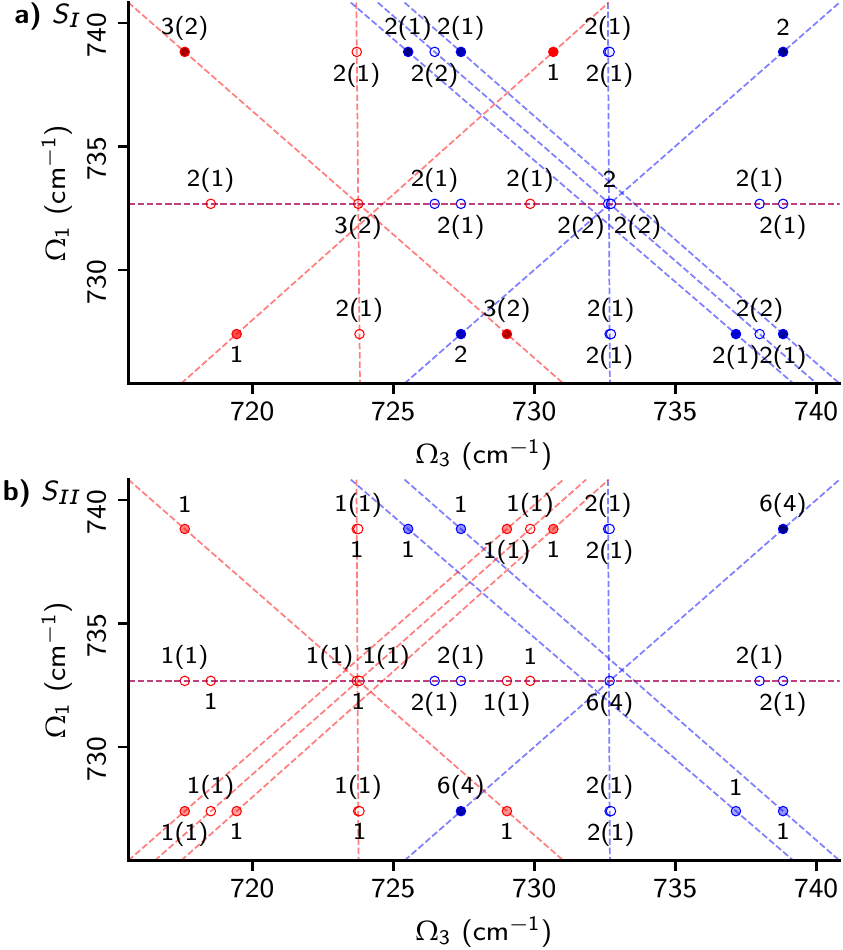}
  \caption{All resonances of CH$_{3}$$^{35}$Cl $\nu_3$ mode starting from $|\nu=0,J_i=6,K_m=1\rangle$ state.
Panel a) shows the resonances for $S_I$ phase-matching direction, panel b) for $S_{II}$ direction.
The labels at each resonance show the number of pathways contributing to the resonance, the numbers in parentheses are the numbers of RC pathways.\label{fig:pws_per_peak}}
\end{figure}

The resonance structure of symmetric tops is more complicated than that of linear rotors, see Fig.~\ref{fig:resonance_structure}.
On top of the double ``x''-shaped structure, there is additional double ``+''-shaped structure due to Q-X and Y-Q branches.
Additionaly, there are two branches located at the band centers, Q-Q and Q-2Q.
Moreover, transitions for different $K_{m}$ values produce locally diagonal structures for each $J_{i}$ value.
The diagonal branches P-P, Q-Q and R-R are the most highly congested in terms of the number of pathways per each peak and the spectral distance between peaks (see Fig.~\ref{fig:resonance_structure}c).
Each resonance in these branches is associated with 6 individual pathways that differ by the state they occupy during waiting time, with 4 of them being RC.
See Fig.~S2 in the Supplemental Material for all pathways contributing to an R-R resonance.

The ESA shifted diagonal branches, P-2P, Q-2Q and R-2R, are similarly spectrally congested, as can be seen in Fig.~\ref{fig:resonance_structure}a) for P-2P and R-2R, and in Fig.~\ref{fig:resonance_structure}e) for Q-2Q.
Compared to linear rotor subbranches, R-2R and P-2P contain additional middle subbranches with $\Omega_{3}=\ketbra{2R}{1Q}$ and \ketbra{2P}{1Q}, respectively.
Both subbranches are RC.
The full Feynman diagrams for R-2R branch are shown in Fig.~S2 in the Supplemental Material.
The new Q-2Q branch is also split into three with $\Omega_{3}$ coherences from lower to higher frequency: $\ketbra{2R}{1R}$, $\ketbra{2Q}{1Q}$, $\ketbra{2P}{1P}$.
Here, the first and the last one are RC.
The spacing between subbranches in Q-2Q is given by the spacing between $Q(J_{i}+1)$, $Q(J_{i})$, $Q(J_{i}-1)$ lines in $2\leftarrow{}1$ hot band, which is due to centrifugal distortion.

In contrast to (shifted) diagonal branches, the antidiagonal P-R, R-P as highlighted in Fig.~\ref{fig:resonance_structure}b), and the ESA P-2R, R-2P branches are least spectrally congested, especially the latter pair which is not split into subbranches.
This is because the locally diagonal $K_{m}$ structures are clearly separated along the antidiagonal branches.
The improved spectral separation off the diagonal shows the potential of RR2DIR spectroscopy to separate components of complex gas mixtures.
Within the ``+''-shaped structure, the clusters of resonances with different $J_{i}$ quantum numbers are also well separated, see Fig.~\ref{fig:resonance_structure}d).
In all cases the branches are split into two subbranches that are only weakly shifted from each other by centrifugal distortion.
For example, for R-2Q branch the $\Omega_{3}$ split is between blue-shifted $\ketbra{2Q}{1Q}$ coherence and red-shifted $\ketbra{2R}{1R}$ coherence, and analogously for P-2Q between blue-shifted $\ketbra{2P}{1P}$ coherence and red-shifted $\ketbra{2Q}{1Q}$ coherence.
For X-2Q, Q-2X branches each resonance corresponds to a single pathway, with the blue-shifted subbranches being RC for Q-2P and R-2Q branches and red-shifted subbranches being RC for P-2Q and Q-2R branches.
In Q-X, X-Q branches each resonance comprises two pathways with one of them being RC.

Figure~\ref{fig:pws_per_peak} shows all resonances of CH$_{3}$$^{35}$Cl $\nu_3$ mode starting from the $|\nu=0,J_i=6,K_m=1\rangle$ state~\footnote{This specific state was selected solely because it produces a pattern of well separated resonances.} for $S_I$ (a) and $S_{II}$ (b) phase-matching directions.
Figure~S3 in the Supplemental Material shows the same for $S_{III}$.
The label at each resonance gives the number of pathways contributing to it and the number of RC pathways among them in parenthesis.
For example, $2(2)$ indicates a purely RC resonance comprising two pathways.
Note that in the cases where the splitting between subbranches is very weak, it may appear that a single resonance has multiple labels.
See for example the $S_{II}$ Q-2Q branch in Fig.~\ref{fig:pws_per_peak}b), which was noted previously to be split into three closely spaced subbranches (Fig. \ref{fig:resonance_structure}e).
Comparing Figs.~\ref{fig:pws_per_peak}a) and b) shows that for both phasematching conditions the number of branches is the same, but the splitting into subbranches and the number of pathways per resonance is different.
For example, for $S_I$ it is the Q-Q branch that is split into three subbranches, whereas for $S_{II}$ it is the Q-2Q branch.
More apparently, in the $S_I$ direction the R-2R branch contains only one subbranch, whereas in the $S_{II}$ direction it contains three.
The figure can also be used to verify that removing all RC pathways would produce the same branch structures in both directions, as was noted when discussing unbalancing of pathways in Sec.~\ref{sec:general-framework}.

As indicated in Sec.~\ref{sec:general-framework}, collisional transfer of coherence will modify the structure of RR2DIR spectra by producing additional off-diagonal resonances.
The contribution of these resonances to the total signal is expected to be low as long as the linewidths are significantly below the spacing between resonances and the experimental delay $t_{2}$ is kept significantly below the coherence relaxation time [$t_2 \ll 1/(\gamma p)$, where $\Gamma=\gamma p$].
For CO in air at $T=296$ K with $B\approx 1.93$ cm$^{-1}$, this implies $p \ll 2B/\gamma \approx  51$ atm and $t_{2} \ll 440$ ps (at $p=1$ atm).
For CH$_{3}$Cl we can distinguish two distinct regimes of coherence transfer.
Because of small spacing between resonances within $K_m$ structures (\textasciitilde{}0.05 cm$^{-1}$) and within Q-Q, Q-2Q branches, see corner panels in Fig.~\ref{fig:resonance_structure}, coherence transfer will be significant even at atmospheric pressure, but only within these separated clusters of closely lying resonances.
On the other hand, coherence transfer between resonances with different $J_i$ numbers outside of Q-Q and Q-2Q branches is expected to be weak as long as  $p \ll 32$ atm and $t_2 \ll 274$ ps (at 1 atm).

\section{Polarization dependence of the molecular response}
\label{sec:fourfold}

In principle, the polarization-dependent response can be obtained for each pathway individually by specifying its rovibrational quantum numbers, choosing beam polarizations and performing the sum over degenerate rotational states in Eq.~\eqref{eq:14}.
However, the expression in its current form obscures the fact that the polarization dependence is not unique to each pathway.
In fact, in the limit of high $J_{i}$ quantum number there are only several distinct polarization response functions, which can be used to separate all the pathways considered here into disjoint sets.
These sets can then be selectively suppressed with specific polarization conditions, as shown in section~\ref{sec:suppression}.
To perform this classification, Eq.~\eqref{eq:14} first needs to be factorized into vibrational, angular momentum and polarization terms.
This task is facilitated by expressing the polarization-dependent molecular response, Eq.~\eqref{eq:14}, in terms of the expectation value of the four-fold dipole interaction operator $O_{ijkl}$:
\begin{equation}
  \label{eq:20}
  \widetilde{\epsilon}_{4} \cdot [\mathbf{R}(t_{3},t_{2},t_{1}): \widetilde{\epsilon}_{3} \widetilde{\epsilon}_{2} \widetilde{\epsilon}_{1}]_{\nu_{i},J_{i}} = i\frac{(-1)^{\lambda}}{\hbar^{3}} \mathcal{I}(t_{1},t_{2},t_{3}) \langle O_{ijkl}\rangle,
\end{equation}
with:
\begin{equation}
  \label{eq:21}
  \langle O_{ijkl} \rangle = \Tr (O_{ijkl}[\rho^{(0)}(-\infty)]_{\nu_{i},J_{i}})
\end{equation}
The $O_{ijkl}$ operator is defined as:
\begin{equation}
  \label{eq:22}
  O_{ijkl} = (\widetilde{\epsilon}_i\cdot\vec{\mu_{i}})P_{j}(\widetilde{\epsilon}_j\cdot\vec{\mu_{j}})P_{k}(\widetilde{\epsilon}_k\cdot\vec{\mu_{k}})P_{l}(\widetilde{\epsilon}_l\cdot\vec{\mu_{l}}),
\end{equation}
where $P_j$ is the normalized projection operator onto \emph{(2J+1)}-dimensional rovibrational subspace:
\begin{equation}
P_{\alpha} = \sum_{M_{\alpha}}|\nu_{\alpha}J_{\alpha}M_{\alpha}\rangle\langle \nu_{\alpha}J_{\alpha}M_{\alpha}|.
\end{equation}
Subsequently, the density matrix and the $O_{ijkl}$ operator are expressed in terms of spherical tensors and spherical tensor operators.
With a judicious use of spherical tensor algebra and angular momentum recoupling, $\langle O_{ijkl} \rangle$ is formally decomposed into a form that enables classification of pathways with regards to their polarization dependence.
This decomposition was previously performed by Williams \textit{et al.} \cite{Williams1994,Williams1997} and Vaccaro \textit{et al.} \cite{Wasserman1998,Murdock2010}, and the details are also provided in Appendix~\ref{app:fourfold}.
The expectation value is expressed as:
\begin{equation}
  \label{eq:23}
  \begin{split}
    \langle O_{ijkl} \rangle = \frac{N_{\nu_{i}J_{i}}}{N} \frac{1}{\sqrt{2J_{i}+1}} \langle\nu_{i}J_{i}\|T^{(0)}(\boldsymbol{\mu})\| \nu_{i}J_{i}\rangle
     R^{(0)}_{0}(\boldsymbol{\epsilon}; \mathbf{J}).
  \end{split}
\end{equation}
This factorization clearly separates the transition dipole amplitude factor, $\langle\nu_{i}J_{i}\|T^{(0)}(\boldsymbol{\mu})\| \nu_{i}J_{i}\rangle$, from the polarization-angular momentum factor, $R^{(0)}_{0}(\boldsymbol{\epsilon}; \mathbf{J})$.
For brevity, dipole operators, polarization vectors and rotational angular momentum numbers were collected into compound arguments:
\begin{equation*}
  \label{eq:24}
  \begin{split}
    \boldsymbol{\mu} = (\vec{\mu_{i}}, \vec{\mu_{j}}, \vec{\mu_{k}}, \vec{\mu_{l}}),\;
    \bm{\varepsilon} = (\widetilde{\epsilon_{i}}, \widetilde{\epsilon_{j}}, \widetilde{\epsilon_{k}}, \widetilde{\epsilon_{l}}),\;
    \mathbf{J} = (J_{i}, J_{j}, J_{k}, J_{l}).
  \end{split}
\end{equation*}
The transition dipole amplitude factor is given by:
\begin{multline}
  \label{eq:25}
    \langle \nu_i J_i\|T^{(0)}(\boldsymbol{\mu})\|\nu_i J_i\rangle = \langle \nu_i J_i\|\vec{\mu}_1\|\nu_1 J_1\rangle \langle \nu_1 J_1\|\vec{\mu}_2\|\nu_2 J_2\rangle\\
    \times \langle \nu_2 J_2\|\vec{\mu}_3\|\nu_3 J_3\rangle\langle \nu_3 J_3\|\vec{\mu}_4\|\nu_i J_i\rangle,
\end{multline}
which is a four-fold product of reduced matrix elements of individual step transitions.
Assuming complete separation of molecular rotational and vibrational degrees of freedom, the reduced matrix element can be expressed as:
\begin{equation}
  \label{eq:26}
  \langle \nu' J'\|\vec{\mu}\|\nu''J'' \rangle = \pm S_{\mrm{pol}} \sqrt{S(J',J'')} \langle \nu'|\vec{\mu}|\nu'' \rangle,
\end{equation}
where $S_{\mrm{pol}} = 1/\sqrt{3}$ for linearly polarized light and $S_{\mrm{pol}} = 1$ for unpolarized light.
$S(J',J'')$ is the H\"onl-London factor and $\langle \nu'|\vec{\mu}|\nu'' \rangle$ is the vibrational band intensity.
The reduced matrix element is positive for R- and Q-branch transitions and negative for P-branch transitions.
Its magnitude is also related to the Einstein $A$ coefficient by:
\begin{equation}
  \label{eq:27}
  |\langle \nu'J'\|\vec{\mu}\|\nu''J''\rangle|^{2} = A_{\nu'J'\to\nu''J''}\frac{3\epsilon_{0}hc^{3}(2J'+1)}{16\pi^{3}\nu^{3}_{\nu'J',\nu''J''}}.
\end{equation}

The polarization-angular momentum factor is given by the sum:
\begin{equation}
  \label{eq:28}
  R^{(0)}_{0}(\boldsymbol{\epsilon}; \mathbf{J}) = \sum_{k=0}^{2} T^{(0)}_{0}(\boldsymbol{\epsilon}; k, k) G(\mathbf{J}; k),
\end{equation}
where $T^{(0)}_{0}(\boldsymbol{\epsilon}; k, k)$ is the scalar component of the polarization tensor composed of beam and detection polarizations, see Eq.~\eqref{eq:44} and Appendix.~\ref{app:comp-polar-tens}.
The $G$-factor in Eq.~\eqref{eq:28} encapsulates the dependence of molecular response on $J$ rotational quantum numbers:
\begin{equation}
  \label{eq:29}
  \begin{split}
    G(&J_i, J_j, J_k, J_l; k)\\ & = (2k+1)\begin{Bmatrix}
      k & k & 0\\
      J_i & J_i & J_k
    \end{Bmatrix}
    \begin{Bmatrix}
      1 & 1 & k\\
      J_k & J_i & J_j
    \end{Bmatrix}
    \begin{Bmatrix}
      1 & 1 & k\\
      J_k & J_i & J_l
    \end{Bmatrix},
  \end{split}
\end{equation}
where the brackets $\{\cdots\}$ denote Wigner 6j coefficients. 
The symmetry of Wigner 6j coefficients makes the quantity invariant under exchange of $J_{l}$ with $J_{j}$.
It is convenient to express $J_{j}$, $J_{k}$, $J_{l}$ arguments as $J_{\alpha}=J_{i} + \Delta J_{\alpha}$.
With this convention and the usual dipole transition selection rules we obtain 19 different argument sequences of $J$ values for each $k$ value.
 Using analytical formulas for simple cases of Wigner 6j symbols, compact formulas for $G$-factors can be derived as a function of $J_{i}$ and $k$, see Tab.~S3 and S4 in Supplemental Material.

An explicit expression for $R^{(0)}_{0}(\boldsymbol{\epsilon}; \mathbf{J})$ is obtained by substituting Eqs.~\eqref{eq:29} and (\ref{eq:56}--\ref{eq:58}) into Eq.~\eqref{eq:28}.
The resulting formula can be factored into a relatively simple and general form shown below:
\begin{equation}
  \label{eq:30}
  \begin{split}
    R^{(0)}_{0}(\boldsymbol{\epsilon}; \mathbf{J}) =& \frac{c_{00}}{60(2J_{i}+1)^{3/2}}\big(
    c_{12} \cos(\theta_{1}+\theta_{2}-\theta_{3}-\theta_{4})\\
    &+c_{13} \cos(\theta_{1}-\theta_{2}+\theta_{3}-\theta_{4})\\
    &+c_{14} \cos(\theta_{1}-\theta_{2}-\theta_{3}+\theta_{4})
    \big).
  \end{split}
\end{equation}
The $c_{\alpha\beta}$ coefficients originate from the $G$-factors in Eq.~\eqref{eq:28} and in general depend on $J_{i}$.
To investigate polarization dependence of third-order rovibrational signals within a single vibrational mode, we considered all the resonant pathways having at its root a rotational state $J_{i}$ in the ground vibrational manifold and assigned to them corresponding $R$-factors.
Collecting the pathways into classes associated with the same $R$-factor, we obtain 39 distinct classes.
The complete table of labels identifying the pathways and $c_{\alpha\beta}$ coefficients is provided as Tab.~S5 in the Supplemental Material.

\begin{table*}
\centering
\caption{\label{tab:Rfactors_highj}Coefficients defining polarization-angular momentum dependence factors in the high-$J$ limit, $\mathcal{R}^{(0)}_{0}(\boldsymbol{\epsilon}; \mathbf{J})$.
See Eqs.~(\ref{eq:30}, \ref{eq:31}).}
\begin{ruledtabular}
  \begin{tabular}{ccccccc}
    \multirow{2}{*}{Class} & \multicolumn{3}{c}{Pathway label}  & \multirow{2}{*}{$c_{12}$} & \multirow{2}{*}{$c_{13}$} & \multirow{2}{*}{$c_{14}$} \\
    & $S_{I}$ & $S_{II}$ & $S_{III}$ \\\midrule
    $\Theta_{1}$ & \textbf{PRP}, \textbf{RPR} & \textbf{PRR}, \textbf{RPP} & \textbf{PPP}, \textbf{RRR} & $6$ & $1$ & $1$\\
    $\Theta_{2}$ & \textbf{PQQ}, \textbf{QPR}, \textbf{QRP}, \textbf{RQQ} & \textbf{PQQ}, \textbf{QPP}, \textbf{QRR}, \textbf{RQQ} & \textbf{PQQ}, \textbf{QPP}, \textbf{QRR}, \textbf{RQQ} & $-3$ & $-3$ & $2$\\
    $\Theta_{3}$ & \textbf{PQP}, \textbf{QPQ}, \textbf{QRQ}, \textbf{RQR} & \textbf{PQR}, \textbf{QPQ}, \textbf{QRQ}, \textbf{RQP} & \textbf{PQP}, \textbf{QPQ}, \textbf{QRQ}, \textbf{RQR} & $-3$ & $2$ & $-3$\\
    $\Theta_{4}$ & \textbf{QQQ} & \textbf{QQQ} & \textbf{QQQ} & $4$ & $4$ & $4$\\
    $\Theta_{5}$ & \textbf{PPR}, \textbf{RRP} & \textbf{PPP}, \textbf{RRR} & \textbf{PRR}, \textbf{RPP} & $1$ & $6$ & $1$\\
    $\Theta_{6}$ & \textbf{PPP}, \textbf{RRR} & \textbf{PPR}, \textbf{RRP} & \textbf{PRP}, \textbf{RPR} & $1$ & $1$ & $6$\\
    $\Theta_{7}$ & \textbf{PPQ}, \textbf{QQP}, \textbf{QQR}, \textbf{RRQ} & \textbf{PPQ}, \textbf{QQP}, \textbf{QQR}, \textbf{RRQ} & \textbf{PRQ}, \textbf{QQP}, \textbf{QQR}, \textbf{RPQ}& $2$ & $-3$ & $-3$
  \end{tabular}
\end{ruledtabular}
\end{table*}

It was shown previously that $c_{\alpha\beta}$ coefficients only weakly depend on $J_{i}$~\cite{Murdock2009}, therefore instead of considering polarization dependence of individual pathways or 2D resonances, we can investigate the dependence of whole branches of rovibrational transitions.
The transition to $J_{i}$-independent molecular response is effected by defining a reduced R-factor and taking the limit:
\begin{equation}
  \label{eq:31}
  \mathcal{R}^{(0)}_{0}(\boldsymbol{\epsilon}; \mathbf{J}) =
  \lim_{J_{i}\to\infty} (2J_{i}+1)^{3/2} R^{(0)}_{0}(\boldsymbol{\epsilon}; \mathbf{J}).
\end{equation}
The high-$J$ limit greatly simplifies classification of pathways by reducing the previous 39 classes to only 7, which are defined by the coefficients in Table~\ref{tab:Rfactors_highj}.
We label these classes \mbox{$\Theta_1$--$\Theta_7$}.
In all cases $c_{00}=1$.
For brevity, in Tab~\ref{tab:Rfactors_highj} we use degenerate pathway labels, where \textbf{X}$\to$ X, \textoverline{X}, 2X, \textoverline{2X}.
For example, P2P2P, P2P\ov{P} $\to$ \textbf{PPP}, P\ov{R}\ov{R} $\to$ \textbf{PRR}, R\ov{P}\ov{2R} $\to$ \textbf{RPR}. 

\begin{figure*}
  \includegraphics{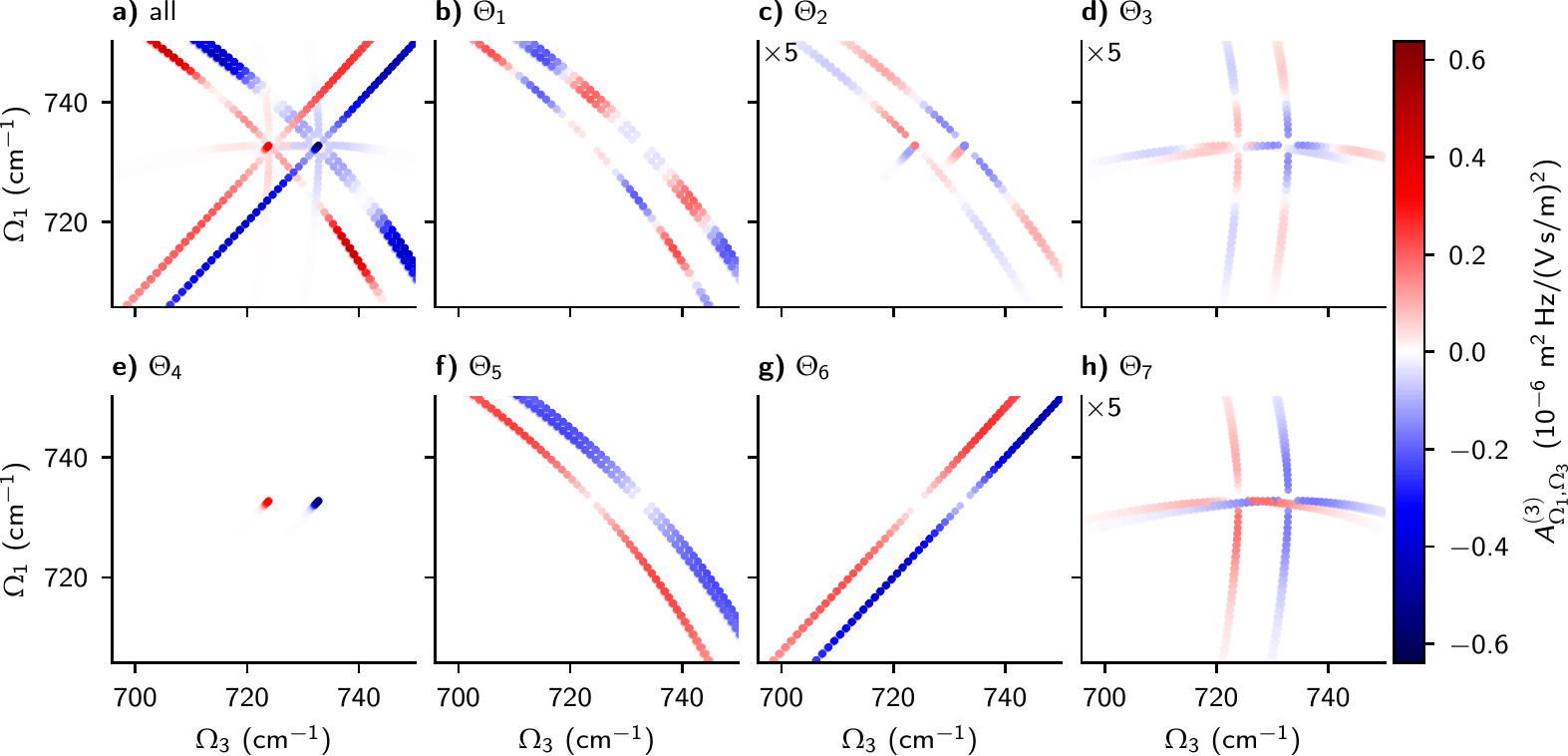}
  \caption{Polarization decomposition of third-order response of methyl chloride $\nu_{3}$ vibrational mode phase-matched in the $S_{I}$ direction.
    The plotted quantity is the 2D resonance amplitude defined in Eq.~\eqref{eq:66}.
a) total response; b--h) subsets of response associated with polarization classes $\Theta_{1}$--$\Theta_{7}$, see Tab.~\ref{tab:Rfactors_highj}.
    \label{fig:symtop_gallery_SI}}
\end{figure*}

\begin{figure*}
  \includegraphics{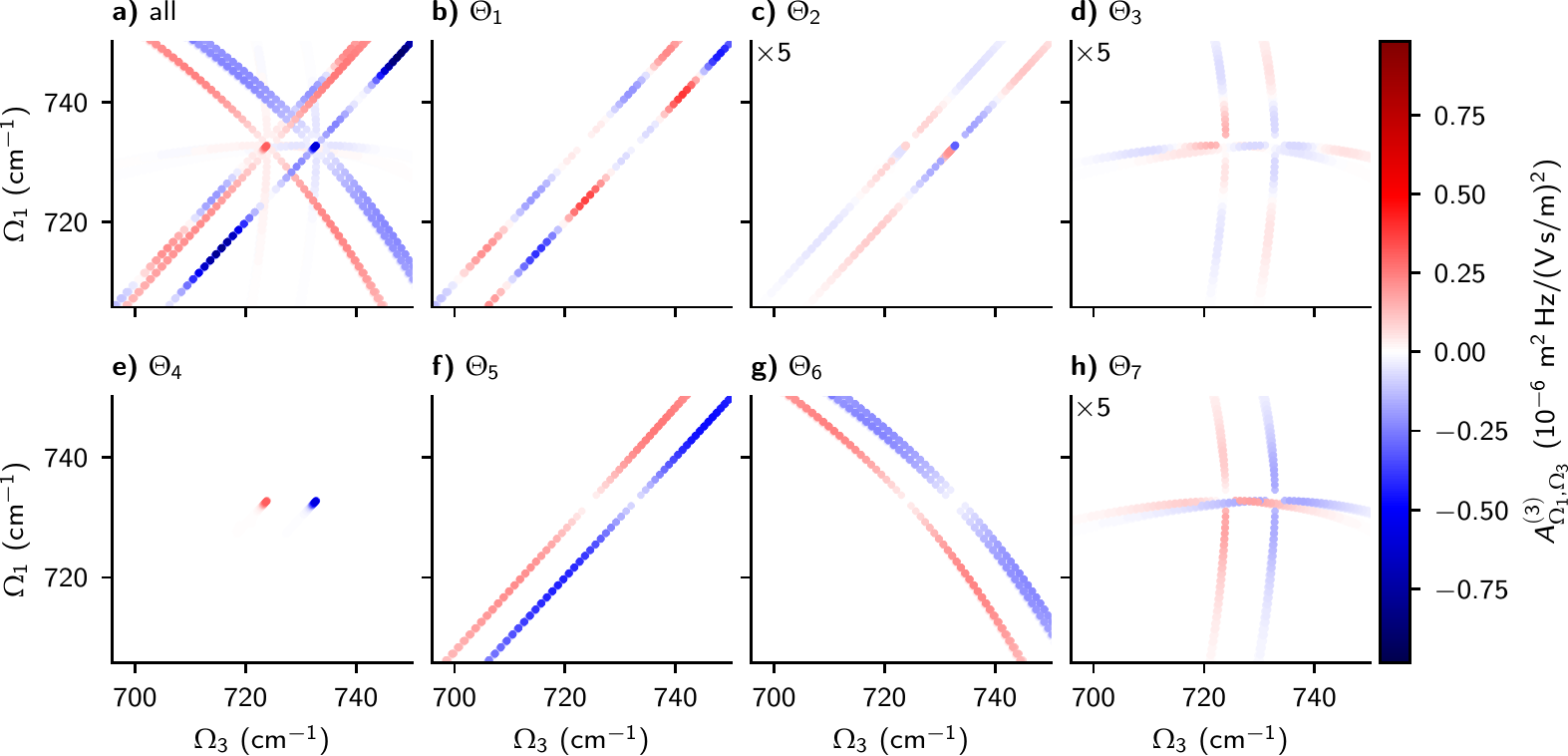}
  \caption{Polarization decomposition of third-order response of methyl chloride $\nu_{3}$ vibrational mode phase-matched in the $S_{II}$ direction.
    The plotted quantity is the 2D resonance amplitude defined in Eq.~\eqref{eq:66}.
a) total response; b--h) subsets of response associated with polarization classes $\Theta_{1}$--$\Theta_{7}$, see Tab.~\ref{tab:Rfactors_highj}.
    \label{fig:symtop_gallery_SII}}
\end{figure*}

We now present the decomposition of CH$_{3}$$^{35}$Cl 2D resonance maps with regards to polarization classes in the high-$J$ limit.
Figures~\ref{fig:symtop_gallery_SI} and \ref{fig:symtop_gallery_SII} show the resonance amplitudes $A^{(3)}_{\Omega_1,\Omega_3}(t_2 =\text{ 1 ps})$, Eq.~\eqref{eq:66}, for $S_I$ and $S_{II}$ directions, respectively. In both figures, panel a) shows the total amplitudes and the remaining panels show contributions limited to the 7 different classes of pathways in Table~\ref{tab:Rfactors_highj}.
The resonances involving Q-type transitions, $\Theta_2$, $\Theta_3$, $\Theta_7$,  are relatively weak in CH$_{3}$Cl, therefore they were multiplied by 5 to make them more clearly visible.

Here we discuss several general features of the polarization decomposition.
A key point is that this decomposition separates all RC pathways from non-RC pathways.
The RC pathways are $\Theta_1$, $\Theta_2$, $\Theta_3$, which is evident from the modulation along the branches for these classes due to the $t_2$-dependent phase factors in Eq.~\eqref{eq:66}.
Comparing classes $\Theta_1$ and $\Theta_2$ in Fig.~\ref{fig:symtop_gallery_SI} and \ref{fig:symtop_gallery_SII}, it is clear that RC resonances mostly do not overlap spectrally for different phase-matching directions, as was discussed in sec.~\ref{sec:general-framework}.
As an exception, $S_I$ and $S_{II}$ pathways in $\Theta_3$ class do overlap, but their pathway intensities are different.
Each of the three phasematching directions contains pathways associated with all polarization classes, therefore pathway selection with phase cycling is orthogonal and complementary to polarization control techniques discussed in the next section.
It is worth pointing out that when Q-type transitions are forbidden, \textit{e.g.} in diatomic molecules or for stretching modes of linear molecules, there are only 3 polarization classes, $\Theta_1$, $\Theta_5$ and $\Theta_6$.

\section{Suppressing pathways with polarization}
\label{sec:suppression}

Given the variety of applications of 2D IR spectroscopy, it would be advantageous to have the ability to suppress any possible subset of polarization classes, in order to measure only the pathways relevant to the physical or chemical phenomena under study.
From another perspective, measurements of the molecular response under different polarization conditions, combined with \textit{a priori} knowledge of the pathways suppressed under these conditions, provide additional constraints for the global analysis~\cite{Stokkum2004,Wilderen2011} performed on the whole dataset.
In Ref.~\cite{Kowzan2022b}, we present several polarization conditions that significantly improve signal separation in RR2DIR spectra, especially when the experimental signal is additionally restricted to pathways phasematched in a particular direction.
These conditions and others that suppress specific pathways can be obtained by substituting appropriate $c_{\alpha\beta}$ coefficients from Tab.~\ref{tab:Rfactors_highj} and finding the root of Eq.~\eqref{eq:30}.
Since $R_{0}^{(0)}$ is a scalar and invariant with respect to rotation, there are only 3 independent angles.
Assuming $\theta_{1}=0$, the root with respect to $\theta_{4}$ is:
\begin{equation}
  \label{eq:32}
  \theta_{4} = -\tan^{-1}\left( \frac{c_{12}\cos \theta^{(-)}_{23} + c_{13}\cos \theta^{(-)}_{23} + c_{14}\cos \theta^{(+)}_{23}}{ c_{12}\cos \theta^{(-)}_{23} - c_{13}\cos \theta^{(-)}_{23} + c_{14}\cos \theta^{(+)}_{23} } \right),
\end{equation}
where $\theta_{23}^{(\pm)}=\theta_{2}\pm\theta_{3}$.
Specific polarization conditions can be obtained by selecting a subset of polarization classes and finding their common root using Eq.~\eqref{eq:32}.
Alternatively, one can plot R-factor values as a function of polarization angles and find the common roots graphically.
The former can be done easily by calling appropriate functions in the \textsc{rotsim2d} library and latter by using the \textsc{polarizations} tool~\cite{Kowzan2022rotsim2d}.
Particularly useful for suppression are two angles, the so-called magic angle already widely used in nonlinear spectroscopy
\begin{equation}
    \theta_{\mathrm{MA}}=\tan^{-1} \sqrt{2} \approx 54.74^{\circ},
\end{equation}
and a new angle we introduce and name the population-alignment cancelling angle
\begin{equation}
  \theta_{\mathrm{PAC}}=\sin^{-1} \frac{2}{\sqrt{7}}\approx 49.11^{\circ}\; .
\end{equation}
In Ref.~\cite{Kowzan2022b} we present several example of third-order amplitude spectra, where we use polarization conditions with these angles to suppress different subsets of polarization classes.
Here we relate the polarization conditions to the 2D resonance maps of Figs.~\ref{fig:symtop_gallery_SI} and \ref{fig:symtop_gallery_SII}.
It is well known that the conventional MA condition $(\theta_1, \theta_2, \theta_3, \theta_4) = (0, 0, \theta_{\mathrm{MA}}, \theta_{\mathrm{MA}})$ suppresses the molecular-orientation dependence of nonlinear spectroscopy signals.
In the current context of rotationally-resolved spectra, this can be better explained by noting that the MA condition zeroes precisely those pathways that include RC, which coherently evolve during $t_{2}$, \textit{i.e.} classes $\Theta_{1}$, $\Theta_{2}$ and $\Theta_{3}$.

In contrast, no polarization condition exactly zeros all non-RC pathways.
However, strong suppression is attainable with several new polarization conditions we introduce.
The polarization-alignment canceling (PAC) condition, $(0,0,\theta_\mathrm{PAC}, -\theta_\mathrm{PAC})$, suppresses $\Theta_{5}$ and $\Theta_{6}$ classes; the middle MA condition, $(0,\theta_\mathrm{MA}, \theta_\mathrm{MA}, 0)$, suppresses $\Theta_{3}$, $\Theta_{6}$ and $\Theta_{7}$; and the middle PAC condition,  $(0,\theta_\mathrm{PAC}, -\theta_\mathrm{PAC}, 0)$, suppresses $\Theta_{1}$ and $\Theta_{5}$ classes.

In appendix~\ref{app:magic-angle} we show that the PAC condition suppresses the orientation part and cancels the population with alignment component of the response for $\Theta_{5}$ and $\Theta_{6}$ classes.
The same pathways can be eliminated by any sequence of angles for which $\theta_{1}=\theta_{2}=0$ and $\tan\theta_{3}=-(4/3)\cot\theta_{4}$, but picking $\theta_{3}=\theta_{\mathrm{PAC}}= \sin^{-1}(2/\sqrt{7})$ maximizes the magnitude of unsuppressed pathways.
Analogously, a generalized MA condition can be defined as $\tan\theta_{4}=2\cot\theta_{3}$, but the conventional condition maximizes the intensity of non-RC pathways.

In addition to the polarization conditions introduced in Ref.~\cite{Kowzan2022b}, here we introduce two additional conditions: the alternating MA condition, $(0,\theta_\mathrm{MA}, 0, \theta_\mathrm{MA})$, and alternating PAC condition, $(0,\theta_\mathrm{PAC}, 0, \theta_\mathrm{PAC})$.
Addition of these two conditions to the previous four enables total control over polarization-dependent response of diatomic molecules and stretching modes of linear molecules, which are limited to classes $\Theta_{1}$, $\Theta_{5}$ and $\Theta_{6}$.
This can be seen in Tab.~\ref{tab:suppression}, where we summarize the effect of polarization conditions on RR2DIR spectra.
The conditions using the MA remove any individual class, whereas the PAC conditions suppress any given two out of these three classes.

\begin{table}
  \centering
  \caption{\label{tab:suppression}Summary of the effects of special polarization conditions on the polarization classes in the high-J limit. The values in the table are the reduced R-factors, Eq.~\eqref{eq:31}, for specified class and condition.}
  \begin{ruledtabular}
    \begin{tabular}{lrrrrrrr}
      Condition  & $\Theta_{1}$ & $\Theta_{5}$ & $\Theta_{6}$ & $\Theta_{2}$ & $\Theta_{3}$ & $\Theta_{4}$ & $\Theta_{7}$ \\
      \midrule
      MA         & $0$          & $1/9$        & $1/9$        & $0$          & $0$          & $1/9$        & $-1/9$       \\
      alt. MA    & $1/9$        & $0$          & $1/9$        & $0$          & $-1/9$        & $1/9$        & $0$          \\
      middle MA  & $1/9$        & $1/9$        & $0$          & $-1/9$        & $0$          & $1/9$        & $0$          \\
      PAC        & $2/21$       & $0$          & $0$          & $-1/21$       & $-1/21$       & $1/21$       & $1/21$       \\
      middle PAC & $0$          & $0$          & $2/21$       & $1/21$      & $-1/21$       & $1/21$       & $-1/21$      \\
      alt. PAC   & $0$          & $2/21$       & $0$          & $-1/21$       & $1/21$      & $1/21$       & $-1/21$      \\
    \end{tabular}
  \end{ruledtabular}
\end{table}

Vaccarro and coworkers previously derived different polarizations conditions of the form $(\pi/2, \pi/4, \pi/2, \theta_{4})$, suppressing parts of the molecular response~\cite{Wasserman1998,Murdock2009}.
The key difference between the cited conditions and those presented here and in Ref.~\cite{Kowzan2022b} is that the former were tailored for degenerate FWM and two-color stimulated-emission pumping (TC-SEP) experiments with narrowband light sources, which did not probe RC pathways.
For example, the $\theta_{4} = -\tan^{-1}(1/3)$ condition suppresses QQQ pathways but does not suppress QPP and QRR pathways, while all of them contribute to Q-Q and Q-2Q branches.
The $\theta_{4} = -\tan^{-1}(3/4)$ condition suppresses $\Theta_{1}$ and $\Theta_{6}$ classes, just as the alternating PAC condition, and $\theta_{4} = \tan^{-1}(1/2)$ suppresses $\Theta_{2}$, $\Theta_{5}$ and $\Theta_{7}$ classes, just as the alternating MA condition.
In this case the advantage of the alternating conditions is that they additionally maximize the amplitude of the remaining pathways.
We can also consider the effect of $(\pi/4, -\pi/4, \pi/2, 0)$ polarization condition commonly used in liquid-phase experiments, which removes single-mode diagonal response~\cite{Zanni2001,Hamm2011a}.
In the gas phase, the condition suppresses classes $\Theta_{1}$, $\Theta_{4}$ and $\Theta_{7}$, which does not simplify the spectrum since all branches remain present.
On the other hand, it ensures that the sum over pathway amplitudes phase-matched in all directions is equal to zero, which intuitively agrees with the liquid-phase effect, where all rotational transitions are collapsed into single vibrational response.

\begin{table}\centering
  \caption{\label{tab:middlema_rel_amps}Amplitudes of $S_{II}$ pathway classes suppressed under $(0, \theta_{\mathrm{MA}}, \theta_{\mathrm{MA}}, 0)$ polarization condition relative to $(0,0,0,0)$ polarization.}
  \begingroup\renewcommand{\arraystretch}{1.2}
  \begin{ruledtabular}
\begin{tabular}{cp{4.5cm}c}
Class & Pathway & Relative amplitude\\
\midrule
  \multirow{4}{0.5cm}{$\Theta_{3}$} & PQR, P\ensuremath{{}\mkern1mu\overline{\mkern-1mu\mbox{Q}}}\ensuremath{{}\mkern1mu\overline{\mkern-1mu\mbox{R}}}, QRQ, QPQ, Q\ensuremath{{}\mkern1mu\overline{\mkern-1mu\mbox{P}}}\ensuremath{{}\mkern1mu\overline{\mkern-1mu\mbox{Q}}}, Q\ensuremath{{}\mkern1mu\overline{\mkern-1mu\mbox{R}}}\ensuremath{{}\mkern1mu\overline{\mkern-1mu\mbox{Q}}}, RQP, R\ensuremath{{}\mkern1mu\overline{\mkern-1mu\mbox{Q}}}\ensuremath{{}\mkern1mu\overline{\mkern-1mu\mbox{P}}} & \multirow{2}{*}{0}\\
 & P\ensuremath{{}\mkern1mu\overline{\mkern-1mu\mbox{Q}}}2R, Q\ensuremath{{}\mkern1mu\overline{\mkern-1mu\mbox{P}}}2Q & \(- 5 / \left(3 \left(J_{i} - 1\right)\right)\)\\
 & Q\ensuremath{{}\mkern1mu\overline{\mkern-1mu\mbox{R}}}2Q, R\ensuremath{{}\mkern1mu\overline{\mkern-1mu\mbox{Q}}}2P & \(5 / \left(3 \left(J_{i} + 2\right)\right)\)\\
\midrule
\multirow{6}{0.5cm}{$\Theta_{7}$} & PPQ, P\ensuremath{{}\mkern1mu\overline{\mkern-1mu\mbox{P}}}\ensuremath{{}\mkern1mu\overline{\mkern-1mu\mbox{Q}}}, QQP, QQR, Q\ensuremath{{}\mkern1mu\overline{\mkern-1mu\mbox{Q}}}\ensuremath{{}\mkern1mu\overline{\mkern-1mu\mbox{R}}}, Q\ensuremath{{}\mkern1mu\overline{\mkern-1mu\mbox{Q}}}\ensuremath{{}\mkern1mu\overline{\mkern-1mu\mbox{P}}}, RRQ, R\ensuremath{{}\mkern1mu\overline{\mkern-1mu\mbox{R}}}\ensuremath{{}\mkern1mu\overline{\mkern-1mu\mbox{Q}}} & \multirow{2}{*}{0}\\
 & P\ensuremath{{}\mkern1mu\overline{\mkern-1mu\mbox{P}}}2Q & \(5 / \left(3 \left(J_{i} + 1\right)\right)\)\\
 & Q\ensuremath{{}\mkern1mu\overline{\mkern-1mu\mbox{Q}}}2P & \(- 5 / \left(3 \left(J_{i} - 1\right)\right)\)\\
 & Q\ensuremath{{}\mkern1mu\overline{\mkern-1mu\mbox{Q}}}2R & \(5 / \left(3 \left(J_{i} + 2\right)\right)\)\\
 & Q\ensuremath{{}\mkern1mu\overline{\mkern-1mu\mbox{Q}}}2R & \(- 5 / 3 J_{i}\)\\
\midrule
\multirow{3}{0.5cm}{$\Theta_{6}$} & PPR, P\ensuremath{{}\mkern1mu\overline{\mkern-1mu\mbox{P}}}\ensuremath{{}\mkern1mu\overline{\mkern-1mu\mbox{R}}}, RRP, R\ensuremath{{}\mkern1mu\overline{\mkern-1mu\mbox{R}}}\ensuremath{{}\mkern1mu\overline{\mkern-1mu\mbox{P}}} & \(0\)\\
 & P\ensuremath{{}\mkern1mu\overline{\mkern-1mu\mbox{P}}}2R & \(5 / \left(3 \left(4 J_{i}^{2} + 1\right)\right)\)\\
 & R\ensuremath{{}\mkern1mu\overline{\mkern-1mu\mbox{R}}}2P & \(5 / \left(3 \left(4 J_{i}^{2} + 8 J_{i} + 5\right)\right)\)\\
\end{tabular}
  \end{ruledtabular}\endgroup
\end{table}

\begin{figure}
  \centering
  \includegraphics{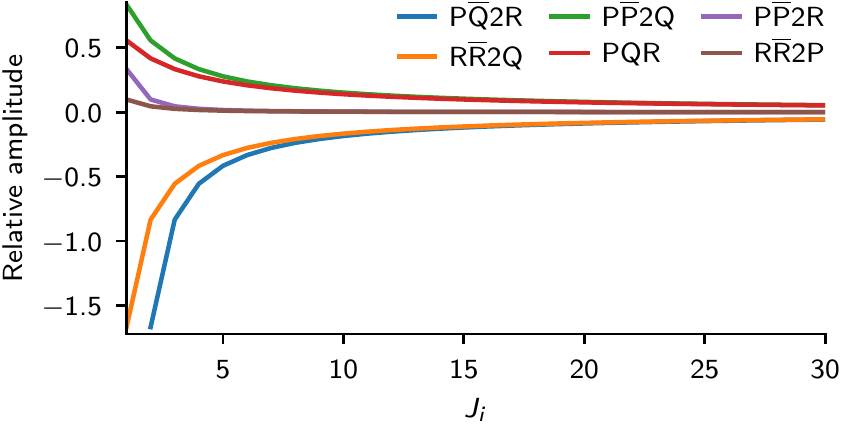}
  \caption{Amplitudes of selected $S_{II}$ pathways under $(0, \theta_{\mathrm{MA}}, \theta_{\mathrm{MA}}, 0)$ polarization condition relative to $(0,0,0,0)$ polarization.
See Tab.~\ref{tab:middlema_rel_amps}.\label{fig:jdependent_amps_middlema}}
\end{figure}

In the preceding, we have discussed classification and suppression of pathways with polarization.
Now we consider the validity of the high-$J$ limit assumed in these results.
The commonly used MA condition exhibits no $J$-dependence, all RC pathways are exactly zeroed under it.
The pathways that are suppressed under the middle MA condition, $(0, \theta_{\mathrm{MA}}, \theta_{\mathrm{MA}}, 0)$, exhibit a varied dependence on the initial state angular momentum, $J_{i}$.
We analyze the dependence by examining the amplitudes of pathways under the specified polarization condition relative to $(0,0,0,0)$ polarization and as a function of $J_{i}$.
Table~\ref{tab:middlema_rel_amps} presents analytical expressions for $R^{(0)}_{0}(\boldsymbol{\epsilon}; \mathbf{J})/R^{(0)}_{0}(\boldsymbol{\epsilon}_{\mathrm{0000}}; \mathbf{J})$, for pathway classes zeroed in the high-J limit.
A subset of pathways has zero amplitude regardless of $J_{i}$ value.
Out of these pathways, the ones belonging to $\Theta_{3}$ and $\Theta_{6}$ classes had no $J_{i}$ dependence to start with, see Tab.~S5.
For $\Theta_{7}$ class, it is the polarization condition that removes the $J_{i}$ dependence.
The $J_{i}$-dependent amplitudes converge to the high-$J$ limit as $\pm 1/J_{i}$ or as $\pm 1/J_{i}^{2}$.
While most pathways quickly converge to the limit, some of them start at relative magnitude above unity and contribute significantly up to $J_{i}\sim 20$, see Fig.~\ref{fig:jdependent_amps_middlema}.
Relative magnitude higher than 1 is caused by the contribution from $k=1$ term in Eq.~\eqref{eq:28}, which is $0$ for $(0,0,0,0)$, maximum for $(0,\pi/2,0,\pi/2)$ [see Eq.~\eqref{eq:57}], and converges to $0$ for $J_{i}\to\infty$.
We refrain from describing the dependence of pathway amplitudes on $J_{i}$ for other polarization conditions, because they exhibit similar behavior:
some subsets of pathways are suppressed for all $J_{i}$, others converge as $\pm 1/J_{i}$ or as $\pm 1/J_{i}^{2}$, and some pathways are enhanced at low $J_{i}$ numbers.

\section{Interstate coherences during waiting time}
\label{sec:interstate-coherences}

A notable feature of RR2DIR spectroscopy with broadband pulses is the variety of interstate coherences produced by the first two excitations.
The type of coherent state produced depends on the phase-matching condition---for S$_{I}$ and S$_{II}$ directions, it will be a low-frequency rotational coherence in the ground or first excited vibrational manifold, while for S$_{III}$ direction, it will be a high-frequency coherence between the ground and second excited vibrational manifolds.
Here, we focus on RC pathways phase-matched in S$_{I}$ and S$_{II}$ directions.
We describe notable features of the coherent evolution during $t_{2}$ and present several ways to exploit it.
A complete list of RC pathways and frequencies categorized with respect to branches and subbranches is given in the Supplemental Material in Tab.~S1 and S2.

\begin{figure}
  \centering
  \includegraphics{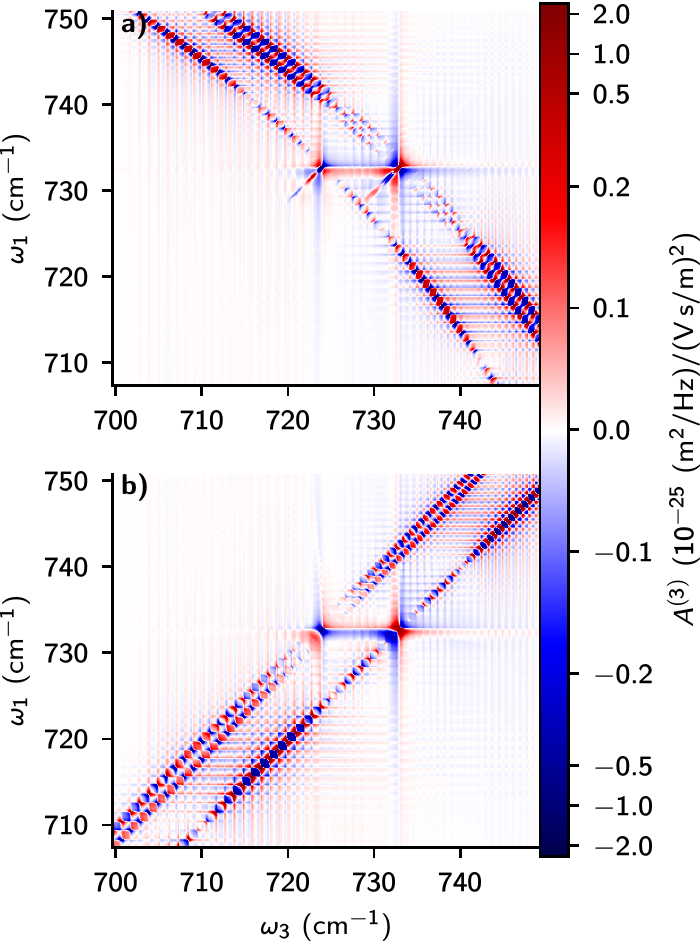}
  \caption{CH$_{3}$$^{35}$Cl difference spectra, $A^{(3)}(\omega_1, t_2=1\text{ ps}, \omega_3)-A^{(3)}(\omega_1, t_2=1.2\text{ ps}, \omega_3)$, phase-matched in (a) $S_{I}$ and (b) $S_{II}$ direction.
    In both cases the polarization condition was $(0, \theta_{\mrm{MA}}, \theta_{\mrm{MA}}, 0)$.
  Logarithmic scale is used for $|A^{(3)}|>0.19$ and linear scale for lower absolute values.\label{fig:exploit}}
\end{figure}

\begin{figure*}
  \centering
  \includegraphics{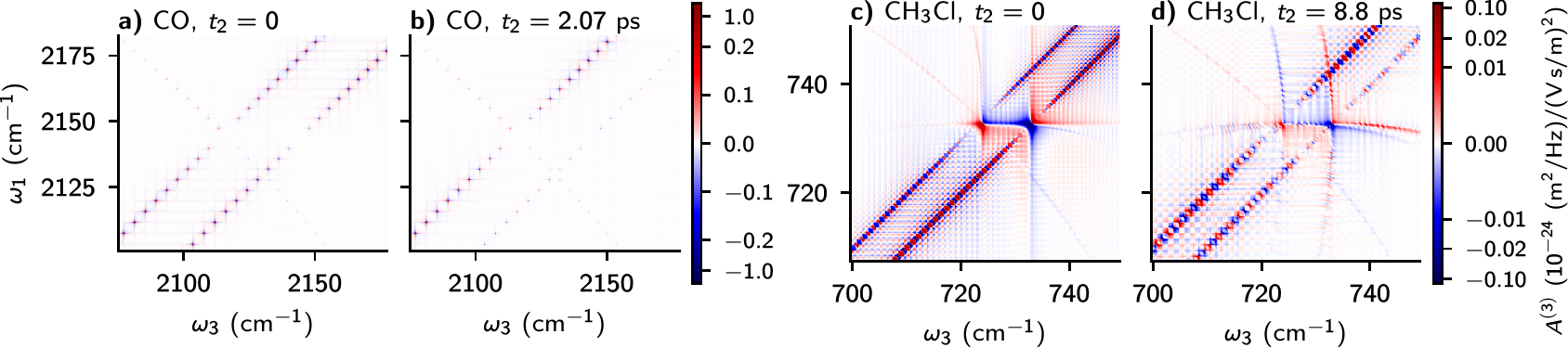}
  \caption{Panels a) and b): CO spectra under $(0, 0, \theta_{\mrm{PAC}}, -\theta_{\mrm{PAC}})$ polarization condition at (a) $t_{2}=0$ and at (b) $t_{2}=2.07$ ps.
    Logarithmic scale is used for $|A^{(3)}|>0.2$ and linear scale for lower absolute values.
    Panels c) and d): CH$_{3}$$^{35}$Cl spectra under $(0, 0, \theta_{\mrm{PAC}}, -\theta_{\mrm{PAC}})$ polarization condition at (c) $t_{2}=0$ and at (d) $t_{2}=8.8$ ps.
    Logarithmic scale is used for $|A^{(3)}|>0.01$ and linear scale for lower absolute values.
    \label{fig:wt_interf}}
\end{figure*}

In Sec.~\ref{sec:suppression} we have found there is no polarization condition that suppresses all non-RC pathways.
However, as long as collisional relaxation occurs on much longer time scale than rotational coherence evolution, RC pathways can be isolated by a differential time measurement.
Figures~\ref{fig:exploit} and \ref{fig:wt_interf} show RR2DIR spectra $A^{(3)}$, including lineshapes, for both CO and CH$_3$$^{35}$Cl.
Transition amplitudes, energy levels, partition functions and pressure-broadening parameters from HITRAN~\cite{Gordon2017,Li2015,Coxon2004,Devi2018} were used to generate the spectra.
Panels a) and b) in Fig.~\ref{fig:exploit} show difference between spectra, $A^{(3)}(\omega_1, t_2=1\text{ ps}, \omega_{3})-A^{(3)}(\omega_1, t_2=1.2\text{ ps}, \omega_{3})$, under the middle MA condition for (a) $S_{I}$ and (b) $S_{II}$ direction.
The subtraction removes non-RC pathways, $\Theta_{4}$, $\Theta_{5}$, $\Theta_{6}$, $\Theta_{7}$, since their amplitudes change minimally on the time scale of 200 fs.
Additionally, the used polarization condition removes the $\Theta_{3}$ RC class, which leaves only classes $\Theta_{1}$ and $\Theta_{2}$.
For the $S_{II}$ signal, these are the diagonal RC pathways and for $S_{I}$ the anti-diagonal RC pathways.

In RR2DIR spectroscopy, most RC pathways do not overlap spectrally with other RC pathways in the $(\Omega_{1}, \Omega_{3})$ plane, as was shown in Fig.~\ref{fig:pws_per_peak}.
This is in contrast to rotational coherence spectroscopies~\cite{Felker1995,Seeger2009}, in which the signal is a sum of contributions from many interfering RC pathways.
Nevertheless, there can be found several subbranches that include multiple RC pathways oscillating at different rotational frequencies and interfering during waiting time.
For the S$_{II}$ direction, these are the diagonal P-P, Q-Q and R-R branches, which are not split into multiple subbranches.
The P-P branch includes only negative RC frequencies, R-R only positive frequencies, and Q-Q mixes both.
In analogy to suppressing parts of molecular response with specific polarization conditions, we considered whether all rotational coherences within a subbranch can be made to interfere destructively by measuring the response at specific waiting time $t_{2}$.

We first consider interference of pathways phasematched in $S_{II}$ direction.
For now we limit the analysis to diatomic molecules and stretching modes of linear molecules, by excluding pathways with Q-type transitions.
In this case, for the P-P subbranch there is only a pair of pathways: PRR oscillating at frequency $-2B_{0}(2J_{i}-1)$ and P\ov{R}\ov{R} at $-2B_{1}(2J_{i}+1)$.
The beat signal between them can be approximately expressed as:
\begin{equation}
  \label{eq:34}
  e^{-\frac{i}{\hbar}[4B_{0}J_{i}-\Delta B(2J_{i}+1)]t_{2}} \cos \left[(2B_{0}-\Delta B (2J_{i}+1))\frac{t_{2}}{\hbar}\right],
\end{equation}
where $\Delta B=B_{0}-B_{1}$, $B_{\nu}$ is the rotational constant for $\nu$ vibrational state, and higher order rotational Hamiltonian terms are ignored.
For the R-R subbranch, the pair of pathways is: RPP at $2B_{0}(2J_i+3)$ and R\ov{P}\ov{P} at $2B_{1}(2J_{i}+1)$.
The beat signal for the R-R subbranch is approximately:
\begin{equation}
  \label{eq:35}
  e^{\frac{i}{\hbar}[4B_{0}(J_{i}+1)-\Delta B(2J_{i}+1)]t_{2}} \cos \left[(2B_{0}+\Delta B (2J_{i}+1))\frac{t_{2}}{\hbar} \right].
\end{equation}
The envelope of the P-P branch is slowly modulated at frequency $[2B_{0}-\Delta B (2J_{i}+1)]/\hbar$ whereas the for the R-R branch it is modulated at $[2B_{0} + \Delta B (2J_{i}+1)]/\hbar$. 
So for one branch the modulation frequency increases with $J_{i}$, and for the other it decreases.
If $\Delta B$ were zero, one could easily cancel all RC pathways in both branches by recording the signal at $t_2 = \hbar \pi / (4 B_0)$,
but non-zero $\Delta B$, such complete destructive interference is not possible.
However, a significant suppression of RC pathways over a broad spectral range is still achievable if $\Delta B J_i t_2/\hbar \ll 1$.
For CO, in the $\pm 25$ cm$^{-1}$ range around the band origin ($J_{i}=-6,\dots,6$) the $t_{2}$ value giving optimal suppression changes by $0.28$ ps~\cite{Liu2020,Huber1979}.
For reference, optimal $t_{2}$ for $J_{i}=0$ is $2.18$ ps.
Comparing panels a) and b) in Fig.~\ref{fig:wt_interf}, we see that this approach is successful for CO.
Panel b) shows 2DIR spectrum at $t_{2}=2.08$ ps under the PAC condition in which P-P and R-R resonances are strongly suppressed by destructive interference and only P-2P and R-2R branches have appreciable amplitude.

We now turn to the more general case of a symmetric top molecule.
For CH$_{3}$$^{35}$Cl, in the same spectral range around the band origin as before ($J_{i}=-29,\dots,29$) the optimal $t_{2}$ changes by $5.11$ ps~\cite{Litz2003}, compared to optimal $t_{2}=9.81$ ps for $J_{i}=0$.
Figure~\ref{fig:wt_interf}d) shows the 2D spectrum of CH$_{3}$$^{35}$Cl under the PAC angle condition at $t_{2}=8.8$ ps.
Compared to Fig.~\ref{fig:wt_interf}c), the R-R branch is partially suppressed but the P-P branch is largely unaffected, besides the change of phase of complex 2D resonances.
Even in the R-R branch destructive interference is not as complete as for CO.
For one, this is caused by uneven amplitude of the two interfering pathways.
The ratio of amplitudes is equal to $(J_{i}-1)/(J_{i}+1)$ for $K_{m}=0$, but for maximum $K_{m}$ and in the limit of $J_{i}\to\infty$ it reaches $1/3$.
Moreover, symmetric tops include an additional pair of RC pathways.
For the P-P branch these are PQQ oscillating at $-2B_{0}J_{i}$ and P\ov{Q}\ov{Q} at $-2B_{1}J_{i}$, and similarly for the R-R branch.
The beat signal between them is approximately given by:
\begin{equation}
  \label{eq:36}
  e^{-\frac{i}{\hbar}(B_{0}+B_{1})t_{2}}\cos \left(\Delta B J_{i} \frac{t_{2}}{\hbar}\right),
\end{equation}
where the period of the real envelope strongly depends on $J_{i}$, which precludes the possibility of destructive interference over the whole subbranch.

The interference of RC pathways in the Q-Q branch is qualitatively different from that in P-P and R-R branches, since the Q-Q branch includes pathways with both positive and negative frequency coherences.
The total of four RC pathways can be split into two pairs, such that the pathway amplitudes are equal within both pairs for all $J_{i}$, $K_{m}$.
This could potentially lead to perfect pairwise destructive interference within both pairs.
The two pairs oscillate at frequencies $2B_{0}(J_{i}+1)$, $-2B_{1}(J_{i}+1)$ and $-2B_{0}J_{i}$, $2B_{1}J_{i}$, allowing us to write the beat signal as:
\begin{equation}
  \label{eq:37}
  \begin{split}
    &A_{J_{i},K_{m}}e^{-\frac{i}{\hbar}\Delta B(J_{i}+1)t_{2}}\cos \left[(B_{0}+B_{1})(J_{i}+1)\frac{t_{2}}{\hbar} \right]\\
    +& B_{J_{i},K_{m}}e^{\frac{i}{\hbar}\Delta BJ_{i}t_{2}}\cos \left[(B_{0}+B_{1})J_{i}\frac{t_{2}}{\hbar} \right].
  \end{split}
\end{equation}
In contrast to Eqs.~(\ref{eq:34}--\ref{eq:36}), here both the complex exponentials and the real cosines strongly depend on $J_{i}$, which will unfortunately prevent us from obtaining destructive interference over multiple resonances.

For the $S_{I}$ direction, there are several subbranches which contain interfering RC pathways.
An antidiagonal subbranch of P-R branch contains a pair coherences at frequencies $-2B_{0}J_{i}$, $-2B_{1}J_{i}$ with interference pattern mimicking Eq.~\eqref{eq:36}.
Analogous interference signal is also present in an R-P subbranch with frequencies $2B_{0}(J_{i}+1)$, $2B_{1}(J_{i}+1)$.
The four RC pathways from S$_{II}$ Q-Q branch, Eq.~\eqref{eq:37}, are also present as $S_{I}$ pathways, but they are essentially split into two Q-Q subbranches phase-matched in S$_{I}$ direction.
The $A_{J_{i},K_{m}}$ pair is associated with $\Omega_{3}=\ketbra{1R}{0R}$ subbranch and the $B_{J_{i},K_{m}}$ pair with $\Omega_{3}=\ketbra{1P}{0P}$ subbranch.
As we noted when describing interference of $S_{II}$ pathways, in none of these case can we expect broadband destructive interference.
Finally, the pair of $2B_{1}J_{i}$ and $-2B_{1}(J_{i}+1)$ coherences in the \ket{2Q}\bra{1Q} subbranch of the Q-2Q branch produces a notable beat pattern:
\begin{equation}
  \label{eq:38}
  e^{-\frac{i}{\hbar}B_{1}t_{2}}\cos \left[B_{1}(2J_{i}+1)\frac{t_{2}}{\hbar} \right].
\end{equation}
Here, the complex low-frequency envelopes of all the resonances in the subbranch oscillate in sync at $B_{1}$ frequency. Therefore, any observed irregularities could serve as sensitive probes for higher order terms of the rotational Hamiltonian, which we omitted in present analysis.

\section{Conclusions}
\label{sec:conclusions}

In this article, we have presented the fundamental background and notation for description of RR2DIR spectra.
We discussed the features of RR2DIR spectroscopy unique to it among third-order spectroscopies, including the band and branch structure and the separation of the molecular response into polarization classes.
The presented theory was used to explain the imbalance of rephasing vs nonrephasing pathways and the effect of various polarization conditions on RR2DIR spectra.
The key new results most immediately applicable to applications are highlighted in a concise letter-style paper~\cite{Kowzan2022b}.
These results were supplemented by two additional polarization conditions which enable complete control over polarization-dependent response of diatomic molecules and stretching modes of linear molecules.
Furthermore, we have discussed the influence of rotational coherence evolution during waiting time on the spectra and the conditions for collective destructive interference over whole 2D branches.
Since most results apply to whole branches, we emphasize that the presented theory is also highly relevant for gas-phase 2DIR measurements that do not resolve individual lines, for example the recent work of Ziegler and co-workers~\cite{Mandal2018,Pack2019}, either because of insufficient resolution, high line density, or large pressure broadening.
While simulations were performed on a simple diatomic molecule and a symmetric top molecule, our results are fully applicable to the more prevalent asymmetric tops.

The results presented in the article were obtained with the help of \textsc{rotsim2d} library and applications, which enables the user to simulate 2DIR spectra, inspect their structure and study their polarization dependence, and derive other polarization conditions.
These computer resources have been made freely available for download here~\cite{Kowzan2022rotsim2d}.
We expect in the future that machine learning will be applied to deciphering RR2DIR spectra, as has been applied to linear multi-species spectra~\cite{Liang2022,Schmidt2021}.
The first-principles framework and computer simulation tools presented here can be used for generating spectra for training machine learning algorithms for interpreting RR2DIR spectra and extracting species concentrations.
With this solid theoretical foundation, rapidly advancing progress in temporally coherent mid-IR and long-wave IR light sources, and advances in computation, rotationally-resolved 2DIR spectroscopy is well poised to become a powerful tool for molecular spectroscopy.

\begin{acknowledgments}
This project has received funding from the European Union's Horizon 2020 research and innovation programme under the Marie Sklodowska-Curie grant agreement No 101028278.
This work was supported by the U.S.
National Science Foundation under award number 1708743 and the U.S. Air Force Office of Scientific Research under grant number FA9550-20-1-0259.
\end{acknowledgments}

\section{Decomposition of $O_{ijkl}$}
\label{app:fourfold}

The expectation value of the four-fold dipole interaction operator, $\langle O_{ijkl}\rangle$, encapsulates the dependence of the response on beam polarization, molecular orientation and alignment, and molecular transition dipoles.
The main task of the current section is to decompose  $\langle O_{ijkl}\rangle$ into polarization and angular momentum part and transition dipole part, as in Eq.~\eqref{eq:23}.
Analyzing the former will allow us to find polarization conditions useful for 2D spectroscopy.
The form of $\langle O_{ijkl} \rangle$ most useful for our purposes is obtained by expressing all involved quantities as spherical tensors or spherical tensor operators, and subsequently by manipulating them with angular momentum algebra techniques~\cite{Brink1968,Zare1991}.

The dot product of a polarization vector and the dipole interaction operator can be expressed in terms of spherical tensor operators as:
\begin{equation}
  \label{eq:39}
  \begin{split}
    \widetilde{\epsilon}_i\cdot\vec{\mu}_i &= -\sqrt{3}[T^{(1)}(\widetilde{\epsilon}_i)\otimes T^{(1)}(\vec{\mu}_i)]^{(0)}_0\\
    &= \sum_q (-1)^q T^{(1)}_q(\widetilde{\epsilon}_i)T^{(1)}_{-q}(\vec{\mu}_i),
  \end{split}
\end{equation}
where the above is a special case of general spherical tensor product:
\begin{equation}
\label{eq:40}
\begin{split}
[T^{(k_1)}\otimes T^{(k_2)}]^{(K)}_Q = &\sum_q (-1)^{k_1-k_2+Q}\sqrt{2K+1}\\
&\times\begin{pmatrix}k_1&k_2&K\\q&Q-q&-Q\end{pmatrix}T^{(k_1)}_q T^{(k_2)}_{Q-q},
\end{split}
\end{equation}
In a similar manner we can express the projection operators $P_{\alpha}$ as rank zero spherical tensor operators:
\begin{equation}
\label{eq:41}
P_{\alpha} = \sqrt{2J_{\alpha}+1}T^{(0)}_0(\nu_{\alpha}J_{\alpha}),
\end{equation}
where:
\begin{equation}
\label{eq:42}
\begin{split}
T^{(K)}_{Q}&(\nu'J';\nu''J'') = \sum_{m'',m'}(-1)^{J'-m'}\sqrt{2K+1}\\
&\times\begin{pmatrix}J'&K&J''\\-m'&Q&m''\end{pmatrix}
|\nu'J'm'\rangle\langle \nu''J''m''|.
\end{split}
\end{equation}
are the so-called state multipoles \cite{Blum2012}.
Subsequently, we collect the polarization tensors into (at most) fourth order spherical tensor, $T^{(K)}(\boldsymbol{\epsilon}; k_1,k_2)$, and separately collect dipole interaction operators with projection operators into another fourth order spherical tensor operator, $T^{(K)}(\boldsymbol{\mu};k_1,k_2)$.
This is effected by using the recoupling transformation for four angular momenta \cite{Williams1994,Zare1991}, which allows us to express the total four-fold interaction operator as:
\begin{equation}
\label{eq:43}
\begin{split}
O_{ijkl} =& \sqrt{(2J_j+1)(2J_k+1)(2J_l+1)}\left[\sum_{K=0}^4\sum_{Q = -K}^{K}(-1)^{K-Q}\right.\\
&\times\left.\sum_{k_1}^2\sum_{k_2}^2 T^{(K)}_Q(\boldsymbol{\epsilon}; k_1,k_2)
T^{(K)}_{-Q}(\boldsymbol{\mu};k_1,k_2)
\right].
\end{split}
\end{equation}
The compound polarization tensor is explicitly given by:
\begin{equation}
  \label{eq:44}
  \begin{split}
  T^{(K)}(\boldsymbol{\epsilon}; k_1,k_2) = \big[&
  [T^{(1)}(\widetilde{\epsilon}_{i}) \otimes T^{(1)}(\widetilde{\epsilon}_{j})]^{(k_{1})}\\ &\otimes
  [T^{(1)}(\widetilde{\epsilon}_{k}) \otimes T^{(1)}(\widetilde{\epsilon}_{l})]^{(k_{2})}\big]^{(K)},
\end{split}
\end{equation}
and the compound dipole tensor operator by:
\begin{equation}
  \label{eq:45}
  \begin{split}
    T^{(K)}&(\boldsymbol{\mu};k_1,k_2)\\= \big[&
    [T^{(1)}(\vec{\mu_{i}}) \otimes T^{(0)}(\nu_{j}J_{j}) \otimes T^{(1)}(\vec{\mu_{j}})]^{(k_{1})} \otimes T^{(0)}(\nu_{k}J_{k})\\ & \otimes
    [T^{(1)}(\vec{\mu_{k}}) \otimes T^{(0)}(\nu_{l}J_{l}) \otimes T^{(1)}(\vec{\mu_{l}})]^{(k_{2})}\big]^{(K)}.
  \end{split}
\end{equation}
The polarization tensor, Eq.~\eqref{eq:44}, cannot be reduced further and it only remains to express it explicitly in terms of polarization angles, Eq.~\eqref{eq:6}, see Appendix~\ref{app:comp-polar-tens}.
The dipole tensor operator in Eq.~\eqref{eq:45} can be further factored into physically meaningful components when taking the expectation value of $O_{ijkl}$.
For this, we express $\rho$ as a sum of state multipoles, Eq.~\eqref{eq:42}:
\begin{equation}
  \label{eq:46}
  \begin{split}
    \rho = \sum_{J',J'',K,Q} \Tr\left[\rho T^{(K)}_Q(\nu'J';\nu''J'')^{\dagger}\right] T^{(K)}_{Q}(\nu'J';\nu''J''),
  \end{split}
\end{equation}
with the expansion coefficients given by:
\begin{equation}
  \label{eq:47}
  \begin{split}
    \Tr [\rho &T^{(K)}_Q(\nu' J';\nu'' J'')^{\dagger}] \\ =&\sum_{M',M''}(-1)^{J'-M'}\sqrt{2K+1}
    \begin{pmatrix}J'&J''&K\\-M'&M''&Q\end{pmatrix}\\ &\times\bra{\nu' J'M'}\rho\ket{\nu'' J''M''}.
  \end{split}
\end{equation}
For the initial density matrix, we restrict the above to $J'=J''$.
In this case, different $K$ orders represent angular distribution anisotropies, with $K=0$ term corresponding to population averaged over $M$ sublevels, odd $K$ terms corresponding to orientation and even $K$ terms to alignment of the population.
Here, orientation implies preferential occupation of some $\ket{J,M}$ vs. $\ket{J,-M}$ state and alignment implies preferential occupation of some pair of $\ket{J,M}$ and $\ket{J,-M}$ states over other pairs of $\ket{J,M'}$ and $\ket{J,-M'}$ states.
Furthermore, we restrict our treatment to initial density matrix in thermal equilibrium, Eq.~\eqref{eq:13}, which implies that only $K=Q=0$ term is nonzero.
In this case, the state multipole expansion of $[\rho^{(0)}(-\infty)]_{\nu_{i},J_{i}}$ is particularly simple:
\begin{equation}
  \label{eq:48}
  [\rho^{(0)}(-\infty)]_{\nu_{i},J_{i}} =
  \frac{N_{\nu_{i}J_{i}}}{N}\frac{1}{\sqrt{2J_{i}+1}}T^{(0)}_0(\nu_{i} J_{i}),
\end{equation}
$T^{(0)}_0(\nu_{i} J_{i})\equiv T^{(0)}_0(\nu_{i} J_{i};\nu_{i} J_{i})$, $N$ is the total concentration of the active molecule and $N_{\nu_{i}J_{i}}$ is the concentration of the molecules in the $\nu_{i}J_{i}$ state.

The expectation value of $O_{ijkl}$ is given by the trace of its product with $[\rho^{(0)}(-\infty)]_{\nu_{i},J_{i}}$, Eq.~\eqref{eq:21}.
Having expressed the density matrix in terms of state multipoles, the evaluation of the trace is simplified by the following equality, which is valid for an arbitrary state multipole $T^{(K_{1})}_{Q_{1}}(\nu'J';\nu''J'')$ and spherical tensor operator $T^{(K_{2})}_{Q_{2}}(\hat{A})$:
\begin{multline}
  \label{eq:49}
  \operatorname{Tr}[T^{(K_{1})}_{Q_{1}}(\nu'J';\nu''J'') T^{(K_{2})}_{Q_{2}}(\hat{A})]
  \\ =\sum_{m'} \bra{\nu'J'm'}
  T^{(K_{1})}_{Q_{1}}(\nu'J';\nu''J'')
  T^{(K_{2})}_{Q_{2}}(\hat{A}) \ket{\nu'J'm'}\\
  = \frac{(-1)^{J'+J''+Q_{1}}}{\sqrt{2K_{1}+1}} \langle \nu''J''\|T^{(K_{2})}(\hat{A})\|\nu' J'\rangle
  \delta_{K_{1},K_{2}} \delta_{Q_{1},-Q_{2}}.
\end{multline}
The latter equality is obtained by inserting expansion Eq.~\eqref{eq:42}, using the Wigner-Eckhart theorem on $T^{(K_{2})}_{Q_{2}}(\hat{A})$ and using the orthogonality of Wigner-3j coefficients to perform the summation.
Note that for $T^{(K_{1})}_{Q_{1}}(\nu'J';\nu''J'')=T^{(0)}_0(\nu_{i} J_{i})$, Eq.~\eqref{eq:49} reduces to:
\begin{equation}
  \label{eq:50}
  \operatorname{Tr}[T^{(0)}_{0}(\nu J) T^{(K_{2})}_{Q_{2}}(\hat{A})] = \langle \nu J\|T^{(K_{2})}(\hat{A})\|\nu J\rangle
  \delta_{0,K_{2}}.
\end{equation}
Substituting the density matrix, Eq.~\eqref{eq:48}, and the four-fold dipole operator, Eq.~\eqref{eq:43}, into Eq.~\eqref{eq:21}, we obtain:
\begin{widetext}
  \begin{equation}
    \label{eq:51}
    \begin{split}
      \langle O_{ijkl} \rangle & = \frac{N_{\nu_{i}J_{i}}}{N}\frac{\sqrt{(2J_j+1)(2J_k+1)(2J_l+1)}}{ \sqrt{2J_{i}+1}} \sum_{K=0}^4\sum_{Q = -K}^{K}(-1)^{K-Q}
\sum_{k_1}^2\sum_{k_2}^2 T^{(K)}_Q(\boldsymbol{\epsilon}; k_1,k_2)
\Tr [T^{(0)}_{0}(\nu_{i} J_{i}) T^{(K)}_{-Q}(\boldsymbol{\mu};k_1,k_2)] \\
      & =\frac{N_{\nu_{i}J_{i}}}{N}\frac{\sqrt{(2J_j+1)(2J_k+1)(2J_l+1)}}{\sqrt{2J_{i}+1}} \sum_{k=0}^{2} T^{(0)}_{0}(\boldsymbol{\epsilon}; k, k) \langle\nu_{i}J_{i}\|T^{(0)}(\boldsymbol{\mu}; k, k)\| \nu_{i}J_{i}\rangle\\
      & = \frac{N_{\nu_{i}J_{i}}}{N} \frac{1}{\sqrt{2J_{i}+1}} \langle\nu_{i}J_{i}\|T^{(0)}(\boldsymbol{\mu})\| \nu_{i}J_{i}\rangle
      \sum_{k=0}^{2} T^{(0)}_{0}(\boldsymbol{\epsilon}; k, k) G(J_{i},J_{j},J_{k},J_{l}; k),
    \end{split}
  \end{equation}
\end{widetext}
which brings us to Eqs.~(\ref{eq:23}, \ref{eq:28}).
The first equality was obtained by applying Eq.~\eqref{eq:50}, which collapses sums over $K$ and $Q$ to a single term.
The sum over $k_{1}$ and $k_{2}$ was collapsed to a single sum over $k$ since contraction of a tensor product to a scalar ($K=Q=0$) requires $k_{1}=k_{2}$.
The second equality was obtained by repeatedly applying the Wigner-Eckhart theorem for tensor product operators~\cite{Zare1991} on $\langle\nu_{i}J_{i}\|T^{(0)}(\boldsymbol{\mu}; k, k)\| \nu_{i}J_{i}\rangle$ [Eq.~\eqref{eq:45}]:
\begin{equation}
  \label{eq:52}
  \begin{split}
  \langle  \nu_{\alpha}J_{\alpha} \|[T^{(k_{1})} \otimes T^{(k_{2})}]^{k}&\|\nu_{\beta}J_{\beta}\rangle =
  (-1)^{k+j_{\alpha}+j_{\beta}} \sqrt{2k+1}\\ \times \sum_{\nu'',J''} \begin{Bmatrix}k_{1}&k_{2}&k\\J_{\beta}&J_{\alpha}&J''\end{Bmatrix}&
  \langle \nu_{\alpha}J_{\alpha}\|T^{(k_{1})}|\nu'' J''\rangle \langle \nu'' J'' \| T^{(k_{2})} \| \nu_{\beta}J_{\beta}\rangle.
  \end{split}
\end{equation}
The presence of $T^{(0)}(\nu_{\alpha}J_{\alpha})$ operators in Eq.~\eqref{eq:45} reduces applications of Eq.~\eqref{eq:52} to only a single term in the sum with $(\nu'',J'')=(\nu_{\alpha},J_{\alpha})$ and cancels the $\sqrt{2J_{\alpha}+1}$ factors in Eq.~\eqref{eq:51}.
The Wigner-6j coefficients are collected into the G-factor, Eq.~\eqref{eq:29}, and the reduced matrix elements into $\langle\nu_{i}J_{i}\|T^{(0)}(\boldsymbol{\mu})\| \nu_{i}J_{i}\rangle$, Eq.~\eqref{eq:25}.

\section{Compound polarization tensor}
\label{app:comp-polar-tens}

The spherical components of an arbitrary Cartesian vector, $\vec{v} = x\hat{e}_{x} + y\hat{e}_{y} + z\hat{e}_{z}$, are given by:
\begin{equation}
  \label{eq:53}
  T^{(1)}_{\pm q}(\vec{v}) = \mp\frac{1}{\sqrt{2}}(x+iy),\qquad
  T^{(1)}_{0}(\vec{v}) = z.
\end{equation}
Given the parametrization of polarization vectors, Eq.~\eqref{eq:6}, the spherical components are given by:
\begin{equation}
  \label{eq:54}
  T^{(1)}_{\pm q}(\hat{\epsilon}) = \mp\frac{1}{\sqrt{2}}e^{\pm i \theta},\qquad
  T^{(1)}_{0}(\hat{\epsilon}) = 0.
\end{equation}
Using Eq.~\eqref{eq:40}, we explicitly expand $T^{(0)}_0(\boldsymbol{\epsilon}; k, k)$, Eq.~\eqref{eq:44}, in terms of the component vectors:
\begin{equation}
  \label{eq:55}
  \begin{split}
    T^{(0)}_0&(\boldsymbol{\epsilon}; k, k) =
    (2k+1)\sum_{q=-k}^{k} \,\sum_{q',q'}
    \begin{pmatrix}k & k & 0\\q & -q & 0\end{pmatrix}\\ \times&
    \begin{pmatrix}1 & 1 & k\\q' & -q' & -q\end{pmatrix}
    \begin{pmatrix}1 & 1 & k\\q'' & -q-q'' & q\end{pmatrix}\\ \times&
    T^{(1)}_{q'}(\widetilde{\epsilon}_i)
    T^{(1)}_{q-q'}(\widetilde{\epsilon}_j)
    T^{(1)}_{q''}(\widetilde{\epsilon}_k)
    T^{(1)}_{-q-q''}(\widetilde{\epsilon}_l).
  \end{split}
\end{equation}
Evaluating the sums for different $k$ components and expressing the results in terms of cosines of four angles obtains:
\begin{equation}
  \label{eq:56}
  \begin{split}
  T^{(0)}_0(\boldsymbol{\epsilon}; 0, 0) =  &\frac{1}{6}\big(
  \cos(\theta_{i}-\theta_{j}-\theta_{k}+\theta_{l})\\
  &+ \cos(\theta_{i}-\theta_{j}+\theta_{k}-\theta_{l})\big)\\
    = &\frac{1}{3}\cos(\theta_{i}-\theta_{j})\cos(\theta_{k}-\theta_{l})
\end{split}
\end{equation}
\begin{equation}
  \label{eq:57}
  \begin{split}
    T^{(0)}_0(\boldsymbol{\epsilon}; 1, 1) =  &\frac{\sqrt{3}}{12}\big(
    \cos(\theta_{i}-\theta_{j}-\theta_{k}+\theta_{l})\\
    &- \cos(\theta_{i}-\theta_{j}+\theta_{k}-\theta_{l})\big)\\
    = &\frac{\sqrt{3}}{6}\sin(\theta_{i}-\theta_{j})\sin(\theta_{k}-\theta_{l})
  \end{split}
\end{equation}
\begin{equation}
  \label{eq:58}
  \begin{split}
    T^{(0)}_0(\boldsymbol{\epsilon}; 2, 2) = & \frac{\sqrt{5}}{60}\big(
    \cos(\theta_{i}-\theta_{j}-\theta_{k}+\theta_{l})\\
    &+ \cos(\theta_{i}-\theta_{j}+\theta_{k}-\theta_{l})\\
    &+ 6\cos(\theta_{i}+\theta_{j}-\theta_{k}-\theta_{l})\big)
  \end{split}
\end{equation}

\section{The magic angle and population-alignment canceling conditions}
\label{app:magic-angle}

The form of $O_{ijkl}$ operator adopted here, Eq.~\eqref{eq:22}, is convenient because it allows us to express the macroscopic polarization in terms of the expectation value of the operator in the molecule's initial state, Eq.~\eqref{eq:21}, and because it involves only rank zero state multipoles, Eq.~\eqref{eq:41}. The zero-rank state multipoles are involved only trivially in angular momentum recoupling and contribute only a factor of $\delta_{J'',J_{\alpha}}/\sqrt{2J_{\alpha}+1}$ to the reduced matrix elements, Eq.~\eqref{eq:52}.

The price for this formal convenience is that the ordering of dipole operators (and polarization vectors) in Eq.~\eqref{eq:22} no longer matches the experimental order of interactions~\cite{Williams1994}.
See for example Eq.~\eqref{eq:19}, where the sum over $M_{\alpha}$ states for an RC pathway was written in this way and where $ijkl\to 2341$.
This means that in general \textit{ijkl} need not be $1234$ or $4321$, except for purely bra-side or ket-side sequences of transitions, respectively.
For RC pathways in particular, the indices are either $2jk1$ or $1jk2$, so the polarization vectors of the first two pump beams, which produce the RC state, are not directly coupled to each other.
While it does not prevent us from deriving the magic angle, Eq.~\eqref{eq:35}, and other polarization conditions, it prevents us from providing physical interpretation of the invididual $k$ components of the total polarization tensor, Eq.~\eqref{eq:44}, and dipole tensor, Eq.~\eqref{eq:45}.

In the case of RC pathways, we can partially regain the experimental ordering by performing a cyclic permutation of indices, $2jk1\to jk12$ or $1jk2\to jk21$, which maintains the convenience of using rank zero state multipoles but at the same time directly couples the first two interactions which produce RC.
For all RC pathways, we obtain matrix elements of the form:
\begin{equation}
  \label{eq:59}
  \bra{J_{j}M_{j}}
  (\widetilde{\epsilon}_j\cdot\vec{\mu_{j}})P_{k}
  (\widetilde{\epsilon}_k\cdot\vec{\mu_{k}})P_{l}
  (\widetilde{\epsilon}_2\cdot\vec{\mu_{2}})P_{i}
  (\widetilde{\epsilon}_1\cdot\vec{\mu_{1}})
  \ket{J_{j}M_{j}}.
\end{equation}
$P_{i}$ is the projection operator onto the initial rotational subspace ($J_{i}$).
The RC is between states with angular momenta $J_{l}$ and $J_{j}$.
The first interaction ($\widetilde{\epsilon_{1}}\cdot\vec{\mu}_{1}$) induces transition to the $J_{j}$ state and the second one ($\widetilde{\epsilon_{2}}\cdot\vec{\mu}_{2}$) to the $J_{l}$ state.
Thanks to the permutation, we only need to consider the coupling between these two interactions to find the magic angle condition.
From Eq.~\eqref{eq:52}, the Wigner-6j coefficient for coupling of $J_{l}$, $J_{i}$, $J_{j}$, $\widetilde{\epsilon}_{2}$, $\widetilde{\epsilon}_{1}$ is:
\begin{equation*}
  \label{eq:60}
  \begin{Bmatrix}1&1&k\\J_{j}&J_{l}&J_{i}\end{Bmatrix},
\end{equation*}
which imposes the triangle condition on $J_{j}$, $J_{l}$ and $k$.
By definintion $J_{j}\ne J_{l}$ for RC pathways, hence the coupling coefficient is non-zero only for $k>0$.
We can therefore conclude that the general magic angle condition is the one which zeroes $k=1,2$ components of the total polarization tensor, Eqs.~(\ref{eq:57}, \ref{eq:58}), under this recoupling.
Solving both equalities gives the condition from Sec.~\ref{sec:suppression}.

The population-alignment cancelling angle condition can be analyzed similarly.
After performing a permutation of indices that couples first two interactions together, we find that pathways in classes suppressed by the PAC condition have the same ratio of $k=0$ (population) to $k=2$ (alignment) component of the G-factor, Eq.~\eqref{eq:29}, equal to $2\sqrt{5}$.
Therefore, the general PAC condition is a condition that suppresses $k=1$ component, which is easily obtained by setting $\theta_{1}=\theta_{2}$ [Eq.~\eqref{eq:57}], and for which:
\begin{equation}
  \label{eq:61}
  \frac{T^{(0)}_{0}(\boldsymbol{\epsilon}; 2, 2)}{T^{(0)}_{0}(\boldsymbol{\epsilon}; 0, 0)} = -2\sqrt{5},
\end{equation}
under the permuted coupling scheme.

\section{Example calculation---PPP pathway under $(0,0,0,0)$ polarization}
\label{sec:example}

The current article and the Supplemental Material contain all the information necessary to calculate the polarization and angular momentum dependence of third-order pathways.
As an example, we show the evaluation of the $R$-factor, Eq.~\eqref{eq:28}, for the PPP pathway and $(0,0,0,0)$ polarization.

The polarization tensor components can be evaluated using Eqs.~(\ref{eq:56}--\ref{eq:58}) to give:
\begin{align}
  \label{eq:62}
  T^{(0)}_0(\boldsymbol{\epsilon}; 0, 0) &= \frac{1}{3},\;\\
  T^{(0)}_0(\boldsymbol{\epsilon}; 1, 1) &=0,\;\\
  T^{(0)}_0(\boldsymbol{\epsilon}; 2, 2) &=\frac{2\sqrt{5}}{15}.
\end{align}
For PPP pathway, $\Delta J_{\alpha}$ values are as follows: $(\Delta J_{j}, \Delta J_{k}, \Delta J_{l}) = (-1, 0, -1)$, hence the $G$-factors can be obtained from the third row of Tab.~S3 in the Supplemental Material.
By summing $k=0$ and $k=2$ components, we obtain:
\begin{equation}
  \label{eq:63}
  R^{(0)}_{0}(\bm{\varepsilon}_{0000}; \mathrm{PPP}) = \frac{4 J_{i}^{2} + 1}{15 J_{i} \sqrt{2 J_{i} + 1} \left(4 J_{i}^{2} - 1\right)}.
\end{equation}

Since $\langle O_{ijkl}\rangle$ is equal to the sum over $M_{\alpha}$ states in Eq.~\eqref{eq:14}, the $R$-factor can also be expressed in terms of the sum.
This is done by using the Wigner-Eckhart theorem on each transition dipole element:
\begin{equation}
  \label{eq:64}
  \begin{split}
  \langle J_{\alpha}M_{\alpha}|& \epsilon_{x}\cdot\mu_{x}|J_{\beta}M_{\beta}\rangle = (-1)^{J_{\beta}-M_{\beta}}
  \langle J_{\beta}\|\mu^{(1)}\|J_{\alpha}\rangle\\
  \times & \sum_{q} (-1)^{q} \epsilon^{(1)}_{q}
  \begin{pmatrix}J_{\beta}&1&J_{\alpha}\\-M_{\beta}&-q&M_{\alpha} \end{pmatrix},
  \end{split}
\end{equation}
and canceling the reduced matrix elements and the zero-order density matrix, $\rho^{(0)}(-\infty)_{\alpha_{i},J_{i},M_{i}}$, both in Eq.~\eqref{eq:14} and in Eq.~\eqref{eq:51} to obtain:
\begin{widetext}
  \begin{equation}
    \label{eq:65}
    \begin{split}
    R^{(0)}_{0}(\boldsymbol{\varepsilon}; \boldsymbol{J}) =& \sum_{\substack{M_i,M_j, M_{k},M_{l}}}
    (-1)^{J_{i}+J_{j}+J_{k}+J_{l}-M_{i}-M_{j}-M_{k}-M_{l}}
     \sum_{\substack{q_{i},q_{j},q_{k},q_{l}}} (-1)^{q_{i}+q_{j}+q_{k}+q_{l}}
     \epsilon^{(1)}_{q_{i}} \epsilon^{(1)}_{q_{j}} \epsilon^{(1)}_{q_{k}} \epsilon^{(1)}_{q_{l}}
      \\
    &\times
    \begin{pmatrix}J_{i}&1&J_{j}\\-M_{i}&-q_{i}&M_{j} \end{pmatrix}
    \begin{pmatrix}J_{j}&1&J_{k}\\-M_{j}&-q_{j}&M_{k} \end{pmatrix}
    \begin{pmatrix}J_{k}&1&J_{l}\\-M_{k}&-q_{k}&M_{l} \end{pmatrix}
    \begin{pmatrix}J_{l}&1&J_{i}\\-M_{l}&-q_{l}&M_{i} \end{pmatrix}.
    \end{split}
  \end{equation}
\end{widetext}
By setting all $q_{\alpha}=0$, $\epsilon^{(1)}_{q_{a}}=1$, substituting appropriate $J_{\alpha}$ values and summing over $M_{\alpha}$ states, it can be verified that Eq.~\eqref{eq:63} and \eqref{eq:65} give the same value.
For example, for $J_{i}=5$ it is $0.004101$.

\end{document}


\maketitle

\section{R-branch diagrams}

\begin{figure*}[h!]
  \centering
  \includegraphics{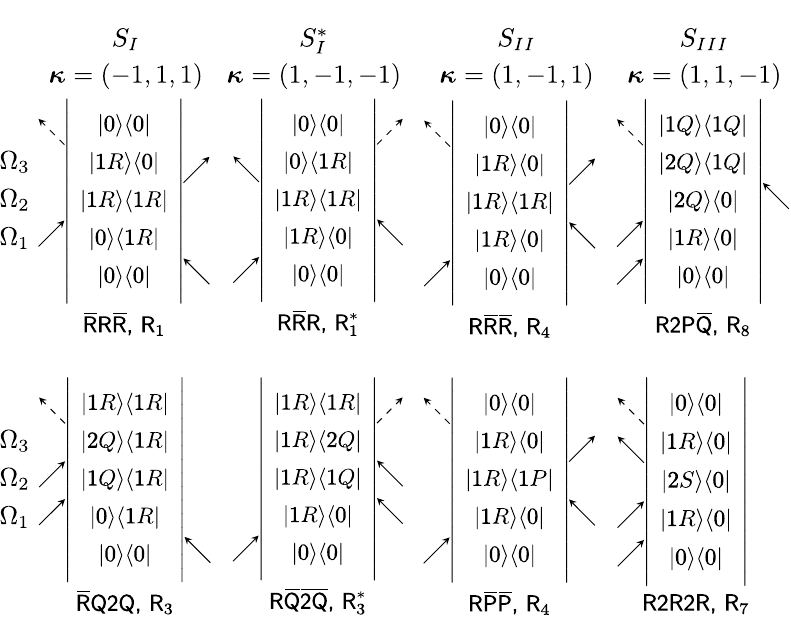}
  \caption{R-branch analogue of Fig.~1 in main article.
    Diagrams of third-order rovibrational pathways phase-matched in direction: $S_{I}$, first column; $S_{I}^{\ast}$, second column; $S_{II}$, third column; $S_{III}$, fourth column.
First row contains pathways not in a rotationally-coherent (RC) state during waiting time.
Second row contains RC pathways.\label{fig:diagrams_Rbranch}}
\end{figure*}

\clearpage\section{Diagonal branch and subbranches}

\begin{figure*}[h!]
  \includegraphics[width=\textwidth]{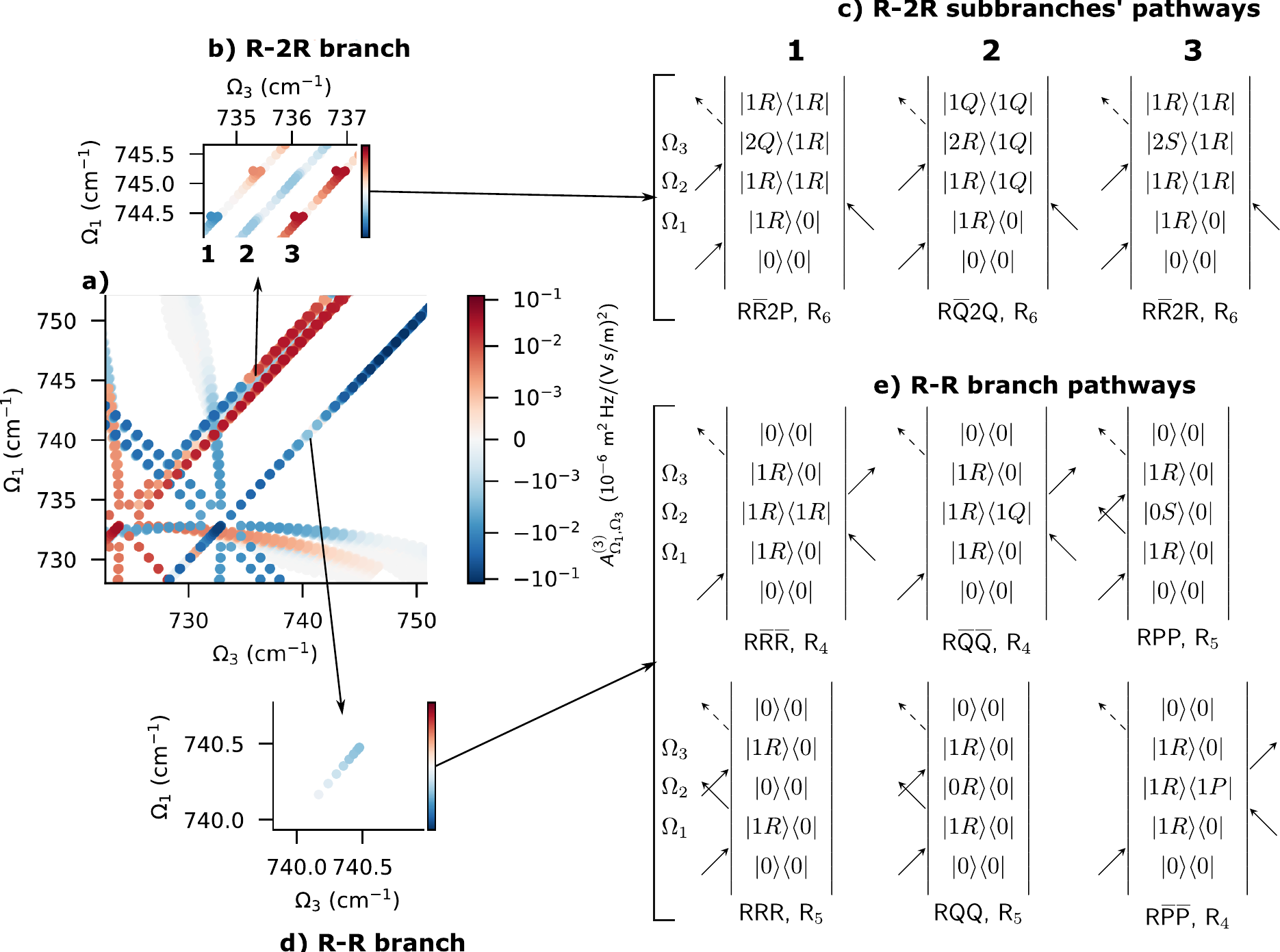}
  \caption{Subbranches in $S_{II}$ direction. a) Zoom in on the top right part of Fig.~4 of the main article. b) R-2R branch is split into three subbranches (1, 2, 3). Subbranches 1 and 2 are rotationally-coherent. c) Each subbranch comprises one third-order pathway. The subbranches can be identified by the $\Omega_{3}$ coherence, which are: $|2Q\rangle\langle 1R|$ for 1, $|2R\rangle\langle 1Q$ for 2 and $|2S\rangle\langle 1R|$ for 3. d) For each $J_{i}$ there are multiple resonances corresponding to different projection of the rotational angular momentum onto molecular axis, $K_{m}$. e) Each R-R resonance is composed of 6 pathways which differ only by $\Omega_{2}$ coherence.\label{fig:subbranches}}
\end{figure*}

\clearpage
\section{Rotational coherence frequencies - $S_{II}$}

\begin{table*}[h!]\centering
  \caption{Frequencies of rotational coherences for pathways phase-matched in the $S_{II}$ direction.\label{tab:rc-SII}}
  \begingroup\def\arraystretch{1.2}
  \begin{tabular}{cccccc}
    \toprule    
    Branch & Subbranch ($\Omega_{1}$, $\Omega_{3}$) & Rot. coherence & Pathway label & Exact frequency & Approx. frequency\\
    \midrule
P-2P & $|0\rangle\langle 1P|$, $|1Q\rangle\langle 2P|$ & $|1Q\rangle\langle 1P|$ & P\ensuremath{{}\mkern1mu\overline{\mkern-1mu\mbox{Q}}}2Q & $- F{\left(\nu_{1},J_{i},K \right)} + F{\left(\nu_{1},J_{i} - 1,K \right)}$ & $- 2 B_{1} J_{i}$\\
P-2P & $|0\rangle\langle 1P|$, $|1R\rangle\langle 2Q|$ & $|1R\rangle\langle 1P|$ & P\ensuremath{{}\mkern1mu\overline{\mkern-1mu\mbox{R}}}2R & $F{\left(\nu_{1},J_{i} - 1,K \right)} - F{\left(\nu_{1},J_{i} + 1,K \right)}$ & $- 2 B_{1} \left(2 J_{i} + 1\right)$\\
P-2Q & $|0\rangle\langle 1P|$, $|1Q\rangle\langle 2Q|$ & $|1Q\rangle\langle 1P|$ & P\ensuremath{{}\mkern1mu\overline{\mkern-1mu\mbox{Q}}}2R & $- F{\left(\nu_{1},J_{i},K \right)} + F{\left(\nu_{1},J_{i} - 1,K \right)}$ & $- 2 B_{1} J_{i}$\\
P-P & $|0\rangle\langle 1P|$, $|0\rangle\langle 1P|$ & $|0\rangle\langle 0O|$ & PRR & $- F{\left(\nu_{0},J_{i},K \right)} + F{\left(\nu_{0},J_{i} - 2,K \right)}$ & $- 2 B_{0} \left(2 J_{i} - 1\right)$\\
P-P & $|0\rangle\langle 1P|$, $|0\rangle\langle 1P|$ & $|0\rangle\langle 0P|$ & PQQ & $- F{\left(\nu_{0},J_{i},K \right)} + F{\left(\nu_{0},J_{i} - 1,K \right)}$ & $- 2 B_{0} J_{i}$\\
P-P & $|0\rangle\langle 1P|$, $|0\rangle\langle 1P|$ & $|1Q\rangle\langle 1P|$ & P\ensuremath{{}\mkern1mu\overline{\mkern-1mu\mbox{Q}}}\ensuremath{{}\mkern1mu\overline{\mkern-1mu\mbox{Q}}} & $- F{\left(\nu_{1},J_{i},K \right)} + F{\left(\nu_{1},J_{i} - 1,K \right)}$ & $- 2 B_{1} J_{i}$\\
P-P & $|0\rangle\langle 1P|$, $|0\rangle\langle 1P|$ & $|1R\rangle\langle 1P|$ & P\ensuremath{{}\mkern1mu\overline{\mkern-1mu\mbox{R}}}\ensuremath{{}\mkern1mu\overline{\mkern-1mu\mbox{R}}} & $F{\left(\nu_{1},J_{i} - 1,K \right)} - F{\left(\nu_{1},J_{i} + 1,K \right)}$ & $- 2 B_{1} \left(2 J_{i} + 1\right)$\\
P-Q & $|0\rangle\langle 1P|$, $|0P\rangle\langle 1P|$ & $|1Q\rangle\langle 1P|$ & P\ensuremath{{}\mkern1mu\overline{\mkern-1mu\mbox{Q}}}\ensuremath{{}\mkern1mu\overline{\mkern-1mu\mbox{R}}} & $- F{\left(\nu_{1},J_{i},K \right)} + F{\left(\nu_{1},J_{i} - 1,K \right)}$ & $- 2 B_{1} J_{i}$\\
P-Q & $|0\rangle\langle 1P|$, $|0\rangle\langle 1Q|$ & $|0\rangle\langle 0P|$ & PQR & $- F{\left(\nu_{0},J_{i},K \right)} + F{\left(\nu_{0},J_{i} - 1,K \right)}$ & $- 2 B_{0} J_{i}$\\
Q-2P & $|0\rangle\langle 1Q|$, $|1R\rangle\langle 2Q|$ & $|1R\rangle\langle 1Q|$ & Q\ensuremath{{}\mkern1mu\overline{\mkern-1mu\mbox{R}}}2Q & $F{\left(\nu_{1},J_{i},K \right)} - F{\left(\nu_{1},J_{i} + 1,K \right)}$ & $- 2 B_{1} \left(J_{i} + 1\right)$\\
Q-2Q & $|0\rangle\langle 1Q|$, $|1P\rangle\langle 2P|$ & $|1P\rangle\langle 1Q|$ & Q\ensuremath{{}\mkern1mu\overline{\mkern-1mu\mbox{P}}}2P & $F{\left(\nu_{1},J_{i},K \right)} - F{\left(\nu_{1},J_{i} - 1,K \right)}$ & $2 B_{1} J_{i}$\\
Q-2Q & $|0\rangle\langle 1Q|$, $|1R\rangle\langle 2R|$ & $|1R\rangle\langle 1Q|$ & Q\ensuremath{{}\mkern1mu\overline{\mkern-1mu\mbox{R}}}2R & $F{\left(\nu_{1},J_{i},K \right)} - F{\left(\nu_{1},J_{i} + 1,K \right)}$ & $- 2 B_{1} \left(J_{i} + 1\right)$\\
Q-2R & $|0\rangle\langle 1Q|$, $|1P\rangle\langle 2Q|$ & $|1P\rangle\langle 1Q|$ & Q\ensuremath{{}\mkern1mu\overline{\mkern-1mu\mbox{P}}}2Q & $F{\left(\nu_{1},J_{i},K \right)} - F{\left(\nu_{1},J_{i} - 1,K \right)}$ & $2 B_{1} J_{i}$\\
Q-P & $|0\rangle\langle 1Q|$, $|0R\rangle\langle 1Q|$ & $|1R\rangle\langle 1Q|$ & Q\ensuremath{{}\mkern1mu\overline{\mkern-1mu\mbox{R}}}\ensuremath{{}\mkern1mu\overline{\mkern-1mu\mbox{Q}}} & $F{\left(\nu_{1},J_{i},K \right)} - F{\left(\nu_{1},J_{i} + 1,K \right)}$ & $- 2 B_{1} \left(J_{i} + 1\right)$\\
Q-P & $|0\rangle\langle 1Q|$, $|0\rangle\langle 1P|$ & $|0\rangle\langle 0P|$ & QRQ & $- F{\left(\nu_{0},J_{i},K \right)} + F{\left(\nu_{0},J_{i} - 1,K \right)}$ & $- 2 B_{0} J_{i}$\\
Q-Q & $|0\rangle\langle 1Q|$, $|0\rangle\langle 1Q|$ & $|0\rangle\langle 0P|$ & QRR & $- F{\left(\nu_{0},J_{i},K \right)} + F{\left(\nu_{0},J_{i} - 1,K \right)}$ & $- 2 B_{0} J_{i}$\\
Q-Q & $|0\rangle\langle 1Q|$, $|0\rangle\langle 1Q|$ & $|0\rangle\langle 0R|$ & QPP & $- F{\left(\nu_{0},J_{i},K \right)} + F{\left(\nu_{0},J_{i} + 1,K \right)}$ & $2 B_{0} \left(J_{i} + 1\right)$\\
Q-Q & $|0\rangle\langle 1Q|$, $|0\rangle\langle 1Q|$ & $|1P\rangle\langle 1Q|$ & Q\ensuremath{{}\mkern1mu\overline{\mkern-1mu\mbox{P}}}\ensuremath{{}\mkern1mu\overline{\mkern-1mu\mbox{P}}} & $F{\left(\nu_{1},J_{i},K \right)} - F{\left(\nu_{1},J_{i} - 1,K \right)}$ & $2 B_{1} J_{i}$\\
Q-Q & $|0\rangle\langle 1Q|$, $|0\rangle\langle 1Q|$ & $|1R\rangle\langle 1Q|$ & Q\ensuremath{{}\mkern1mu\overline{\mkern-1mu\mbox{R}}}\ensuremath{{}\mkern1mu\overline{\mkern-1mu\mbox{R}}} & $F{\left(\nu_{1},J_{i},K \right)} - F{\left(\nu_{1},J_{i} + 1,K \right)}$ & $- 2 B_{1} \left(J_{i} + 1\right)$\\
Q-R & $|0\rangle\langle 1Q|$, $|0P\rangle\langle 1Q|$ & $|1P\rangle\langle 1Q|$ & Q\ensuremath{{}\mkern1mu\overline{\mkern-1mu\mbox{P}}}\ensuremath{{}\mkern1mu\overline{\mkern-1mu\mbox{Q}}} & $F{\left(\nu_{1},J_{i},K \right)} - F{\left(\nu_{1},J_{i} - 1,K \right)}$ & $2 B_{1} J_{i}$\\
Q-R & $|0\rangle\langle 1Q|$, $|0\rangle\langle 1R|$ & $|0\rangle\langle 0R|$ & QPQ & $- F{\left(\nu_{0},J_{i},K \right)} + F{\left(\nu_{0},J_{i} + 1,K \right)}$ & $2 B_{0} \left(J_{i} + 1\right)$\\
R-2Q & $|0\rangle\langle 1R|$, $|1Q\rangle\langle 2Q|$ & $|1Q\rangle\langle 1R|$ & R\ensuremath{{}\mkern1mu\overline{\mkern-1mu\mbox{Q}}}2P & $- F{\left(\nu_{1},J_{i},K \right)} + F{\left(\nu_{1},J_{i} + 1,K \right)}$ & $2 B_{1} \left(J_{i} + 1\right)$\\
R-2R & $|0\rangle\langle 1R|$, $|1P\rangle\langle 2Q|$ & $|1P\rangle\langle 1R|$ & R\ensuremath{{}\mkern1mu\overline{\mkern-1mu\mbox{P}}}2P & $- F{\left(\nu_{1},J_{i} - 1,K \right)} + F{\left(\nu_{1},J_{i} + 1,K \right)}$ & $2 B_{1} \left(2 J_{i} + 1\right)$\\
R-2R & $|0\rangle\langle 1R|$, $|1Q\rangle\langle 2R|$ & $|1Q\rangle\langle 1R|$ & R\ensuremath{{}\mkern1mu\overline{\mkern-1mu\mbox{Q}}}2Q & $- F{\left(\nu_{1},J_{i},K \right)} + F{\left(\nu_{1},J_{i} + 1,K \right)}$ & $2 B_{1} \left(J_{i} + 1\right)$\\
R-Q & $|0\rangle\langle 1R|$, $|0R\rangle\langle 1R|$ & $|1Q\rangle\langle 1R|$ & R\ensuremath{{}\mkern1mu\overline{\mkern-1mu\mbox{Q}}}\ensuremath{{}\mkern1mu\overline{\mkern-1mu\mbox{P}}} & $- F{\left(\nu_{1},J_{i},K \right)} + F{\left(\nu_{1},J_{i} + 1,K \right)}$ & $2 B_{1} \left(J_{i} + 1\right)$\\
R-Q & $|0\rangle\langle 1R|$, $|0\rangle\langle 1Q|$ & $|0\rangle\langle 0R|$ & RQP & $- F{\left(\nu_{0},J_{i},K \right)} + F{\left(\nu_{0},J_{i} + 1,K \right)}$ & $2 B_{0} \left(J_{i} + 1\right)$\\
R-R & $|0\rangle\langle 1R|$, $|0\rangle\langle 1R|$ & $|0\rangle\langle 0R|$ & RQQ & $- F{\left(\nu_{0},J_{i},K \right)} + F{\left(\nu_{0},J_{i} + 1,K \right)}$ & $2 B_{0} \left(J_{i} + 1\right)$\\
R-R & $|0\rangle\langle 1R|$, $|0\rangle\langle 1R|$ & $|0\rangle\langle 0S|$ & RPP & $- F{\left(\nu_{0},J_{i},K \right)} + F{\left(\nu_{0},J_{i} + 2,K \right)}$ & $2 B_{0} \left(2 J_{i} + 3\right)$\\
R-R & $|0\rangle\langle 1R|$, $|0\rangle\langle 1R|$ & $|1P\rangle\langle 1R|$ & R\ensuremath{{}\mkern1mu\overline{\mkern-1mu\mbox{P}}}\ensuremath{{}\mkern1mu\overline{\mkern-1mu\mbox{P}}} & $- F{\left(\nu_{1},J_{i} - 1,K \right)} + F{\left(\nu_{1},J_{i} + 1,K \right)}$ & $2 B_{1} \left(2 J_{i} + 1\right)$\\
R-R & $|0\rangle\langle 1R|$, $|0\rangle\langle 1R|$ & $|1Q\rangle\langle 1R|$ & R\ensuremath{{}\mkern1mu\overline{\mkern-1mu\mbox{Q}}}\ensuremath{{}\mkern1mu\overline{\mkern-1mu\mbox{Q}}} & $- F{\left(\nu_{1},J_{i},K \right)} + F{\left(\nu_{1},J_{i} + 1,K \right)}$ & $2 B_{1} \left(J_{i} + 1\right)$\\
    \bottomrule
    \end{tabular}
  \endgroup
\end{table*}

\clearpage
\section{Rotational coherence frequencies - $S_{I}$}
\begin{table*}[h!]\centering
  \caption{Frequencies of rotational coherences for pathways phase-matched in the $S_{I}$ direction.\label{tab:rc-SI}}
  \begingroup\def\arraystretch{1.2}
  \begin{tabular}{cccccc}
    \toprule
    Branch & Subbranch ($\Omega_{1}$, $\Omega_{3}$) & Rot. coherence & Pathway label & Exact frequency & Approx. frequency\\
    \midrule
P-2Q & $|0\rangle\langle 1P|$, $|2P\rangle\langle 1P|$ & $|1Q\rangle\langle 1P|$ & P\ensuremath{{}\mkern1mu\overline{\mkern-1mu\mbox{Q}}}\ensuremath{{}\mkern1mu\overline{\mkern-1mu\mbox{2P}}} & $- F{\left(\nu_{1},J_{i},K \right)} + F{\left(\nu_{1},J_{i} - 1,K \right)}$ & $- 2 B_{1} J_{i}$\\
P-2R & $|0\rangle\langle 1P|$, $|2Q\rangle\langle 1P|$ & $|1Q\rangle\langle 1P|$ & P\ensuremath{{}\mkern1mu\overline{\mkern-1mu\mbox{Q}}}\ensuremath{{}\mkern1mu\overline{\mkern-1mu\mbox{2Q}}} & $- F{\left(\nu_{1},J_{i},K \right)} + F{\left(\nu_{1},J_{i} - 1,K \right)}$ & $- 2 B_{1} J_{i}$\\
P-2R & $|0\rangle\langle 1P|$, $|2Q\rangle\langle 1P|$ & $|1R\rangle\langle 1P|$ & P\ensuremath{{}\mkern1mu\overline{\mkern-1mu\mbox{R}}}\ensuremath{{}\mkern1mu\overline{\mkern-1mu\mbox{2P}}} & $F{\left(\nu_{1},J_{i} - 1,K \right)} - F{\left(\nu_{1},J_{i} + 1,K \right)}$ & $- 2 B_{1} \left(2 J_{i} + 1\right)$\\
P-Q & $|0\rangle\langle 1P|$, $|1P\rangle\langle 0P|$ & $|0\rangle\langle 0P|$ & PQ\ensuremath{{}\mkern1mu\overline{\mkern-1mu\mbox{P}}} & $- F{\left(\nu_{0},J_{i},K \right)} + F{\left(\nu_{0},J_{i} - 1,K \right)}$ & $- 2 B_{0} J_{i}$\\
P-Q & $|0\rangle\langle 1P|$, $|1Q\rangle\langle 0|$ & $|1Q\rangle\langle 1P|$ & P\ensuremath{{}\mkern1mu\overline{\mkern-1mu\mbox{Q}}}P & $- F{\left(\nu_{1},J_{i},K \right)} + F{\left(\nu_{1},J_{i} - 1,K \right)}$ & $- 2 B_{1} J_{i}$\\
P-R & $|0\rangle\langle 1P|$, $|1P\rangle\langle 0O|$ & $|0\rangle\langle 0O|$ & PR\ensuremath{{}\mkern1mu\overline{\mkern-1mu\mbox{P}}} & $- F{\left(\nu_{0},J_{i},K \right)} + F{\left(\nu_{0},J_{i} - 2,K \right)}$ & $- 2 B_{0} \left(2 J_{i} - 1\right)$\\
P-R & $|0\rangle\langle 1P|$, $|1Q\rangle\langle 0P|$ & $|0\rangle\langle 0P|$ & PQ\ensuremath{{}\mkern1mu\overline{\mkern-1mu\mbox{Q}}} & $- F{\left(\nu_{0},J_{i},K \right)} + F{\left(\nu_{0},J_{i} - 1,K \right)}$ & $- 2 B_{0} J_{i}$\\
P-R & $|0\rangle\langle 1P|$, $|1Q\rangle\langle 0P|$ & $|1Q\rangle\langle 1P|$ & P\ensuremath{{}\mkern1mu\overline{\mkern-1mu\mbox{Q}}}Q & $- F{\left(\nu_{1},J_{i},K \right)} + F{\left(\nu_{1},J_{i} - 1,K \right)}$ & $- 2 B_{1} J_{i}$\\
P-R & $|0\rangle\langle 1P|$, $|1R\rangle\langle 0|$ & $|1R\rangle\langle 1P|$ & P\ensuremath{{}\mkern1mu\overline{\mkern-1mu\mbox{R}}}P & $F{\left(\nu_{1},J_{i} - 1,K \right)} - F{\left(\nu_{1},J_{i} + 1,K \right)}$ & $- 2 B_{1} \left(2 J_{i} + 1\right)$\\
Q-2P & $|0\rangle\langle 1Q|$, $|2P\rangle\langle 1Q|$ & $|1P\rangle\langle 1Q|$ & Q\ensuremath{{}\mkern1mu\overline{\mkern-1mu\mbox{P}}}\ensuremath{{}\mkern1mu\overline{\mkern-1mu\mbox{2Q}}} & $F{\left(\nu_{1},J_{i},K \right)} - F{\left(\nu_{1},J_{i} - 1,K \right)}$ & $2 B_{1} J_{i}$\\
Q-2Q & $|0\rangle\langle 1Q|$, $|2Q\rangle\langle 1Q|$ & $|1P\rangle\langle 1Q|$ & Q\ensuremath{{}\mkern1mu\overline{\mkern-1mu\mbox{P}}}\ensuremath{{}\mkern1mu\overline{\mkern-1mu\mbox{2R}}} & $F{\left(\nu_{1},J_{i},K \right)} - F{\left(\nu_{1},J_{i} - 1,K \right)}$ & $2 B_{1} J_{i}$\\
Q-2Q & $|0\rangle\langle 1Q|$, $|2Q\rangle\langle 1Q|$ & $|1R\rangle\langle 1Q|$ & Q\ensuremath{{}\mkern1mu\overline{\mkern-1mu\mbox{R}}}\ensuremath{{}\mkern1mu\overline{\mkern-1mu\mbox{2P}}} & $F{\left(\nu_{1},J_{i},K \right)} - F{\left(\nu_{1},J_{i} + 1,K \right)}$ & $- 2 B_{1} \left(J_{i} + 1\right)$\\
Q-2R & $|0\rangle\langle 1Q|$, $|2R\rangle\langle 1Q|$ & $|1R\rangle\langle 1Q|$ & Q\ensuremath{{}\mkern1mu\overline{\mkern-1mu\mbox{R}}}\ensuremath{{}\mkern1mu\overline{\mkern-1mu\mbox{2Q}}} & $F{\left(\nu_{1},J_{i},K \right)} - F{\left(\nu_{1},J_{i} + 1,K \right)}$ & $- 2 B_{1} \left(J_{i} + 1\right)$\\
Q-P & $|0\rangle\langle 1Q|$, $|1P\rangle\langle 0|$ & $|1P\rangle\langle 1Q|$ & Q\ensuremath{{}\mkern1mu\overline{\mkern-1mu\mbox{P}}}Q & $F{\left(\nu_{1},J_{i},K \right)} - F{\left(\nu_{1},J_{i} - 1,K \right)}$ & $2 B_{1} J_{i}$\\
Q-P & $|0\rangle\langle 1Q|$, $|1Q\rangle\langle 0R|$ & $|0\rangle\langle 0R|$ & QP\ensuremath{{}\mkern1mu\overline{\mkern-1mu\mbox{Q}}} & $- F{\left(\nu_{0},J_{i},K \right)} + F{\left(\nu_{0},J_{i} + 1,K \right)}$ & $2 B_{0} \left(J_{i} + 1\right)$\\
Q-Q & $|0\rangle\langle 1Q|$, $|1P\rangle\langle 0P|$ & $|0\rangle\langle 0P|$ & QR\ensuremath{{}\mkern1mu\overline{\mkern-1mu\mbox{P}}} & $- F{\left(\nu_{0},J_{i},K \right)} + F{\left(\nu_{0},J_{i} - 1,K \right)}$ & $- 2 B_{0} J_{i}$\\
Q-Q & $|0\rangle\langle 1Q|$, $|1P\rangle\langle 0P|$ & $|1P\rangle\langle 1Q|$ & Q\ensuremath{{}\mkern1mu\overline{\mkern-1mu\mbox{P}}}R & $F{\left(\nu_{1},J_{i},K \right)} - F{\left(\nu_{1},J_{i} - 1,K \right)}$ & $2 B_{1} J_{i}$\\
Q-Q & $|0\rangle\langle 1Q|$, $|1R\rangle\langle 0R|$ & $|0\rangle\langle 0R|$ & QP\ensuremath{{}\mkern1mu\overline{\mkern-1mu\mbox{R}}} & $- F{\left(\nu_{0},J_{i},K \right)} + F{\left(\nu_{0},J_{i} + 1,K \right)}$ & $2 B_{0} \left(J_{i} + 1\right)$\\
Q-Q & $|0\rangle\langle 1Q|$, $|1R\rangle\langle 0R|$ & $|1R\rangle\langle 1Q|$ & Q\ensuremath{{}\mkern1mu\overline{\mkern-1mu\mbox{R}}}P & $F{\left(\nu_{1},J_{i},K \right)} - F{\left(\nu_{1},J_{i} + 1,K \right)}$ & $- 2 B_{1} \left(J_{i} + 1\right)$\\
Q-R & $|0\rangle\langle 1Q|$, $|1Q\rangle\langle 0P|$ & $|0\rangle\langle 0P|$ & QR\ensuremath{{}\mkern1mu\overline{\mkern-1mu\mbox{Q}}} & $- F{\left(\nu_{0},J_{i},K \right)} + F{\left(\nu_{0},J_{i} - 1,K \right)}$ & $- 2 B_{0} J_{i}$\\
Q-R & $|0\rangle\langle 1Q|$, $|1R\rangle\langle 0|$ & $|1R\rangle\langle 1Q|$ & Q\ensuremath{{}\mkern1mu\overline{\mkern-1mu\mbox{R}}}Q & $F{\left(\nu_{1},J_{i},K \right)} - F{\left(\nu_{1},J_{i} + 1,K \right)}$ & $- 2 B_{1} \left(J_{i} + 1\right)$\\
R-2P & $|0\rangle\langle 1R|$, $|2Q\rangle\langle 1R|$ & $|1P\rangle\langle 1R|$ & R\ensuremath{{}\mkern1mu\overline{\mkern-1mu\mbox{P}}}\ensuremath{{}\mkern1mu\overline{\mkern-1mu\mbox{2R}}} & $- F{\left(\nu_{1},J_{i} - 1,K \right)} + F{\left(\nu_{1},J_{i} + 1,K \right)}$ & $2 B_{1} \left(2 J_{i} + 1\right)$\\
R-2P & $|0\rangle\langle 1R|$, $|2Q\rangle\langle 1R|$ & $|1Q\rangle\langle 1R|$ & R\ensuremath{{}\mkern1mu\overline{\mkern-1mu\mbox{Q}}}\ensuremath{{}\mkern1mu\overline{\mkern-1mu\mbox{2Q}}} & $- F{\left(\nu_{1},J_{i},K \right)} + F{\left(\nu_{1},J_{i} + 1,K \right)}$ & $2 B_{1} \left(J_{i} + 1\right)$\\
R-2Q & $|0\rangle\langle 1R|$, $|2R\rangle\langle 1R|$ & $|1Q\rangle\langle 1R|$ & R\ensuremath{{}\mkern1mu\overline{\mkern-1mu\mbox{Q}}}\ensuremath{{}\mkern1mu\overline{\mkern-1mu\mbox{2R}}} & $- F{\left(\nu_{1},J_{i},K \right)} + F{\left(\nu_{1},J_{i} + 1,K \right)}$ & $2 B_{1} \left(J_{i} + 1\right)$\\
R-P & $|0\rangle\langle 1R|$, $|1P\rangle\langle 0|$ & $|1P\rangle\langle 1R|$ & R\ensuremath{{}\mkern1mu\overline{\mkern-1mu\mbox{P}}}R & $- F{\left(\nu_{1},J_{i} - 1,K \right)} + F{\left(\nu_{1},J_{i} + 1,K \right)}$ & $2 B_{1} \left(2 J_{i} + 1\right)$\\
R-P & $|0\rangle\langle 1R|$, $|1Q\rangle\langle 0R|$ & $|0\rangle\langle 0R|$ & RQ\ensuremath{{}\mkern1mu\overline{\mkern-1mu\mbox{Q}}} & $- F{\left(\nu_{0},J_{i},K \right)} + F{\left(\nu_{0},J_{i} + 1,K \right)}$ & $2 B_{0} \left(J_{i} + 1\right)$\\
R-P & $|0\rangle\langle 1R|$, $|1Q\rangle\langle 0R|$ & $|1Q\rangle\langle 1R|$ & R\ensuremath{{}\mkern1mu\overline{\mkern-1mu\mbox{Q}}}Q & $- F{\left(\nu_{1},J_{i},K \right)} + F{\left(\nu_{1},J_{i} + 1,K \right)}$ & $2 B_{1} \left(J_{i} + 1\right)$\\
R-P & $|0\rangle\langle 1R|$, $|1R\rangle\langle 0S|$ & $|0\rangle\langle 0S|$ & RP\ensuremath{{}\mkern1mu\overline{\mkern-1mu\mbox{R}}} & $- F{\left(\nu_{0},J_{i},K \right)} + F{\left(\nu_{0},J_{i} + 2,K \right)}$ & $2 B_{0} \left(2 J_{i} + 3\right)$\\
R-Q & $|0\rangle\langle 1R|$, $|1Q\rangle\langle 0|$ & $|1Q\rangle\langle 1R|$ & R\ensuremath{{}\mkern1mu\overline{\mkern-1mu\mbox{Q}}}R & $- F{\left(\nu_{1},J_{i},K \right)} + F{\left(\nu_{1},J_{i} + 1,K \right)}$ & $2 B_{1} \left(J_{i} + 1\right)$\\
R-Q & $|0\rangle\langle 1R|$, $|1R\rangle\langle 0R|$ & $|0\rangle\langle 0R|$ & RQ\ensuremath{{}\mkern1mu\overline{\mkern-1mu\mbox{R}}} & $- F{\left(\nu_{0},J_{i},K \right)} + F{\left(\nu_{0},J_{i} + 1,K \right)}$ & $2 B_{0} \left(J_{i} + 1\right)$\\
    \bottomrule
  \end{tabular}
  \endgroup
\end{table*}

\clearpage
\section{Pathways per resonance}

\begin{figure*}[h!]
  \centering
  \includegraphics{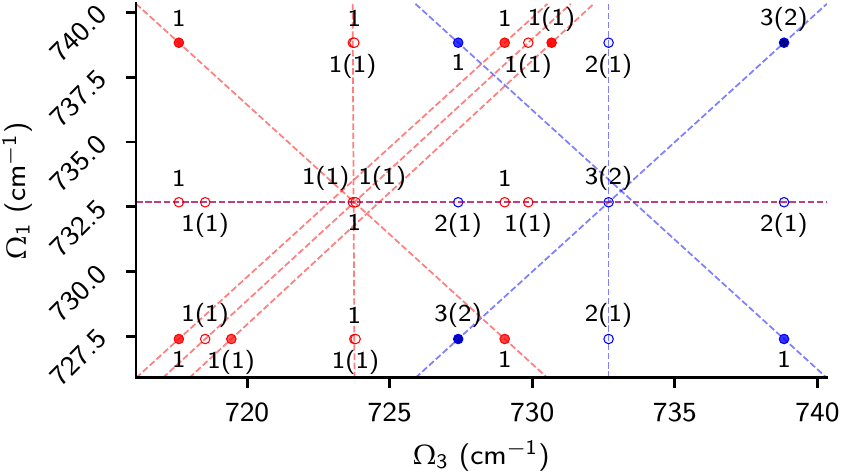}
  \caption{\label{fig:pws_per_peak}All resonances produced by third-order, time-ordered excitation starting in $|\nu=0,J_{i}=6,K_{m}=1\rangle$ state for $S_{III}$ direction. Resonances are labeled with number of pathways contributing to each resonance and number of RC pathways in parentheses.}
\end{figure*}

\clearpage
\section{Decomposition of molecular response into polarization classes}

\begin{figure*}[h!]
  \includegraphics{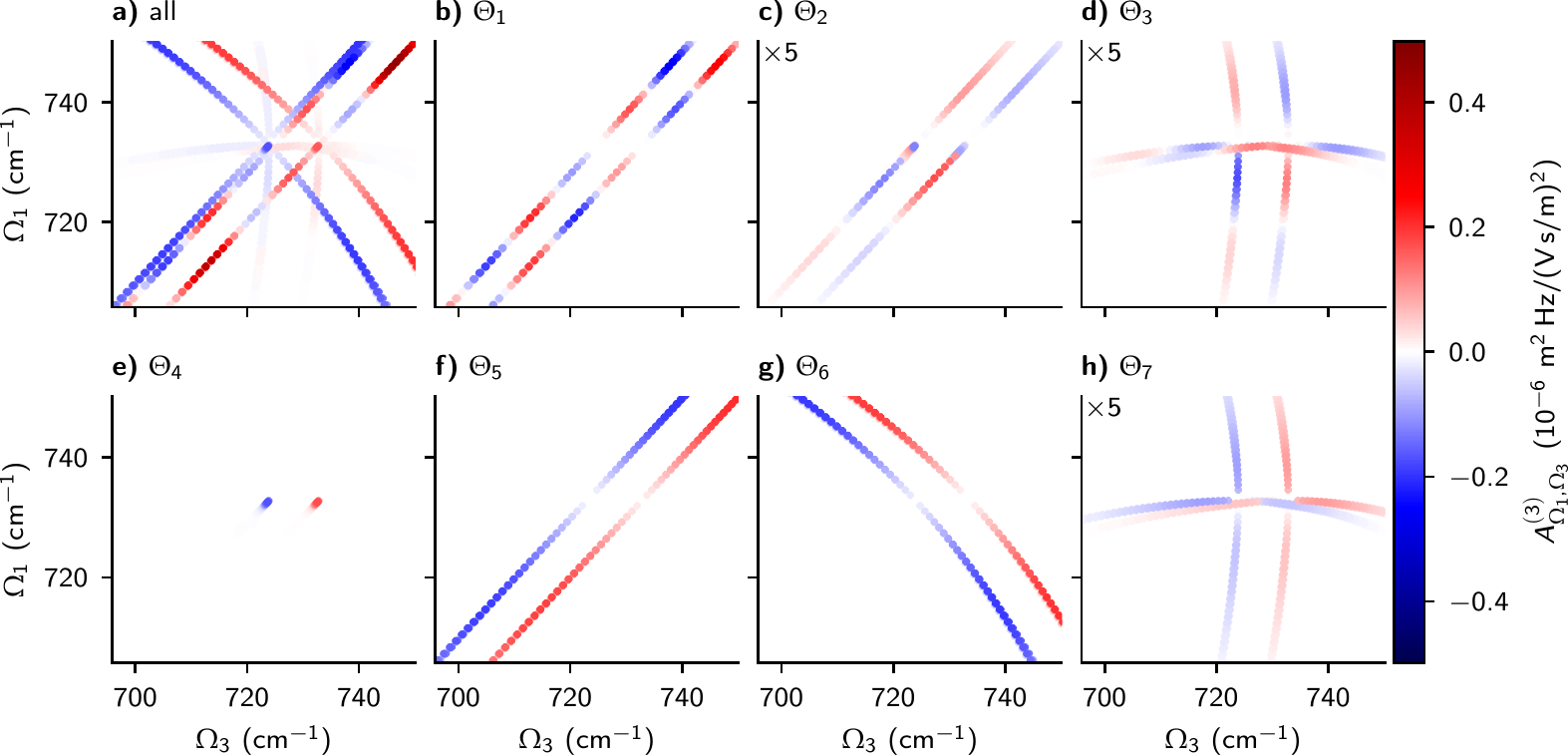}
  \caption{Polarization decomposition of third-order response of methyl chloride $\nu_{3}$ vibrational mode phase-matched in the $S_{III}$ direction.
    The plotted quantity is the 2D resonance amplitude defined in Eq.~22.
a) total response; b--h) subsets of response associated with polarization classes $\Theta_{1}$--$\Theta_{7}$, see Tab.~I.
\label{fig:symtop_gallery_SIII}}
\end{figure*}

\clearpage
\section{Table of G-factors}

\begin{table*}[h!]\centering
  \caption{\label{tab:Gfactors}Reduced G-factors $(2J_{i}+1)^{3/2}G(J_i, J_i+\Delta{}J_{j}, J_i+\Delta{}J_{k}, J_i+\Delta{}J_l; k)$ for different sequences of arguments, see Eq.~(29).}
  \begin{tabular}{cccc}\toprule
    $(\Delta{}J_{j}, \Delta{}J_{k}, \Delta{}J_{l})$ & $k=0$ & $k=1$ & $k=2$\\\midrule
$\left( -1, \  -2, \  -1\right)$ & $0$ & $0$ & $\sqrt{5} \left(2 J_{i} + 1\right) / \left(5 \left(2 J_{i} - 1\right)\right)$\\
$\left( 1, \  2, \  1\right)$ & $0$ & $0$ & $\sqrt{5} \left(2 J_{i} + 1\right) / \left(5 \left(2 J_{i} + 3\right)\right)$\\
$\left( -1, \  0, \  -1\right)$ & $1 / 3$ & $- \sqrt{3} \left(J_{i} + 1\right) / 6 J_{i}$ & $\sqrt{5} \left(J_{i} + 1\right) \left(2 J_{i} + 3\right) / \left(30 J_{i} \left(2 J_{i} - 1\right)\right)$\\
$\left( 1, \  0, \  1\right)$ & $1 / 3$ & $- \sqrt{3} J_{i} / \left(6 \left(J_{i} + 1\right)\right)$ & $\sqrt{5} J_{i} \left(2 J_{i} - 1\right) / \left(30 \left(J_{i} + 1\right) \left(2 J_{i} + 3\right)\right)$\\
$\left( 1, \  0, \  -1\right)$ & $1 / 3$ & $\sqrt{3} / 6$ & $\sqrt{5} / 30$\\
$\left( -1, \  0, \  1\right)$ & $1 / 3$ & $\sqrt{3} / 6$ & $\sqrt{5} / 30$\\
$\left( -1, \  -1, \  -1\right)$ & $0$ & $\sqrt{3} \left(J_{i} - 1\right) \left(2 J_{i} + 1\right) / \left(6 J_{i} \left(2 J_{i} - 1\right)\right)$ & $- \sqrt{5} \left(J_{i} + 1\right) \left(2 J_{i} + 1\right) / \left(10 J_{i} \left(2 J_{i} - 1\right)\right)$\\
$\left( 0, \  -1, \  -1\right)$ & $0$ & $- \sqrt{3} \sqrt{J_{i} - 1} \sqrt{J_{i} + 1} \sqrt{2 J_{i} + 1} / \left(6 J_{i} \sqrt{2 J_{i} - 1}\right)$ & $- \sqrt{5} \sqrt{J_{i} - 1} \sqrt{J_{i} + 1} \sqrt{2 J_{i} + 1} / \left(10 J_{i} \sqrt{2 J_{i} - 1}\right)$\\
$\left( 0, \  -1, \  0\right)$ & $0$ & $\sqrt{3} \left(J_{i} + 1\right) / 6 J_{i}$ & $- \sqrt{5} \left(J_{i} - 1\right) / 10 J_{i}$\\
$\left( 0, \  0, \  -1\right)$ & $- 1 / 3$ & $\sqrt{3} / 6 J_{i}$ & $\sqrt{5} \left(2 J_{i} + 3\right) / 30 J_{i}$\\
$\left( 0, \  0, \  0\right)$ & $1 / 3$ & $- \sqrt{3} / \left(6 J_{i} \left(J_{i} + 1\right)\right)$ & $\sqrt{5} \left(2 J_{i} - 1\right) \left(2 J_{i} + 3\right) / \left(30 J_{i} \left(J_{i} + 1\right)\right)$\\
$\left( 0, \  1, \  0\right)$ & $0$ & $\sqrt{3} J_{i} / \left(6 \left(J_{i} + 1\right)\right)$ & $- \sqrt{5} \left(J_{i} + 2\right) / \left(10 \left(J_{i} + 1\right)\right)$\\
$\left( 1, \  0, \  0\right)$ & $- 1 / 3$ & $- \sqrt{3} / \left(6 \left(J_{i} + 1\right)\right)$ & $\sqrt{5} \left(2 J_{i} - 1\right) / \left(30 \left(J_{i} + 1\right)\right)$\\
$\left( 1, \  1, \  0\right)$ & $0$ & $- \sqrt{3} \sqrt{J_{i}} \sqrt{J_{i} + 2} \sqrt{2 J_{i} + 1} / \left(6 \left(J_{i} + 1\right) \sqrt{2 J_{i} + 3}\right)$ & $- \sqrt{5} \sqrt{J_{i}} \sqrt{J_{i} + 2} \sqrt{2 J_{i} + 1} / \left(10 \left(J_{i} + 1\right) \sqrt{2 J_{i} + 3}\right)$\\
$\left( 1, \  1, \  1\right)$ & $0$ & $\sqrt{3} \left(J_{i} + 2\right) \left(2 J_{i} + 1\right) / \left(6 \left(J_{i} + 1\right) \left(2 J_{i} + 3\right)\right)$ & $- \sqrt{5} J_{i} \left(2 J_{i} + 1\right) / \left(10 \left(J_{i} + 1\right) \left(2 J_{i} + 3\right)\right)$\\
$\left( -1, \  -1, \  0\right)$ & $0$ & $- \sqrt{3} \sqrt{J_{i} - 1} \sqrt{J_{i} + 1} \sqrt{2 J_{i} + 1} / \left(6 J_{i} \sqrt{2 J_{i} - 1}\right)$ & $- \sqrt{5} \sqrt{J_{i} - 1} \sqrt{J_{i} + 1} \sqrt{2 J_{i} + 1} / \left(10 J_{i} \sqrt{2 J_{i} - 1}\right)$\\
$\left( -1, \  0, \  0\right)$ & $- 1 / 3$ & $\sqrt{3} / 6 J_{i}$ & $\sqrt{5} \left(2 J_{i} + 3\right) / 30 J_{i}$\\
$\left( 0, \  0, \  1\right)$ & $- 1 / 3$ & $- \sqrt{3} / \left(6 \left(J_{i} + 1\right)\right)$ & $\sqrt{5} \left(2 J_{i} - 1\right) / \left(30 \left(J_{i} + 1\right)\right)$\\
$\left( 0, \  1, \  1\right)$ & $0$ & $- \sqrt{3} \sqrt{J_{i}} \sqrt{J_{i} + 2} \sqrt{2 J_{i} + 1} / \left(6 \left(J_{i} + 1\right) \sqrt{2 J_{i} + 3}\right)$ & $- \sqrt{5} \sqrt{J_{i}} \sqrt{J_{i} + 2} \sqrt{2 J_{i} + 1} / \left(10 \left(J_{i} + 1\right) \sqrt{2 J_{i} + 3}\right)$\\ \bottomrule
  \end{tabular}
\end{table*}

\begin{table*}\centering
  \caption{\label{tab:Gfactors_highj}Reduced G-factors $(2J_{i}+1)^{3/2}G(J_i, J_i+\Delta{}J_{j}, J_i+\Delta{}J_{k}, J_i+\Delta{}J_l; k)$ in the high-$J_{i}$ limit for different sequences of arguments, see Eq.~(29).}
  \begin{tabular}{cccc}\toprule
    $(\Delta{}J_{j}, \Delta{}J_{k}, \Delta{}J_{l})$ & $k=0$ & $k=1$ & $k=2$\\\midrule
$\left( -1, \  -2, \  -1\right)$ & $0$ & $0$ & $\sqrt{5} / 5$\\
$\left( 1, \  2, \  1\right)$ & $0$ & $0$ & $\sqrt{5} / 5$\\
$\left( -1, \  0, \  -1\right)$ & $1 / 3$ & $- \sqrt{3} / 6$ & $\sqrt{5} / 30$\\
$\left( 1, \  0, \  1\right)$ & $1 / 3$ & $- \sqrt{3} / 6$ & $\sqrt{5} / 30$\\
$\left( 1, \  0, \  -1\right)$ & $1 / 3$ & $\sqrt{3} / 6$ & $\sqrt{5} / 30$\\
$\left( -1, \  0, \  1\right)$ & $1 / 3$ & $\sqrt{3} / 6$ & $\sqrt{5} / 30$\\
$\left( -1, \  -1, \  -1\right)$ & $0$ & $\sqrt{3} / 6$ & $- \sqrt{5} / 10$\\
$\left( 0, \  -1, \  -1\right)$ & $0$ & $- \sqrt{3} / 6$ & $- \sqrt{5} / 10$\\
$\left( 0, \  -1, \  0\right)$ & $0$ & $\sqrt{3} / 6$ & $- \sqrt{5} / 10$\\
$\left( 0, \  0, \  -1\right)$ & $- 1 / 3$ & $0$ & $\sqrt{5} / 15$\\
$\left( 0, \  0, \  0\right)$ & $1 / 3$ & $0$ & $2 \sqrt{5} / 15$\\
$\left( 0, \  1, \  0\right)$ & $0$ & $\sqrt{3} / 6$ & $- \sqrt{5} / 10$\\
$\left( 1, \  0, \  0\right)$ & $- 1 / 3$ & $0$ & $\sqrt{5} / 15$\\
$\left( 1, \  1, \  0\right)$ & $0$ & $- \sqrt{3} / 6$ & $- \sqrt{5} / 10$\\
$\left( 1, \  1, \  1\right)$ & $0$ & $\sqrt{3} / 6$ & $- \sqrt{5} / 10$\\
$\left( -1, \  -1, \  0\right)$ & $0$ & $- \sqrt{3} / 6$ & $- \sqrt{5} / 10$\\
$\left( -1, \  0, \  0\right)$ & $- 1 / 3$ & $0$ & $\sqrt{5} / 15$\\
$\left( 0, \  0, \  1\right)$ & $- 1 / 3$ & $0$ & $\sqrt{5} / 15$\\
$\left( 0, \  1, \  1\right)$ & $0$ & $- \sqrt{3} / 6$ & $- \sqrt{5} / 10$\\ \bottomrule    
  \end{tabular}
\end{table*}

\clearpage
\section{Table of R-factor coefficients}
\begin{table*}[h!]\centering
\caption{\label{tab:Rfactors}Coefficients defining polarization-angular momentum dependence factors, $R^{(0)}_{0}(\widetilde{\epsilon}; \widetilde{J})$. See Eq.~(40).}
  \begin{tabular}{p{4cm}cccc}\toprule
    Label & $c_{00}$ & $c_{12}$ & $c_{13}$ & $c_{14}$\\\midrule

P2P2P, P2P\ensuremath{{}\mkern1mu\overline{\mkern-1mu\mbox{P}}}, PR\ensuremath{{}\mkern1mu\overline{\mkern-1mu\mbox{P}}}, PRR & $\left(2 J_{i} + 1\right) / \left(2 J_{i} - 1\right)$ & $6$ & $1$ & $1$\\
P2Q2P, P\ensuremath{{}\mkern1mu\overline{\mkern-1mu\mbox{Q}}}\ensuremath{{}\mkern1mu\overline{\mkern-1mu\mbox{2P}}}, PQR, P\ensuremath{{}\mkern1mu\overline{\mkern-1mu\mbox{Q}}}\ensuremath{{}\mkern1mu\overline{\mkern-1mu\mbox{R}}}, Q2P2Q, Q\ensuremath{{}\mkern1mu\overline{\mkern-1mu\mbox{P}}}\ensuremath{{}\mkern1mu\overline{\mkern-1mu\mbox{Q}}}, Q\ensuremath{{}\mkern1mu\overline{\mkern-1mu\mbox{P}}}\ensuremath{{}\mkern1mu\overline{\mkern-1mu\mbox{2Q}}}, QRQ & $- \sqrt{J_{i} - 1} \sqrt{J_{i} + 1} \sqrt{2 J_{i} + 1} / \left(J_{i} \sqrt{2 J_{i} - 1}\right)$ & $3$ & $-2$ & $3$\\
PPP, P2R2R & $1 / \left(J_{i} \left(2 J_{i} - 1\right)\right)$ & $2 J_{i}^{2} + 5 J_{i} + 3$ & $12 J_{i}^{2} - 2$ & $2 J_{i}^{2} - 5 J_{i} + 3$\\
PPQ, P2R2Q, QQP, Q\ensuremath{{}\mkern1mu\overline{\mkern-1mu\mbox{Q}}}\ensuremath{{}\mkern1mu\overline{\mkern-1mu\mbox{2P}}}, Q2Q2R, Q\ensuremath{{}\mkern1mu\overline{\mkern-1mu\mbox{Q}}}\ensuremath{{}\mkern1mu\overline{\mkern-1mu\mbox{R}}} & $1 / J_{i}$ & $2 J_{i} + 3$ & $- 3 J_{i} - 2$ & $3 - 3 J_{i}$\\
PPR, P2R2P, R2P2R, RRP & $1$ & $1$ & $1$ & $6$\\
PP\ensuremath{{}\mkern1mu\overline{\mkern-1mu\mbox{P}}}, P2R\ensuremath{{}\mkern1mu\overline{\mkern-1mu\mbox{P}}} & $1 / \left(J_{i} \left(2 J_{i} - 1\right)\right)$ & $2 J_{i}^{2} + 5 J_{i} + 3$ & $2 J_{i}^{2} - 5 J_{i} + 3$ & $12 J_{i}^{2} - 2$\\
PP\ensuremath{{}\mkern1mu\overline{\mkern-1mu\mbox{Q}}}, P2R\ensuremath{{}\mkern1mu\overline{\mkern-1mu\mbox{Q}}}, QQ\ensuremath{{}\mkern1mu\overline{\mkern-1mu\mbox{P}}}, Q2Q\ensuremath{{}\mkern1mu\overline{\mkern-1mu\mbox{P}}}, Q\ensuremath{{}\mkern1mu\overline{\mkern-1mu\mbox{Q}}}2P, Q\ensuremath{{}\mkern1mu\overline{\mkern-1mu\mbox{Q}}}R & $1 / J_{i}$ & $2 J_{i} + 3$ & $3 - 3 J_{i}$ & $- 3 J_{i} - 2$\\
PP\ensuremath{{}\mkern1mu\overline{\mkern-1mu\mbox{R}}}, P2R\ensuremath{{}\mkern1mu\overline{\mkern-1mu\mbox{R}}}, R2P\ensuremath{{}\mkern1mu\overline{\mkern-1mu\mbox{P}}}, RR\ensuremath{{}\mkern1mu\overline{\mkern-1mu\mbox{P}}} & $1$ & $1$ & $6$ & $1$\\
PQQ, P2Q2Q & $- \left(2 J_{i} + 1\right) / \left(J_{i} \left(2 J_{i} - 1\right)\right)$ & $3 J_{i} + 3$ & $3 J_{i} - 2$ & $3 - 2 J_{i}$\\
PQ\ensuremath{{}\mkern1mu\overline{\mkern-1mu\mbox{P}}}, P2Q\ensuremath{{}\mkern1mu\overline{\mkern-1mu\mbox{P}}} & $- \left(2 J_{i} + 1\right) / \left(J_{i} \left(2 J_{i} - 1\right)\right)$ & $3 J_{i} + 3$ & $3 - 2 J_{i}$ & $3 J_{i} - 2$\\
PQ\ensuremath{{}\mkern1mu\overline{\mkern-1mu\mbox{Q}}}, P2Q\ensuremath{{}\mkern1mu\overline{\mkern-1mu\mbox{Q}}}, P\ensuremath{{}\mkern1mu\overline{\mkern-1mu\mbox{Q}}}Q, P\ensuremath{{}\mkern1mu\overline{\mkern-1mu\mbox{Q}}}2Q, Q2P\ensuremath{{}\mkern1mu\overline{\mkern-1mu\mbox{P}}}, Q\ensuremath{{}\mkern1mu\overline{\mkern-1mu\mbox{P}}}2P, Q\ensuremath{{}\mkern1mu\overline{\mkern-1mu\mbox{P}}}R, QR\ensuremath{{}\mkern1mu\overline{\mkern-1mu\mbox{P}}} & $- \sqrt{J_{i} - 1} \sqrt{J_{i} + 1} \sqrt{2 J_{i} + 1} / \left(J_{i} \sqrt{2 J_{i} - 1}\right)$ & $3$ & $3$ & $-2$\\
P\ensuremath{{}\mkern1mu\overline{\mkern-1mu\mbox{P}}}2P, P\ensuremath{{}\mkern1mu\overline{\mkern-1mu\mbox{P}}}R & $\left(2 J_{i} + 1\right) / \left(2 J_{i} - 1\right)$ & $1$ & $6$ & $1$\\
P\ensuremath{{}\mkern1mu\overline{\mkern-1mu\mbox{P}}}P, P\ensuremath{{}\mkern1mu\overline{\mkern-1mu\mbox{P}}}2R & $1 / \left(J_{i} \left(2 J_{i} - 1\right)\right)$ & $2 J_{i}^{2} - 5 J_{i} + 3$ & $2 J_{i}^{2} + 5 J_{i} + 3$ & $12 J_{i}^{2} - 2$\\
P\ensuremath{{}\mkern1mu\overline{\mkern-1mu\mbox{P}}}Q, P\ensuremath{{}\mkern1mu\overline{\mkern-1mu\mbox{P}}}2Q & $\left(2 J_{i} + 1\right) / \left(J_{i} \left(2 J_{i} - 1\right)\right)$ & $2 J_{i} - 3$ & $- 3 J_{i} - 3$ & $2 - 3 J_{i}$\\
P\ensuremath{{}\mkern1mu\overline{\mkern-1mu\mbox{P}}}\ensuremath{{}\mkern1mu\overline{\mkern-1mu\mbox{2P}}}, P\ensuremath{{}\mkern1mu\overline{\mkern-1mu\mbox{P}}}\ensuremath{{}\mkern1mu\overline{\mkern-1mu\mbox{R}}} & $\left(2 J_{i} + 1\right) / \left(2 J_{i} - 1\right)$ & $1$ & $1$ & $6$\\
P\ensuremath{{}\mkern1mu\overline{\mkern-1mu\mbox{P}}}\ensuremath{{}\mkern1mu\overline{\mkern-1mu\mbox{P}}}, P\ensuremath{{}\mkern1mu\overline{\mkern-1mu\mbox{P}}}\ensuremath{{}\mkern1mu\overline{\mkern-1mu\mbox{2R}}} & $1 / \left(J_{i} \left(2 J_{i} - 1\right)\right)$ & $2 J_{i}^{2} - 5 J_{i} + 3$ & $12 J_{i}^{2} - 2$ & $2 J_{i}^{2} + 5 J_{i} + 3$\\
P\ensuremath{{}\mkern1mu\overline{\mkern-1mu\mbox{P}}}\ensuremath{{}\mkern1mu\overline{\mkern-1mu\mbox{Q}}}, P\ensuremath{{}\mkern1mu\overline{\mkern-1mu\mbox{P}}}\ensuremath{{}\mkern1mu\overline{\mkern-1mu\mbox{2Q}}} & $\left(2 J_{i} + 1\right) / \left(J_{i} \left(2 J_{i} - 1\right)\right)$ & $2 J_{i} - 3$ & $2 - 3 J_{i}$ & $- 3 J_{i} - 3$\\
P\ensuremath{{}\mkern1mu\overline{\mkern-1mu\mbox{Q}}}P, P\ensuremath{{}\mkern1mu\overline{\mkern-1mu\mbox{Q}}}2R, Q2P\ensuremath{{}\mkern1mu\overline{\mkern-1mu\mbox{Q}}}, Q\ensuremath{{}\mkern1mu\overline{\mkern-1mu\mbox{P}}}Q, Q\ensuremath{{}\mkern1mu\overline{\mkern-1mu\mbox{P}}}2Q, QR\ensuremath{{}\mkern1mu\overline{\mkern-1mu\mbox{Q}}} & $- 1 / J_{i}$ & $3 J_{i} - 3$ & $- 2 J_{i} - 3$ & $3 J_{i} + 2$\\
P\ensuremath{{}\mkern1mu\overline{\mkern-1mu\mbox{Q}}}\ensuremath{{}\mkern1mu\overline{\mkern-1mu\mbox{Q}}}, P\ensuremath{{}\mkern1mu\overline{\mkern-1mu\mbox{Q}}}\ensuremath{{}\mkern1mu\overline{\mkern-1mu\mbox{2Q}}}, Q2P2P, Q\ensuremath{{}\mkern1mu\overline{\mkern-1mu\mbox{P}}}\ensuremath{{}\mkern1mu\overline{\mkern-1mu\mbox{P}}}, Q\ensuremath{{}\mkern1mu\overline{\mkern-1mu\mbox{P}}}\ensuremath{{}\mkern1mu\overline{\mkern-1mu\mbox{2R}}}, QRR & $- 1 / J_{i}$ & $3 J_{i} - 3$ & $3 J_{i} + 2$ & $- 2 J_{i} - 3$\\
P\ensuremath{{}\mkern1mu\overline{\mkern-1mu\mbox{R}}}P, P\ensuremath{{}\mkern1mu\overline{\mkern-1mu\mbox{R}}}\ensuremath{{}\mkern1mu\overline{\mkern-1mu\mbox{2P}}}, P\ensuremath{{}\mkern1mu\overline{\mkern-1mu\mbox{R}}}2R, P\ensuremath{{}\mkern1mu\overline{\mkern-1mu\mbox{R}}}\ensuremath{{}\mkern1mu\overline{\mkern-1mu\mbox{R}}}, R\ensuremath{{}\mkern1mu\overline{\mkern-1mu\mbox{P}}}2P, R\ensuremath{{}\mkern1mu\overline{\mkern-1mu\mbox{P}}}\ensuremath{{}\mkern1mu\overline{\mkern-1mu\mbox{P}}}, R\ensuremath{{}\mkern1mu\overline{\mkern-1mu\mbox{P}}}R, R\ensuremath{{}\mkern1mu\overline{\mkern-1mu\mbox{P}}}\ensuremath{{}\mkern1mu\overline{\mkern-1mu\mbox{2R}}} & $1$ & $6$ & $1$ & $1$\\
Q2Q2P, Q\ensuremath{{}\mkern1mu\overline{\mkern-1mu\mbox{Q}}}\ensuremath{{}\mkern1mu\overline{\mkern-1mu\mbox{P}}}, QQR, Q\ensuremath{{}\mkern1mu\overline{\mkern-1mu\mbox{Q}}}\ensuremath{{}\mkern1mu\overline{\mkern-1mu\mbox{2R}}}, R2P2Q, RRQ & $1 / \left(J_{i} + 1\right)$ & $2 J_{i} - 1$ & $- 3 J_{i} - 1$ & $- 3 J_{i} - 6$\\
QPP, Q\ensuremath{{}\mkern1mu\overline{\mkern-1mu\mbox{R}}}\ensuremath{{}\mkern1mu\overline{\mkern-1mu\mbox{2P}}}, Q2R2R, Q\ensuremath{{}\mkern1mu\overline{\mkern-1mu\mbox{R}}}\ensuremath{{}\mkern1mu\overline{\mkern-1mu\mbox{R}}}, R\ensuremath{{}\mkern1mu\overline{\mkern-1mu\mbox{Q}}}\ensuremath{{}\mkern1mu\overline{\mkern-1mu\mbox{Q}}}, R\ensuremath{{}\mkern1mu\overline{\mkern-1mu\mbox{Q}}}\ensuremath{{}\mkern1mu\overline{\mkern-1mu\mbox{2Q}}} & $- 1 / \left(J_{i} + 1\right)$ & $3 J_{i} + 6$ & $3 J_{i} + 1$ & $1 - 2 J_{i}$\\
QPQ, Q2R2Q, Q\ensuremath{{}\mkern1mu\overline{\mkern-1mu\mbox{R}}}\ensuremath{{}\mkern1mu\overline{\mkern-1mu\mbox{Q}}}, Q\ensuremath{{}\mkern1mu\overline{\mkern-1mu\mbox{R}}}\ensuremath{{}\mkern1mu\overline{\mkern-1mu\mbox{2Q}}}, RQP, R\ensuremath{{}\mkern1mu\overline{\mkern-1mu\mbox{Q}}}\ensuremath{{}\mkern1mu\overline{\mkern-1mu\mbox{P}}}, R2Q2R, R\ensuremath{{}\mkern1mu\overline{\mkern-1mu\mbox{Q}}}\ensuremath{{}\mkern1mu\overline{\mkern-1mu\mbox{2R}}} & $- \sqrt{J_{i}} \sqrt{J_{i} + 2} \sqrt{2 J_{i} + 1} / \left(\left(J_{i} + 1\right) \sqrt{2 J_{i} + 3}\right)$ & $3$ & $-2$ & $3$\\
QP\ensuremath{{}\mkern1mu\overline{\mkern-1mu\mbox{Q}}}, Q2R\ensuremath{{}\mkern1mu\overline{\mkern-1mu\mbox{Q}}}, Q\ensuremath{{}\mkern1mu\overline{\mkern-1mu\mbox{R}}}Q, Q\ensuremath{{}\mkern1mu\overline{\mkern-1mu\mbox{R}}}2Q, R\ensuremath{{}\mkern1mu\overline{\mkern-1mu\mbox{Q}}}2P, R\ensuremath{{}\mkern1mu\overline{\mkern-1mu\mbox{Q}}}R & $- 1 / \left(J_{i} + 1\right)$ & $3 J_{i} + 6$ & $1 - 2 J_{i}$ & $3 J_{i} + 1$\\
QP\ensuremath{{}\mkern1mu\overline{\mkern-1mu\mbox{R}}}, Q\ensuremath{{}\mkern1mu\overline{\mkern-1mu\mbox{R}}}P, Q2R\ensuremath{{}\mkern1mu\overline{\mkern-1mu\mbox{R}}}, Q\ensuremath{{}\mkern1mu\overline{\mkern-1mu\mbox{R}}}2R, RQ\ensuremath{{}\mkern1mu\overline{\mkern-1mu\mbox{Q}}}, R2Q\ensuremath{{}\mkern1mu\overline{\mkern-1mu\mbox{Q}}}, R\ensuremath{{}\mkern1mu\overline{\mkern-1mu\mbox{Q}}}Q, R\ensuremath{{}\mkern1mu\overline{\mkern-1mu\mbox{Q}}}2Q & $- \sqrt{J_{i}} \sqrt{J_{i} + 2} \sqrt{2 J_{i} + 1} / \left(\left(J_{i} + 1\right) \sqrt{2 J_{i} + 3}\right)$ & $3$ & $3$ & $-2$\\
QQQ, Q2Q2Q, Q\ensuremath{{}\mkern1mu\overline{\mkern-1mu\mbox{Q}}}\ensuremath{{}\mkern1mu\overline{\mkern-1mu\mbox{Q}}}, Q\ensuremath{{}\mkern1mu\overline{\mkern-1mu\mbox{Q}}}\ensuremath{{}\mkern1mu\overline{\mkern-1mu\mbox{2Q}}} & $1 / \left(J_{i} \left(J_{i} + 1\right)\right)$ & $4 J_{i}^{2} + 4 J_{i} - 3$ & $4 J_{i}^{2} + 4 J_{i} + 2$ & $4 J_{i}^{2} + 4 J_{i} - 3$\\
QQ\ensuremath{{}\mkern1mu\overline{\mkern-1mu\mbox{Q}}}, Q2Q\ensuremath{{}\mkern1mu\overline{\mkern-1mu\mbox{Q}}}, Q\ensuremath{{}\mkern1mu\overline{\mkern-1mu\mbox{Q}}}Q, Q\ensuremath{{}\mkern1mu\overline{\mkern-1mu\mbox{Q}}}2Q & $1 / \left(J_{i} \left(J_{i} + 1\right)\right)$ & $4 J_{i}^{2} + 4 J_{i} - 3$ & $4 J_{i}^{2} + 4 J_{i} - 3$ & $4 J_{i}^{2} + 4 J_{i} + 2$\\
Q\ensuremath{{}\mkern1mu\overline{\mkern-1mu\mbox{Q}}}P, QQ\ensuremath{{}\mkern1mu\overline{\mkern-1mu\mbox{R}}}, Q2Q\ensuremath{{}\mkern1mu\overline{\mkern-1mu\mbox{R}}}, Q\ensuremath{{}\mkern1mu\overline{\mkern-1mu\mbox{Q}}}2R, R2P\ensuremath{{}\mkern1mu\overline{\mkern-1mu\mbox{Q}}}, RR\ensuremath{{}\mkern1mu\overline{\mkern-1mu\mbox{Q}}} & $1 / \left(J_{i} + 1\right)$ & $2 J_{i} - 1$ & $- 3 J_{i} - 6$ & $- 3 J_{i} - 1$\\
R2P2P, RRR & $1 / \left(\left(J_{i} + 1\right) \left(2 J_{i} + 3\right)\right)$ & $2 J_{i}^{2} - J_{i}$ & $12 J_{i}^{2} + 24 J_{i} + 10$ & $2 J_{i}^{2} + 9 J_{i} + 10$\\
R2P\ensuremath{{}\mkern1mu\overline{\mkern-1mu\mbox{R}}}, RR\ensuremath{{}\mkern1mu\overline{\mkern-1mu\mbox{R}}} & $1 / \left(\left(J_{i} + 1\right) \left(2 J_{i} + 3\right)\right)$ & $2 J_{i}^{2} - J_{i}$ & $2 J_{i}^{2} + 9 J_{i} + 10$ & $12 J_{i}^{2} + 24 J_{i} + 10$\\
RPP, RP\ensuremath{{}\mkern1mu\overline{\mkern-1mu\mbox{R}}}, R2R2R, R2R\ensuremath{{}\mkern1mu\overline{\mkern-1mu\mbox{R}}} & $\left(2 J_{i} + 1\right) / \left(2 J_{i} + 3\right)$ & $6$ & $1$ & $1$\\
RQQ, R2Q2Q & $- \left(2 J_{i} + 1\right) / \left(\left(J_{i} + 1\right) \left(2 J_{i} + 3\right)\right)$ & $3 J_{i}$ & $3 J_{i} + 5$ & $- 2 J_{i} - 5$\\
RQ\ensuremath{{}\mkern1mu\overline{\mkern-1mu\mbox{R}}}, R2Q\ensuremath{{}\mkern1mu\overline{\mkern-1mu\mbox{R}}} & $- \left(2 J_{i} + 1\right) / \left(\left(J_{i} + 1\right) \left(2 J_{i} + 3\right)\right)$ & $3 J_{i}$ & $- 2 J_{i} - 5$ & $3 J_{i} + 5$\\
R\ensuremath{{}\mkern1mu\overline{\mkern-1mu\mbox{R}}}2P, R\ensuremath{{}\mkern1mu\overline{\mkern-1mu\mbox{R}}}R & $1 / \left(\left(J_{i} + 1\right) \left(2 J_{i} + 3\right)\right)$ & $2 J_{i}^{2} + 9 J_{i} + 10$ & $2 J_{i}^{2} - J_{i}$ & $12 J_{i}^{2} + 24 J_{i} + 10$\\
R\ensuremath{{}\mkern1mu\overline{\mkern-1mu\mbox{R}}}P, R\ensuremath{{}\mkern1mu\overline{\mkern-1mu\mbox{R}}}2R & $\left(2 J_{i} + 1\right) / \left(2 J_{i} + 3\right)$ & $1$ & $6$ & $1$\\
R\ensuremath{{}\mkern1mu\overline{\mkern-1mu\mbox{R}}}Q, R\ensuremath{{}\mkern1mu\overline{\mkern-1mu\mbox{R}}}2Q & $\left(2 J_{i} + 1\right) / \left(\left(J_{i} + 1\right) \left(2 J_{i} + 3\right)\right)$ & $2 J_{i} + 5$ & $- 3 J_{i}$ & $- 3 J_{i} - 5$\\
R\ensuremath{{}\mkern1mu\overline{\mkern-1mu\mbox{R}}}\ensuremath{{}\mkern1mu\overline{\mkern-1mu\mbox{2P}}}, R\ensuremath{{}\mkern1mu\overline{\mkern-1mu\mbox{R}}}\ensuremath{{}\mkern1mu\overline{\mkern-1mu\mbox{R}}} & $1 / \left(\left(J_{i} + 1\right) \left(2 J_{i} + 3\right)\right)$ & $2 J_{i}^{2} + 9 J_{i} + 10$ & $12 J_{i}^{2} + 24 J_{i} + 10$ & $2 J_{i}^{2} - J_{i}$\\
R\ensuremath{{}\mkern1mu\overline{\mkern-1mu\mbox{R}}}\ensuremath{{}\mkern1mu\overline{\mkern-1mu\mbox{P}}}, R\ensuremath{{}\mkern1mu\overline{\mkern-1mu\mbox{R}}}\ensuremath{{}\mkern1mu\overline{\mkern-1mu\mbox{2R}}} & $\left(2 J_{i} + 1\right) / \left(2 J_{i} + 3\right)$ & $1$ & $1$ & $6$\\
R\ensuremath{{}\mkern1mu\overline{\mkern-1mu\mbox{R}}}\ensuremath{{}\mkern1mu\overline{\mkern-1mu\mbox{Q}}}, R\ensuremath{{}\mkern1mu\overline{\mkern-1mu\mbox{R}}}\ensuremath{{}\mkern1mu\overline{\mkern-1mu\mbox{2Q}}} & $\left(2 J_{i} + 1\right) / \left(\left(J_{i} + 1\right) \left(2 J_{i} + 3\right)\right)$ & $2 J_{i} + 5$ & $- 3 J_{i} - 5$ & $- 3 J_{i}$ \\ \bottomrule
\end{tabular}
\end{table*}